\documentclass[11pt]{article}

\usepackage{setspace,graphicx,amssymb,amsmath,mathtools,latexsym,amsfonts,amscd,amsthm,multirow,ctable,mathdots,caption,array,dsfont}
\usepackage{fancyhdr,tabularx,cite,mathrsfs}
\usepackage{stmaryrd}
\usepackage{rotating}
\usepackage{authblk}
\usepackage{hyperref}
\usepackage{comment,enumitem}
\usepackage{geometry}
\usepackage{arydshln}

\newcommand{\PSL}{\operatorname{\textsl{PSL}}}    
\newcommand{\SL}{\operatorname{\textsl{SL}}}      
\newcommand{\GL}{\operatorname{\textsl{GL}}}      
\newcommand{\SO}{\operatorname{\textsl{SO}}}      
\newcommand{\UU}{\operatorname{\textsl{U}}}      
\newcommand{\Co}{\operatorname{\textsl{Co}}}      
\newcommand{\Th}{\operatorname{\textsl{Th}}}      
\newcommand{\End}{\operatorname{\text{End}}}      
\newcommand{\Aut}{\operatorname{\text{Aut}}}      

\newcommand{\eg}{{\bf EG}}

\newcommand*{\bcdot}{\raisebox{-0.6ex}{\scalebox{2}{$\cdot$}}}
\newcommand{\myparagraph}[1]{\paragraph{#1}\mbox\\}

\newcommand{\wk}{{\rm wk}}
\newcommand{\CC}{\mathbb{C}}
\newcommand{\RR}{\mathbb{R}}
\newcommand{\QQ}{\mathbb{Q}}

\newcommand{\HH}{\mathbb{H}}
\newcommand{\ZZ}{\mathbb{Z}}

\newcommand{\PP}{\mathbb{P}}
\newcommand{\MM}{\mathbb{M}}

\newcommand{\Cl}{\text{Cl}}

\newcommand{\C}{\mathcal{C}}
\newcommand{\E}{\mathcal{E}}

\newcommand{\M}{\mathcal{M}}

\newcommand{\Z}{\mathbb{Z}}
\newcommand{\xmod}{{\rm \;mod\;}}

\newcommand{\RS}{\widehat{\CC}}
\newcommand{\vir}{\mathfrak{Vir}}
\newcommand{\witt}{\mathfrak{Witt}}

\def\a{\alpha}
\def\b{\beta}
\def\c{\gamma}
\def\d{\delta}

\def\e{\epsilon}
\def\f{\phi}

\def\im{\mathrm{Im}}

\def\inf{\infty}

\def\l{\lambda}
\def\m{\mu}
\def\n{\nu}

\def\p{\pi}
\def\pa{\partial}

\def\r{\rho}
\def\rr{\rho}
\def\s{\sigma}
\def\t{\tau}

\def\th{\theta}
\def\til{\tilde}

\def\x{\chi}

\def\D{\Delta}

\def\L{\Lambda}
\def\O{\Omega}

\def\Tr{\tr}

\newcommand{\normord}[1]{\vcentcolon\mathrel{#1}\vcentcolon}

\newcommand{\ex}{\operatorname{e}} 
\newcommand{\sk}{{\rm sk}}

\newcommand{\tr}{\operatorname{Tr}}
\newcommand{\str}{\operatorname{Str}}

\newtheorem{thm}{Theorem}[section]
\newtheorem{conj}[thm]{Conjecture}

\newcommand{\bea}{\begin{eqnarray}}
\newcommand{\eea}{\end{eqnarray}}
\newcommand{\bee}{\begin{eqnarray*}}
\newcommand{\eee}{\end{eqnarray*}}
\newcommand{\al}{\begin{align*}}
\newcommand{\eal}{\end{align*}}
\newcommand{\be}{\begin{equation}}
\newcommand{\ee}{\end{equation}}
\newcommand{\eq}[1]{(\ref{#1})}
\newcommand{\bem}{\begin{pmatrix}}
\newcommand{\eem}{\end{pmatrix}}

\numberwithin{equation}{section}

\begin{document}

\title{
\vspace{-35pt}
    \textsc{\huge{ TASI Lectures on Moonshine
    }  }
}

\author[1]{Vassilis Anagiannis\thanks{v.anagiannis@uva.nl}}
\author[1,2]{Miranda C. N. Cheng\thanks{mcheng@uva.nl}}

\vspace{35pt}

\affil[1]{Institute of Physics, University of Amsterdam, Amsterdam, the Netherlands}
\affil[2]{Korteweg-de Vries Institute for Mathematics, University of Amsterdam, Amsterdam, the Netherlands}

\date{}

\maketitle

\begin{abstract}
\normalsize
\vspace{10pt}

The word moonshine refers to unexpected relations between the two distinct mathematical structures: finite group representations and modular objects. It is believed that the key to understanding moonshine is through physical theories with special symmetries. Recent years have seen a varieties of new ways in which finite group representations and modular objects can be connected to each other, and these developments have brought promises and also puzzles into the string theory community. 

These lecture notes aim to bring graduate students in theoretical physics and mathematical physics to the forefront of this active research area. 
In Part II of this note, we review the various cases of moonshine connections, ranging from the classical monstrous moonshine established in the last century to the most recent findings. In Part III, we discuss the relation between the moonshine connections and physics, especially string theory. After briefly reviewing a recent physical realisation of monstrous moonshine, we will describe in some details the mystery of the physical aspects of umbral moonshine, and also mention some other setups where string theory black holes can be connected to moonshine. 

To make the exposition self-contained, we also provide the relevant background knowledge in Part I, including sections on finite groups, modular objects, and two-dimensional conformal field theories. This part  occupies half of the pages of this set of notes and can be skipped by readers who are already familiar with the relevant concepts and techniques.  

\end{abstract}

\newpage

\section*{Introduction}

The word moonshine is employed in mathematics to refer to an unexpected relationship between modular objects and representations of finite groups. 
The study of moonshine phenomenon has seen rapid developments in the past five years. 
While the relation is between two mathematical structures, it is expected that the existence of this surprising relation has its origin in  physics,  and in string theory in particular. 
 
\vspace{2pt}
\begin{figure}[h!]
\begin{center}
\includegraphics[scale=0.4]{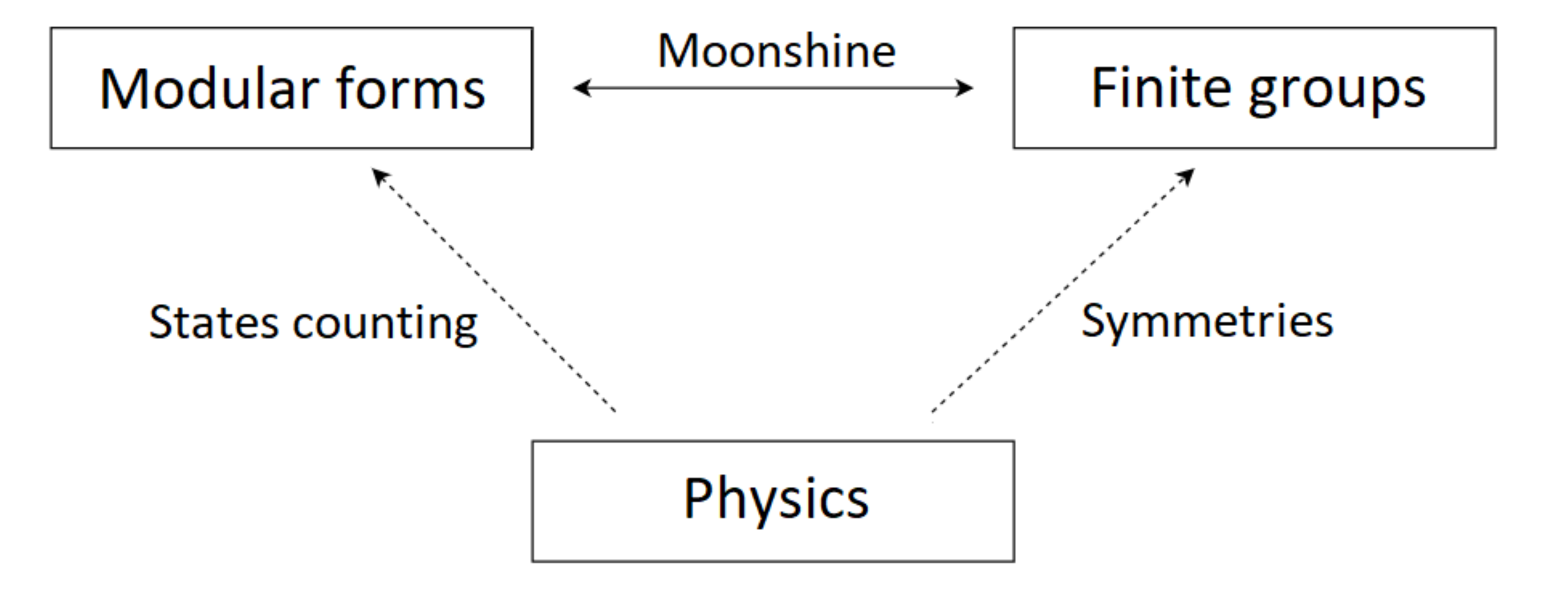}
\end{center}
\end{figure}
\vspace{-8pt}

\subsection*{Moonshine: what, how and why?}

{\bf \em Q: What is moonshine?}
\\\noindent
The structures of modular forms and that of finite group representations have a priori nothing to do with each other. The word moonshine refers to an unexpected relation between them. 
But this simple answer calls for more questions. 
Where does moonshine occur? 
What types of modular objects feature in moonshine relations? 
What types of finite groups can be represented by (mock) modular forms? In what ways? 
Is there a classification possible of such moonshine relations?
Before satisfying answers to these questions are found, the existence of moonshine relation remains to a large extent a mysterious phenomenon. 

\vspace{10pt}

\noindent
{\bf \em Q: How do moonshine relations arise?}\\\noindent
In the classical case of monstrous (and similarly Conway) moonshine, reviewed in \S\ref{cha:weight0}, the relevant group representation has the structure of a vertex operator algebra (VOA). In physical terms, these cases of moonshine can be can be thought of as being ``explained'' by the existence of certain special 2-dimensional conformal field theories (CFTs) that have the relevant finite groups as symmetry groups. 
 
It would be very gratifying to have a similar physical construction for the finite group modules underlying umbral and other recently discovered moonshine relations. In certain ``simpler'' cases this has been achieved (cf. \S\ref{sec:umbral}). Nevertheless, a uniform construction of the umbral moonshine module, reflecting the uniform structure among the twenty-three instances of umbral moonshine, is still absent. Similarly, the modules underlying the various other cases of new moonshines have not been constructed so far.

Apart from the question ``what are these finite group modules?'' there is also a deeper question of ``what kind of (algebraic) structure do these moonshine modules possess?''. 
The {\it mock} modular nature of the modular objects involved in the new cases of  moonshine seems to suggest that a departure from the familiar unitary conformal field theories with discrete spectrum is necessary in order to accommodate the corresponding moonshine modules.  However, the existence of generalised umbral moonshine (cf. \S\ref{sec:umbral}) suggests that certain features resembling those of holomorphic orbifold CFTs should still be present in these new modules.  

Last but not least, the first case of umbral moonshine, often referred to as the Mathieu moonshine, was uncovered in the context of string theory in the background of $K3$ surfaces. Subsequent development established the intimate relation between $K3$ string theory and all twenty-three instances of umbral moonshine. This relation to $K3$ string theory is the topic of \S\ref{chap:moonshineK3}.

To sum up, the origin of the classical monstrous moonshine can be understood as lying in the existence of certain special physical theory -- certain 2d chiral CFTs and the corresponding string theory embedding to be more precise. See \S\ref{cha:weight0} and \S\ref{sec:physics_monstrous}.
The search for the origin of the newer cases of moonshine is still largely a challenging puzzle. 
However, one does expect the answer to again lie in certain physical theories, and most probably arises from the framework of string theory.

\vspace{10pt} 
\noindent
{\bf \em Q: Why should I  care about moonshine? }\\\noindent
Here are a few reasons why the authors, and many other mathematicians and physicists, care about moonshine phenomenon. 
First, theoretical physics is to a large extent about symmetries, and moonshine is to a large extent about hidden symmetries, which (conjecturally) take place in a physical context. There are therefore very good reasons for physicists to care about what is going on in moonshine. More generally speaking, the broad and interdisciplinary nature of the question and its connection to different areas in mathematics and theoretical physics suggests  that the understanding of this mathematical paradigm is likely to spur novel developments in these areas, as was clearly true in the case of the classical monstrous moonshine.
Here we highlight a few examples of such connections, in the context of recent developments: 

\begin{itemize}
\item{
The discovery of Mathieu and umbral moonshine was initiated in the context of  $K3$ string theory, arguably one of the most important examples of string theory compactifications. 
The development of umbral moonshine has always led to new results in the study of automorphic forms and $K3$ geometry, especially in the context of string theory. 
We expect it to also shed light on the structure of BPS states,  non-perturbative black hole quantum states,and on other interesting aspects of the string landscape in the future. See \S\ref{subsec:Mathieu}, \ref{sec:umbral} and \S\ref{chap:moonshineK3}. } 

\item{The discovery of various new moonshine examples (cf. \S\ref{ch:moonshineweight32}), involving various types of finite groups including an infinite family of them, poses a fascinating challenge for theoretical physics to accommodate the corresponding group representations. 
The solution of this (completely well-posed) puzzle is likely to point to novel structures in physics.} 
\item{The connection between certain new examples of moonshine and arithmetic geometry is interesting and constitutes a potentially fruitful path towards new perspectives in number theory. See \S\ref{ch:moonshineweight32}. }
\end{itemize}

Moreover, it is fascinating and puzzling why the two distinct structures of modular objects and finite groups are so deeply connected.
An understanding of the true nature of these connections has the potential to change the way mathematicians think about these objects in a profound way.

\subsection*{About these lecture notes} 

The lecture note is written for the occasion of the TASI 2017 lectures on Moonshine given by the second author.

\subsection*{Acknowledgement}
The second author would like to thank  the organisers of the TASI 2017 school for giving her the opportunity to lecture in this very nice school. We would like to thank Sarah Harrison, Justin Kaidi, Natalie Paquette and Roberto Volpato for comments on an earlier version of the draft. 
Last but not least, the second author would like to thank  Sarah Harrison, Jeff Harvey, Shamit Kachru, Natalie Paquette, Brandon Rayhaun, Roberto Volpato, Max Zimet, and in particular John Duncan for enjoyable collaborations on related topics, and for having taught me so much about this really fun subject! 

\newpage 
\tableofcontents
\newpage 

\part{Background}
In the first part we  briefly summarise the relevant background knowledge in moonshine study, including sections on finite groups, modular objects, and two-dimensional conformal field theories. This part can be safely skipped by readers who are familiar with these topics.

\section{Finite groups and their representations}
In this chapter we briefly review of basic notions of finite groups and  their representations. We mainly follow  \cite{HamermeshMorton1962Gtai,JonesH.F.HughF.1998Grap,webb_2016} (see also \cite{2009arXiv0901.0827E}).

\subsection{Groups}
\label{subsec:Groups}
A group is a set $G$, together with a ``multiplication" operation $\bcdot:G\times G\rightarrow G$, formally denoted as $(G,\bcdot)$. The symbol for this operation is usually implicit, and we often write $ab$  for $a\bcdot b$. A group must satisfy the following axioms:
\begin{enumerate}
\item \textbf{Closure:} $ab=c\in G$ for any $a,b\in G$.
\item \textbf{Associativity:} $(ab)c=a(bc)$ for any $a,b,c\in G$.
\item \textbf{Identity:} There exists a {unique} identity element $e\in G$, such that $eg=ge=g$ for any $g\in G$.
\item \textbf{Inverses:} For every $g\in G$, there exists a  unique inverse element $g^{-1}\in G$, such that $gg^{-1}=g^{-1}g=e$. We also have that $e^{-1}=e$.
\end{enumerate}
Notice that $ab\neq ba$ in general. In the case that $ab=ba$ for every $a,b\in G$, i.e. the group operation is commutative, the group is called \emph{Abelian}. The number of elements of $G$ is called the \emph{order} of $G$, and it can be either finite or infinite. We also define the \emph{order of an element}, $|g|$, to be the minimum number of times we need to multiply it with itself in order to reach the identity, i.e. $g^{|g|}=e$ (the order can also be infinite).

~\\
Next we give a brief summary of a few important notions of group theory.

\myparagraph{Group homomorphisms.}
We say that  a map $\phi:G\rightarrow F$ between two groups $(G,\bcdot)$ and $(F,\star)$ is a {\em group homomorphism} if it preserves the group structure of $G$. In other words, $\phi$ must satisfy
\begin{equation}
\phi(a\bcdot b)=\phi(a)\star\phi(b)~,
\end{equation}
for all $a,b\in G$. If there also exists the inverse homomorphic map $\phi^{-1}:F\rightarrow G$, then $G$ and $F$ are isomorphic; such groups are abstractly the same, but they may still have different realisations. An isomorphism $G\rightarrow G$ is called \emph{automorphism}, and is often called a symmetry of $G$. The set of all automorphisms of $G$, denoted $\Aut G$, forms a group.

\myparagraph{Conjugacy classes.}
Two group elements $a,b\in G$ are said to be conjugate to each other if there exists an element $g\in G$ such that $gag^{-1}=b$. In this case, we symbolically write $a\sim b$. Conjugation is an equivalence relation, since it is reflective ($a\sim a$), symmetric ($a\sim b$ iff $b\sim a$) and transitive (if $a\sim b$ and $b\sim c$ then $a\sim c$). Such a relation implies that $G$ can be split into disjoint subsets $[a]\subset G$, called \emph{conjugacy classes}, each containing all elements that are conjugate to each other:
\begin{equation}
[a]=\left\lbrace b\in G~|~gag^{-1}=b\text{~~for some~~}g\in G\right\rbrace~.
\end{equation}
Obviously, a conjugacy class can be represented by any one of its elements, i.e. $[a]=[b]$ for all $b\sim a$. The number of distinct conjugacy classes is referred to as the \emph{class number} of $G$, denoted here as $\Cl(G)$. All elements of a class have the same order. It is easy to see that  an element  constitutes a conjugacy class of its own if it commutes with all other elements of the group. As a result, in an Abelian group each class contains only one element and the class number equals the order of the group.

A common notation for conjugacy classes is to write the order of its elements, followed by an alphabetical letter. For example, $4A$ denotes a class of order four, $4B$  a different class of order four, $6A$  a class of order six, and so on. The identity is always a class of its own, namely $1A$, the unique class of order one.

\myparagraph{Subgroups.}
A subgroup $H$ is a subset $H\subset G$ which  is itself a group, with  the group structure inherited from $G$. Note that the identity element $e$ always forms a subgroup $\lbrace e\rbrace$ of order $1$, called the trivial subgroup. Subgroups $H$ other than the trivial subgroup and $G$ itself are called \emph{proper subgroups} of $G$, and the notation $H<G$ is used for them (we use the notation $H\leq G$ if we can have $H=G$).

A \emph{normal subgroup} $N$, also denoted as $N\triangleleft G$, is a subgroup of $G$ that is invariant under conjugation by all elements of $G$:
\begin{equation}
N\triangleleft G~\Leftrightarrow~gNg^{-1}=N\text{ for all }g\in G~.
\end{equation}
As such, $N$ is necessarily a union of conjugacy classes. A \emph{maximal normal subgroup} of $G$ is a normal subgroup which is not contained in any other normal subgroup of $G$, apart from $G$ itself. Normal subgroups play a prominent role in quotient groups and group extensions (see below).

The centre $Z(G)$ of a group $G$ is the set of all elements that commute with every other element, i.e.
\begin{equation}
Z(G):=\lbrace a\in G\,\lvert\, ab=ba\text{~~for all } b\in G\rbrace~.
\end{equation}
The centre is always a normal subgroup of $G$. The \emph{centralizer} of an element $g\in G$ is similarly defined by
\begin{equation}
C_G(g)=\lbrace a\in G\,\lvert\, ag=ga\rbrace,~
\end{equation}
being the set of all elements that commute with $g$. Clearly, the centralizer of an element is always a subgroup of $G$.

\myparagraph{Cosets.}
Let $H$ be a subgroup of $G$, and take $g\in G$. We define the \emph{left coset} of $H$ in $G$ with respect to $g$ as the subset
\begin{equation}
gH=\lbrace gh~|~h\in H\rbrace~.
\end{equation}
The set of all left cosets of $H$ in $G$ is denoted by $G/H:=\lbrace gH~|~g\in G\rbrace$. Similarly, the \emph{right coset} of $H$ in $G$ with respect to $g$ is defined as
\begin{equation}
Hg:=\lbrace hg~|~h\in H\rbrace~,
\end{equation}
and the set of all right cosets of $H$ in $G$ is denoted by $H\backslash G:=\lbrace Hg~|~g\in G\rbrace$. 

One can more intuitively define left cosets in terms of an equivalence relation on $G$ (not to be confused with conjugation); namely, for  $a,b\in G$ we set $a\sim b$ iff $ah=b$ for some $h\in H$, i.e. $a$ and $b$ are related by multiplication of an element in $H$ to the right. Then $a,b$ represent the same equivalence class, which is exactly the coset $aH=bH$. All such classes make up $G/H$, which is viewed as a disjoint partition of $G$, as a set. The corresponding statements also hold for right cosets. Some useful facts about cosets include:
\begin{itemize}
\item The number of left cosets is always equal to the number of right cosets, and is known as the \emph{index} of $H$ in $G$, denoted by $[G:H]$.
\item If $G$ is a finite group, then Lagrange's theorem states that the index equals the quotient of the order	of $G$ over the order of $H$, i.e. $[G:H]{|H|}={|G|}$. This is indicatory of how $G$ is partitioned under the coset equivalence relation associated with $H$.
\item  The left and the right cosets of $H$ have the same number of elements, which is equal to the order of $H$.
\item The left and right cosets of a normal subgroup coincide, as can be easily seen from its definition.
\end{itemize}

\myparagraph{Quotient groups and group extensions.}
Cosets, like conjugacy classes, are in general not subgroups. 
However, given a normal subgroup $N$, 
the set $G/N$ of right cosets (which coincides with the set of left cosets) inherits the group structure of $G$, and is  called the \emph{quotient group}. This can be seen from $(aN)(bN) = (ab)N$. The normal subgroup $N$ can then be viewed as the kernel of the homomorphism $\psi:G\rightarrow G/N$.
Note that  in general $G/N$ is not isomorphic to any subgroup of $G$. Moreover, the order of $G/N$ is equal to the index $[G:N]=|G|/|N|$. 

Consider now a \emph{short exact sequence} of groups
\begin{equation}
1\rightarrow M\xrightarrow{\phi} G\xrightarrow{\psi} Q\rightarrow1~. 
\end{equation}
This means that $\phi(M)$, the embedding of $M$ inside $G$, by $\phi$ is the kernel of the homomorphism $\psi$; in other words $M$ is isomorphic to a normal subgroup $N\triangleleft G$, and $Q\cong G/N$. We then say that $G$ is an \emph{extension} of $Q$ by $M$. An extension, as well as the corresponding sequence, is called \emph{split} if  there exists a homomorphism (embedding) $\widetilde{\psi}:Q\rightarrow G$ such that $\psi\circ \widetilde{\psi}=\text{id}_Q$, the identity map on $Q$. We use the semi-direct product to denote such a split extension, $G=N\rtimes Q$. Otherwise, the extension is called \emph{non-split}, and we write $G=N.Q$~.

\subsection{Classification of finite groups}
From now on we focus on finite groups, which are groups with a finite number of elements. The problem of classifying such groups can be reduced to the classification of a finite simple groups. 
A group  is said to be \emph{simple} if it has no proper normal subgroups. 
If $G$ is not simple, then it can always be ``decomposed" into a series of smaller groups, by considering quotients by maximal normal subgroups. To be precise, one can consider the \emph{composition series}, which has the form
\begin{equation}
1\triangleleft N_1\triangleleft N_2\triangleleft\cdots \triangleleft N_{n-1}\triangleleft N_n=G~.
\end{equation}
Here $1$ denotes the trivial group, and every step of the series involves a maximal normal subgroup $N_{i-1}$ of $N_i$, as well as the implied quotient group $N_i/N_{i-1}$. 
It can be shown that all the resulting quotient groups are simple, and the Jordan--H{\"o}lder theorem guarantees that for given $G$, two different composition series lead to the same simple groups. As a result, studying finite simple groups is to a large extent sufficient to understand general finite groups. 

After a heroic effort spanning over half a century and involving more than 100 mathematicians leading to tens of thousands of pages of proof,  
all finite simple groups have been classified (see \cite{GorensteinDaniel1982Fsg:, RonaldSolomon2001Abho} for historical remarks). 
They  belong to one of the following four categories:  cyclic groups $\ZZ_p$ for prime $p$, alternating groups $\mathcal{A}_n$ ($n\geq5$), 16 families of Lie type and 26 {\em sporadic groups}. Unlike the rest of finite simple groups, the 26 sporadic groups appear ``sporadically'' and are not part of infinite families. We will say more about the sporadic groups in the following section.

\subsection{Sporadic groups and lattices}
\label{sec:sporadic_lattices}

The largest sporadic group is the Fischer--Griess  Monster group $\mathbb M$, which gets its name from its enormous size. 
The number of its element is 
\[
|\mathbb M| =  2^{46} \cdot 3^{20} \cdot 5^9 \cdot 7^6 \cdot 11^2 \cdot 13^3 \cdot 17 \cdot 19 \cdot 23 \cdot 29 \cdot 31 \cdot 41 \cdot 47 \cdot 59 \cdot 71 \approx 8 \times 10^{53}~, 
\]
which isroughly the same as the number of atoms in the solar system. 
The Monster contains 20 of the 26 sporadic groups as its subgroups or quotients of subgroups, and these 20 is said to form 3 generations of a {\em happy family} by Robert Griess. In particular, the happy family includes the five Mathieu groups $M_{11}, M_{12}, M_{22}, M_{23}, M_{24}$. They are all subgroups of $M_{24}$, which is  in turn a subgroup of the permutation group $S_{24}$,  and are the  the first sporadic groups that were discovered. 
The rest 6 which are not involved in the Monster are called the {\em pariahs} of sporadic groups.

The sporadic nature of the sporadic groups makes their existence somewhat mysterious and one might wonder what their ``natural'' representations are. 
An important hint is that many of the sporadic groups, especially those connected to the Monster, arise as subgroups of quotients of the automorphic groups of various special lattices. 
The appearance of moonshine involving sporadic groups sheds important light on the question, and the construction of moonshine often relies on the existence of these special lattices. As a result, in what follows we will briefly review the definition of lattices and their root systems, and introduce the special lattices relevant for moonshine.

Let $V$ be a finite-dimensional real vector space of dimension $r$, equipped with an inner product $\langle\cdot,\cdot\rangle$. A finite subset $X\subset V$ of non-zero vectors is called a \emph{root system of rank $r$}, if the following conditions are satisfied
\begin{itemize}
\item $X$ spans $V$.
\item  $X$ is closed under reflections. Namely, $\beta-2\dfrac{\langle\alpha,\beta\rangle}{\langle\alpha,\alpha\rangle}\alpha \in X$  for all $\a, \beta\in X$.
\item The only multiples of $\alpha\in X$ that belong to $X$ are $\alpha$ and $-\alpha$.
\item For all $\alpha,\beta\in X$, we  have  $\dfrac{2\langle\alpha,\beta\rangle}{\langle\alpha,\alpha\rangle}\in\ZZ$.
\end{itemize}
The elements $\alpha\in X$ of a root system are called \emph{roots}.  A root system $X$ is said to be  irreducible if it cannot be partitioned into proper orthogonal subsets $X=X_1\cup X_2$. 
It turns out that the roots of such a system can have at most two possible lengths. If all roots have the same length, then the irreducible root system is called \emph{simply-laced}. 
One can choose a subset $\Phi$ of roots  $f_i\in X$ with $i=1,\dots,r$, such that each root can be written as an integral combination of $f_i\in\Phi$ with either all negative or all positive coefficients. Such a subset is called a set of \emph{simple roots}, and is unique up to the action of the group generated by reflections with respect to all roots, called the {\em Weyl group}  of $X$ and denoted by Weyl$(X)$. 

\begin{figure}
\begin{center}
\caption{The ADE Dynkin diagrams.
\label{fig:dynkin}
}
\vspace{5pt}
\includegraphics[scale=0.25]{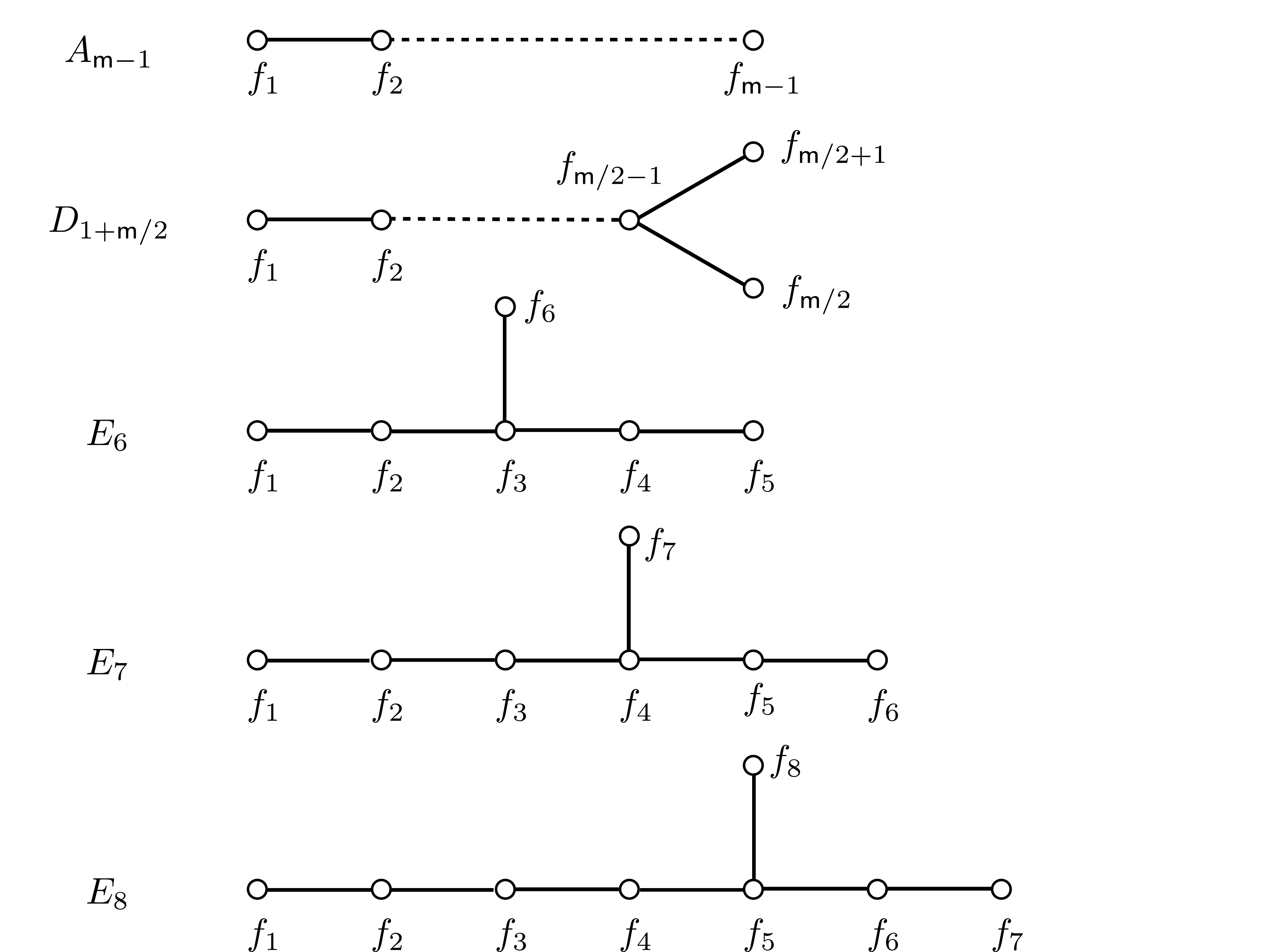}
\end{center}
\end{figure}

To each irreducible root system we can attach a connected \emph{Dynkin diagram}. Each simple root is associated with a node, and nodes associated to two distinct simple roots $f_i,f_j$ are connected with $N_{ij}$ lines, where
\begin{equation}
N_{ij}=\frac{2\langle f_i,f_j\rangle}{\langle f_i,f_i\rangle}\frac{2\langle f_j,f_i\rangle}{\langle f_j,f_j\rangle}\in\lbrace0,1,2,3\rbrace~.
\end{equation}
For simply-laced root systems we only have $N_{ij}\in\lbrace0,1\rbrace$. These  correspond to the Dynkin diagrams of type $A_n,D_n,  E_6, E_7, E_8$ with the subscript denoting the rank of the associated root system, as shown in figure \ref{fig:dynkin}.

Each irreducible root system contains a unique \emph{highest  root} $\theta$ with respect to a given set $\Phi$ of simple roots, whose decomposition
\begin{equation}
\theta=\sum_{i=1}^r a_i f_i
\end{equation}
maximizes the sum $\sum a_i$. The \emph{Coxeter number} of $X$ is then defined by
\begin{equation}
\text{Cox}(X):=1+\sum_{i=1}^ra_i~.
\end{equation}
The Coxeter number can also be defined in terms of Weyl$(X)$. The product of reflections with respect to all simple roots $w=r_{f_1}r_{f_2}\cdots r_{f_r}\in W^X$ is called \emph{Coxeter element}, and its order equals the Coxeter number $m$.

A \emph{lattice} $L$ of rank $n$ is a free Abelian group isomorphic to the additive group $\ZZ^n$, equipped with a symmetric bilinear form $\langle\cdot,\cdot\rangle$. Embedding $L$ into $\RR^n$ gives the picture of a set of points inside the vector space $\RR^n$. A few properties some lattices have that will be useful for us include the following:
\begin{itemize}
\item \emph{Positive-definite}: the bilinear form induces a positive-definite inner product on $\RR^n$.
\item \emph{Integral}: $\langle\lambda,\mu\rangle\in\ZZ$ for all $\lambda,\mu\in L$.
\item \emph{Even}: $\langle\lambda,\lambda\rangle\in2\ZZ$ for all $\lambda\in L$.
\item \emph{Unimodular}: the dual lattice, defined by $L^*:=\lbrace \lambda\in L \otimes_\ZZ \RR |~\langle\lambda,L\rangle\subset\ZZ\rbrace$, is isomorphic to the lattice itself.
\end{itemize}
All elements $\lambda\in L$ such that $\langle\lambda,\lambda\rangle=2$ are called the \emph{roots} of $L$.

Even, unimodular, positive-definite lattices in $24$ dimensions play a distinguished role in several instances of moonshine, as we will discuss in the Part II and Part III of this note.  It was proven by H. V. Niemeier in 1973 that there are only 24 inequivalent such lattices  \cite{Nie_DefQdtFrm24}. One of them, first discovered by J. Leech in 1967 and named the \emph{Leech lattice}, is the only one of the 24 that has no root vectors \cite{Lee_SphPkgs,Con_ChrLeeLat,Lee_SphPkgHgrSpc}.  
The other 23, which we refer to as the {\em Niemeier lattices},  have non-trivial root systems. In fact, one useful construction of the Niemeier lattices is by combining the root lattices with the appropriate ``glue vectors" \cite{sphere_packing}. 
It turns out that the 23 Niemeier lattices are uniquely labelled by the root systems $X$, called the {\em Niemeier root systems}, which are precisely one of the 23 unions of simply-laced (ADE) root systems $X= \cup_i  Y_i$ satisfying the following conditions: 1) All components have the same Coxeter number, ${\rm Cox}(Y_i)={\rm Cox}(Y_j)$; 2) the total rank equals the rank of the lattice $\sum_i {\rm rk} (Y_i)=24$. Some examples out of the 23 include $24A_1$, $2A_{12}$, $8E_8$ and $D_{16}E_8$.

For each of these 24 even, unimodular, positive-definite lattices of rank 24 $N$ we define a finite group 
\be\label{def:niemeier_grp}
G_{N} := {\rm Aut}(N)/{\rm Weyl } (N), 
\ee
where ${\rm Weyl } (N)$ denotes the Weyl group of the root system of $N$. In particular, when $N=\Lambda$ is the Leech lattice, the Weyl group ${\rm Weyl } (\Lambda)$ is the trivial group and $G_{\Lambda} \cong \Co_0$ is the Conway group $\Co_0$. 
By considering the quotients of this group and subgroups stabilising various structures we can obtain many of the sporadic groups. For instance, the sporadic simple group $\Co_1$ is given by the quotient by the centre   $\Co_1\cong \Co_0/\{\pm1\}$, and the Mathieu group $M_{23}$ arises as the subgroup fixing a specific rank-2 sublattice.  See Chapter 10 of \cite{sphere_packing}  for a detailed discussion. 
If instead we choose $N$ to be the Niemeier lattice with root system $24A_1$, for instance, the finite group $G_{N}\cong M_{24}$ is given by the largest Mathieu group. For the Niemeier lattice with root system $12A_2$, the finite group is $2.M_{12}$, the non-trivial extension of the Mathieu group $M_{12}$. 
These groups will play an important role in moonshine (cf. \S\ref{sec:conway_moonshine}, \S\ref{sec:M24moonshine},  \S\ref{sec:umbral}) and in the discussion of their physical context  (cf. \S\ref{chap:moonshineK3}).

\subsection{Representations of finite groups}
\label{subsec:rep}

The structure of groups is just what we need to describe symmetries. To put this into use we need the concept of representations of groups. In what follows we limit our discussion to {\em complex representations}, namely we will consider the group action on a complex vector space $V$. More precisely, consider the group homomorphism $\r:G\rightarrow\GL(V)$. We can think of the images $\r(g)$ as invertible $n\times n$ complex matrices. In particular we have $\rr\left(g^{-1}\right)=(\rr(g))^{-1}$. 
The vector space together with the map $(V,\rr)$ is called a \emph{representation} of dimension $n$. Often one refers to either $V$ or $\rr$ as the representation, while implicitly referring to the full data. The vector space $V$ is also called a $G$-module in this context, and is said to carry a $G$-action. We say that the $G$-action is \emph{faithful}, if no two distinct elements $g,g'\in G$ lead to $\rho(g)=\rho(g')$ (the corresponding representation is also called faithful).

\myparagraph{Irreducible representations and dual representations. }
Given two representations $(V,\rr)$ and $(V',\rr')$ one can define their direct sum and their tensor product in a straightforward way, which leads to new representations $V\oplus V'$ and $V\otimes V'$. 

Two representations $\rr,\rr'$ are \emph{equivalent} if there exists an invertible $n\times n$ matrix $M$ such that $M\rr'(g)=\rr(g)M$ for all $g\in G$. A \emph{subrepresentation} of a representation $(V,\rr)$ is a representation $(U,\rr')$, where $U$ is a subspace $U\subset V$ that is preserved by the action of $G$, and $\rr'$ is the restriction of $\rr$ to $U$. A representation $V$ is said to be \emph{irreducible} if it does not contain any proper subrepresentation, and \emph{indecomposable} if it cannot be written as a direct sum of two (or more) non-zero subrepresentations. For finite groups, these two notions coincide. A representation is called \emph{completely reducible} if it is a direct sum of finitely many irreducible representations, i.e. if it can be fully decomposed into irreducible pieces. An irreducible representation of $G$ can become reducible if we restrict to a subgroup $H<G$, and its decomposition into irreducible representations of $H$ is given by the so-called \emph{branching rules}.

The so-called Maschke's theorem states that all (finite-dimensional) representations of a finite group are always completely reducible. There are two steps for proving this. First we show that a unitary representation is always completely reducible, by using the fact that given an inner product $\lbrace\cdot,\cdot\rbrace: V\times V \to \CC$, the orthogonal complement of $U$ in $V$ is also a subrepresentation if $U$ itself is a subrepresentation of $V$. Next we show that any representation is unitary with respect to the group-invariant inner product
\begin{equation}
\lbrace v,w\rbrace\coloneqq\frac{1}{| G|}\sum_{g\in G}\langle\rr(g)v,\rr(g)w\rangle,~~v,w\in V~,
\end{equation}
which then completes the proof.

We also mention the \emph{dual representation} $\rr^*$ of a representation $\rr$, defined by
\begin{equation}
\label{eq:dual_rep}
\rr^*(g):=\left(\rr\left(g^{-1}\right)\right)^T~,~~g\in G~, 
\end{equation}
which is the natural group action on the dual space $V^*=\End(V,\CC)$. Taking $\rho$ to be unitary, we have   $\rho^*(g)=\overline{\rho(g)}$.  In other words, the dual representation is equivalent to the complex conjugate representation.

\myparagraph{Characters.}
The character $\x_\rr$ of a representation $(V,\rr)$, with $V$ a vector space over $\CC$, is a map $G\rightarrow\CC$ defined by the trace of the representation matrices,
\begin{equation}
\x_\rr(g)\coloneqq\tr(\rr(g))~,~~g\in G.
\end{equation}
We will also often denote this trace by $\tr_V g$.
If $\rr$ is irreducible, $\x_\rr$ is called an irreducible character. Some properties of characters (for finite groups) are summarized below:
\begin{itemize}

\item The character is a \emph{class function}, i.e. $\x_\rr\left(hgh^{-1}\right)=\x_\rr(g)~~\forall~g,h\in G$. This follows directly from the cyclic property of the trace.

\item Two complex representations for a finite group have the same characters if and only if they are equivalent, which can be shown using the orthogonality property discussed below. 

\item The restriction of a character of $G$ to a subgroup $H<G$ is a character of $H$.

\item $\x_\rr\left(g^{-1}\right)=\overline{\x_\rr(g)}$, as follows from the fact that 
all eigenvalues of $\rr(g)$ are $|g|$-th roots of unity.

\item For two representations  $\rr,\rr'$ of $G$ and $g\in G$,  we have:
\begin{equation}
\x_{\rr\oplus\rr'}(g)=\x_{\rr}(g)+\x_{\rr'}(g)~,~~\x_{\rr\otimes\rr'}(g)=\x_{\rr}(g)\x_{\rr'}(g)~,~~\x_{\rr^*}(g)=\overline{\x_{\rr}(g)}~.
\end{equation}
This means that the characters form a commutative and associative algebra.
\end{itemize}

\myparagraph{Orthogonality.}
Due to Schur's orthogonality relations (e.g. \S 4 of \cite{JonesH.F.HughF.1998Grap}), characters of unitary representations are equipped with a Hermitian inner product,
\begin{equation}
\label{eq:orthog_char}
\langle\x_\rr,\x_{\rr'}\rangle=\frac{1}{|G|}\sum_{g\in G}\x_\rr(g)\overline{\x_{\rr'}(g)}~.
\end{equation}
When $\rho$ and $\rho'$ are irreducible representations, one can show that $\langle\x_\rr,\x_{\rr'}\rangle=1$ if the two irreducible representations are equivalent, and it vanishes otherwise. As a result, characters of irreducible representations are orthonormal vectors in the space of class functions. In fact, it is possible to show that they span this space, from which one can conclude the important fact that the number of (inequivalent) irreducible representations equals the number of conjugacy classes  (see for instance \S 3-7 of \cite{HamermeshMorton1962Gtai}). 
Moreover, one can show that there is another orthonormality property,
\begin{equation}
\sum_{\rr}^{}\x_{\rr}(g)~\overline{\x_{\rr}(h)}=\left\lbrace\begin{matrix} |C_G(g)|, & h\in [g] \\ 0, & \text{otherwise} \end{matrix}\right.~,
\end{equation}
where the sum is over inequivalent irreducible representations, and $|C_G(g)|$ denotes the order of the centralizer of $g\in G$, which is equal to the order of the group divided by the number of elements in the conjugacy class $[g]$.

\myparagraph{Character table.}
We have already mentioned that the number of irreducible representations of a finite group $G$ is equal to the number of conjugacy classes of $G$. We can group all characters of $G$ into its \emph{character table}, which is a square table of size $\Cl(G)\times\Cl(G)$, with rows labelling the different irreducible representations and columns labelling the different conjugacy classes. In other words, the $(i,j)$ component of the character table is the character $\x_i(g)$ of the $i$-th irreducible representation, evaluated at any $g$ in the $j$-th conjugacy class . As an example, the character table for the alternating group $\mathcal{A}_6$ is displayed in Table \ref{tbl:char_table_ex}.
\begin{table}
\centering
\begin{tabular}{|c|c|c|c|c|c|c|c|c|}
\hline    &    & 1A & 2A & 3A & 3B & 4A & 5A & 5B \\ \hline \hline
          & 2P & 1A & 1A & 3A & 3B & 2A & 5B & 5A \\ \hline
          & 3P & 1A & 2A & 1A & 1A & 4A & 5B & 5A \\ \hline
          & 5P & 1A & 2A & 3A & 3B & 4A & 1A & 1A \\ \hline  \hline
$\x_1$    &    &  1 &  1 &  1 &  1 & 1 & 1 & 1 \\ \hline
$\x_2$    &    &  5 &  1 &  2 & -1 &-1 & 0 & 0 \\ \hline
$\x_3$    &    &  5 &  1 & -1 &  2 &-1 & 0 & 0 \\ \hline
$\x_4$    &    &  8 &  0 & -1 & -1 & 0 &$A$& $^*A$ \\ \hline
$\x_5$    &    &  8 &  0 & -1 & -1 & 0 &$^*A$& $A$ \\ \hline
$\x_6$    &    &  9 &  1 &  0 &  0 & 1 &-1 &-1 \\ \hline
$\x_7$    &    & 10 & -2 &  1 &  1 & 0 & 0 & 0 \\ \hline
\end{tabular}
\caption{Character table for $\mathcal{A}_6$, where $A=\frac{1-\sqrt{5}}{2}$ and $^*A=\frac{1+\sqrt{5}}{2}$.}
\label{tbl:char_table_ex}
\end{table}
Note that there is an additional piece of information in the above table, the so-called power map. The row starting with $s$P gives the conjugacy classes $[g^s]$.

\myparagraph{Supermodules.}
We say that a $G$-module on a superspace ($\ZZ_2$-graded vector space) is a $G$-supermodule. Explicitly, if $V$ is a $G$-supermodule it has the structure 
\be\label{def:supermodule}
V  = V_{\underline{0}} \oplus V_{\underline{1}}
\ee
where $V_{\underline{0}}$ and $V_{\underline{1}}$ are both $G$-modules. We will sometimes refer to $V$ as a {\em virtual representation} of $G$. 
 The supertrace ${\str}$ is defined to act with a minus sign on the odd subspaces: ${\str}_V g := \tr_{V_{\underline{0}}} g-\tr_{V_{\underline{1}}}g$.

\myparagraph{Cycle shapes and Frame shapes.} 
As the name suggests, an $N$-dimensional \emph{permutation representation} $\rr_p$ of a group $G$ has as its representation matrices $N\times N$ permutation matrices (all elements zero, apart from a single entry of $1$ in each row and column). Given such a representation, to each conjugacy class in such a representation we can associate a \emph{cycle shape}, which encodes the number and type of permutation cycles that elements of this class correspond to. A cycle shape has the general form
\begin{equation}
n_1^{\ell_1}n_2^{\ell_2}\cdots n_r^{\ell_r}~,~~\sum_{s=1}^r\ell_s n_s=N~, 
\end{equation}
where $n_s, \ell_s$ are all positive integers, and $n$ denotes  an \emph{$n$-cycle}, i.e. it represents a permutation of $n$ elements. The exponents  $\ell_s$ count the number of $n_s$-cycles. Clearly, an order $k$ element can only have cycles of size $n_s$ which divides $k$. Note that the cycle shapes can be read directly off the character table, including the power map. 

More generally, we can define the \emph{Frame shape} of $g\in G$ given any representation $\rho$, provided that all characters of $\rho$ are rational numbers. 
Their rationality ensures that if $\lambda$ is an eigenvalue of $g$, then $\lambda^k$ is also an eigenvalue when $k$ is co-prime to $|g|$ \cite{MasonGeoffrey1984Farc}. 
Denoting by  $\l_1, \l_2,\dots ,\l_N$ the $g$-eigenvalues,  then there exists a set of positive integers $i_1,i_2,\dots,i_s$ and a set of non-zero integers $\ell_1,\ell_2,\dots,\ell_s$ with the same cardinality such that 
 \begin{equation}
\det(\mathbf{1}-t\rho(g))=\prod_{i=1}^N(1-t\lambda_i)=\prod_{r=1}^s(1-t^{i_r})^{\ell_r}~.
\end{equation}
Clearly one must have $\sum_{s=1}^r\ell_s i_s=N$ and we call 
$
i_1^{\ell_1}i_2^{\ell_2}\cdots i_r^{\ell_r}
$
the Frame shape of the conjugacy class $[g]$ for the representation $\rho$.

\section{Modular objects}
\label{sec:modular}

In this chapter we will briefly introduce the concept of modular forms and their extensions including  mock modular forms, Jacobi forms, and mock Jacobi forms.

\subsection{Modular forms}
\label{subsec:modular_forms}
One of the standard references on modular forms, which we partially follow here,  is \cite{2008T1om}. 
It is well-known that $\SL_2(\RR)$ acts on the upper-half plane $\HH:=\{\t\in \CC\lvert~ \Im(\t) > 0\}$ by a fractional linear (M{\"o}bius) transformation:
\be
\gamma = \begin{pmatrix} a &b \\ c&d\end{pmatrix} : \HH \to \HH~,~~ \t \mapsto \gamma\t := {a\t+b\over c\t+d}~. 
\ee

In defining  modular forms we consider discrete subgroups of $\SL_2(\RR)$, an important example of which is the modular group $\SL_2(\Z)$.  
It is generated by 
\be
T = \bem 1& 1\\ 0&1\eem~~{\rm and}~~ S = \bem 0& 1\\ -1&0 \eem
\ee
satisfying $(ST)^3=1$ and $S^2=-1$. We will also often work with $\PSL_2(\ZZ)\cong \SL_2(\ZZ)/\{\pm 1\}$, which is also the mapping class group of the torus (cf. \S\ref{sec:CFT}).
We will often also consider the upper-half plane extended by adding the {\em cusps} $\{i\infty\} \cup \QQ$, which $\SL_2(\ZZ)$ acts transitively on as one can see from $\gamma\infty = {a\over c}$.

We start by defining weight zero modular forms on the modular group $\SL_2(\Z)$, which are simply holomorphic functions on $\HH$ that are invariant under the action of $\SL_2(\Z)$: 
\be \label{def0_mmf}
f(\t)  =  f(\gamma\t)\quad \forall~ \gamma \in \SL_2(\Z).
\ee
In particular, $f$ has to be holomorphic as $\t$ approaches the boundary of $\HH$ at the cusps $\{i\infty\} \cup \QQ$. But this  turns out to be too restrictive: basic complex analysis tells us that constants are the only such functions.  
As a result, we would like to further generalise the above definition in the following directions:
\begin{enumerate}
\item {\it Analyticity}:  the function is allowed to have exponential growth near the cusps. Such functions are said to be {\it weakly holomorphic modular forms}. 
\item {\it Weights}: one allows for a scaling factor in the transformation rule. See (\ref{def1_mmf}).  
\item {\it Other Groups}: one replaces $\SL_2(\Z)$ by a general  
$\Gamma < \SL_2(\RR)$ in the transformation property (\ref{def0_mmf}). 
\item {\it Multipliers}: one modifies the transformation rule  (\ref{def0_mmf}) by allowing for a non-trivial character  $\psi : \SL_2(\Z) \to \CC^\ast$. See \eqref{def1_mmf}.
\item {\it Vector-Valued}: instead of $f: \HH \to \CC$ we consider a vector-valued function $f: \HH \to \CC^n$ with $n$ components. 
\end{enumerate}

Of course, the above generalisations can be combined. For instance one can consider a vector-valued modular form with multipliers for a subgroup $\Gamma$ of $\SL_2(\RR)$. Obviously, in the vector-valued case the character $\psi$ is no longer a phase but a matrix. 
Also, the above concepts are not entirely independent. For instance, a component of a vector-valued modular form for $\SL_2(\Z)$ can  be considered as a 
(single-valued) modular form for a subgroup of $\SL_2(\Z)$, and vice versa.

We will first start with the first generalisation and introduce the concept of {\it modular functions}. 
We say that $f:\HH\to \CC$ is a modular function if $f$ is meromorphic in $\HH$, satisfies the transformation rule (\ref{def0_mmf}), and grows like $e^{2\pi i \t m}$ for some \({m> -\infty}\). 
In fact, modular functions form a function field with a single generator, called the {\it Hauptmodul} or {\it principal modulus}. 
This is because the fundamental domain $\SL_2(\Z)\backslash \HH$ is a genus zero Riemann surface when finitely many points are added. 
Writing the upper-half plane with the cusps attached as $\widehat \HH = \HH \cup\{i\infty\} \cup \QQ $, the  Hauptmodul has the property that it is an isomorphism 
between the two spheres $\SL_2(\Z)\backslash \widehat\HH$ and $\widehat \CC$. Clearly, such a Hauptmodul is unique up to M{\"o}bius transformations, or the choice of three points on the sphere. As a result, there is a unique Hauptmodul with the expansion
\be
J(\t) = q^{-1} + O(q)
\ee
near $\t \to i\infty$. Here and in what follows we will write $q:= \ex(\t)$, where $\ex(x):=e^{2\pi i x}$ for $x\in \CC$.  
In terms of the  Eisenstein series and Dedekind eta function (cf. \eq{def:E4E6} and \eq{def:dedeta}), the $J$-function is given by
\be\label{eq:def_J}
J(\tau) = j(\tau) -744 =  {E_4^3(\t)\over  \eta^{24}(\t)}-744. 
\ee
In general, a Hauptmodul can be defined as the generator of the field of modular functions for $\Gamma\leq \SL_2(\RR)$ whenever $\Gamma\backslash \widehat\HH$ is genus zero. 
These Hauptmoduls play an important role in moonshine. 

Apart from the definition given above, there are three other equivalent ways of viewing modular functions. First, due to  (\ref{def0_mmf}) we can view $f$ as a function from the suitably compactified fundamental domain $\SL_2(\Z) \backslash \HH$ to the Riemann sphere $\widehat\CC=\CC\cup \{\infty\}$. Second, due to the relation between  $\SL_2(\Z)$ and rank two lattices we can associate to each $\t$ a complex lattice $\Lambda_\t :=1\cdot \ZZ+\t\cdot \ZZ$, and identify $f$ as a function that associates to each such lattice $\Lambda_\t$ a complex function $f(\t)$, which is moreover invariant under a rescaling of the lattice. The third way, which plays an important role in the a relation between modular forms and 2-dimensional conformal field theories, stems from the interpretation of $\SL_2(\Z)\backslash \HH$ as the complex structure moduli space of a Riemann surface of genus one. 
This can be easily understood from the fact that a torus can be described as the complex plane modulo a rank two lattice, and is therefore up to a scale given by 
$\CC/\Lambda_\t$ for some $\t\in \HH$.
The modular function can then be thought of as associating to each torus a complex number which only depends on its complex structure modulus $\t$. 
In this context, the group $\PSL_2(\Z) := \SL_2(\Z)/\{{\mathds{1}},-{\mathds{1}}\}$ plays the role of the mapping class group of a torus (cf. \S\ref{sec:CFT}), where the $\Z_2=\{{\mathds{1}},-{\mathds{1}}\}$ central subgroup acts trivially on $\HH$.

Next we turn to the second generalisation and introduce modular forms on the modular group $\SL_2(\Z)$ of a general weight $k$. They  are defined as holomorphic functions on $\HH$ that transform under the action of $\SL_2(\Z)$ as: 
\be\label{def1_mmf}
f(\t)  =  ( c\t+d)^{-k} f\left(\frac{a\t+b}{c\t+d}\right)\quad \forall~\bem a& b\\ c&d\eem \in \SL_2(\Z)~.
\ee
From the lattice point of view, we consider complex functions $f$ associated to a lattice $\Lambda$ that scale like $f\mapsto \lambda^{-k}f$ under a rescaling $\Lambda \mapsto \lambda \Lambda$, $\lambda\in\CC$, of the lattice. 
We will consider integral and half-integral weight $k$.\footnote{Clearly, special care needs to be taken when $k$ is half-integral. Strictly speaking, one should work with the metaplectic double cover of $\SL_2(\Z)$. However  we will avoid discussing the subtleties here as they will not cause any difficulty for us. We will refer the reader to \cite{MR1773561} for more details.}

With this definition we start to get some non-trivial examples, even when holomorphicity at the cusps is required. 
For instance the following Eisenstein series
\begin{gather}\label{def:E4E6}
\begin{split}
E_4(\t) &= 1+240 \sum_{n=1}^\infty \frac{n^3 q^n}{1-q^n}= 1+240 \,q +2160\,q^2+\dots \\ 
E_6(\t) &= 1-504 \sum_{n=1}^\infty \frac{n^5 q^n}{1-q^n}= 1-504 \,q -16632\,q^2+\dots
\end{split}
\end{gather}
are examples of modular forms of weight 4 and weight 6, respectively. 
But the definition is still somewhat too restrictive as these two Eisenstein series are all there is:  the ring of modular forms on $\SL_2(\Z)$ is generated freely by $E_4$ and $E_6$. Namely, any modular form of integral weight $k$ can be written (uniquely) as a sum of monomials $E_4^\a E_6^\b$ with $k=4\a+6\b$. We denote the space of modular forms of weight $k$ for group $\Gamma$ by $M_k(\Gamma)$. Among modular forms, the so-called cusp forms are often of special interest. 
We say that a modular form $f$ of weight $k$ is a cusp form if $y^{k/2} f(x+i y)$ is bounded as $y\to \infty$. This condition guarantees that $f$ has vanishing constants in its Fourier coefficients at all cusps. 

In the third type of generalisation, we often encounter the $\SL_2(\ZZ)$ subgroups defined by the following congruences. For a positive integer $N$, we define 
\be\label{def:cong_subgr}
\Gamma_0(N) := \left\{ \bem a& b \\ c&d \eem \in \SL_2(\ZZ) ~\mid c\equiv 0 \xmod{N}  \right\}~. 
\ee
Below we will illustrate the generalisations above with some examples.  

First we consider the Jacobi theta functions.
Consider a 1-dimensional lattice with bilinear form $\langle x,x \rangle =x^2$. The associated theta function is 
\be
\theta_3(\tau)  = \sum_{n\in \Z} q^{n^2/2} \, . 
\ee
This simple function turns out to admit an expression in terms of infinite products 
\be
\theta_3(\tau)  = \sum_{n\in \Z} q^{n^2/2} = \prod_{n=1}^\infty (1-q^n)(1+q^{n-1/2})^2~, 
\ee
and has nice modular properties. 
To describe the modular properties, it is most natural to introduce another two theta functions,
\begin{equation}
\begin{split}
\theta_2(\tau) & = \sum_{n+{1\over 2}\in \Z} q^{n^2/2} =2q^{1/8} \prod_{n=1}^\infty (1-q^n)(1+q^{n})^2~, \\
\theta_4(\tau) & = \sum_{n\in \Z} (-1)^n q^{n^2/2} = \prod_{n=1}^\infty (1-q^n)(1-q^{n-1/2})^2~.
\end{split}
\end{equation}
It turns out that they are the three components of a vector-valued modular form for $\SL_2(\Z)$
\be\label{eqn:transf_theta}
\Theta(\t) := \bem  \theta_2(\tau) \\ \theta_3(\tau) \\ \theta_4(\tau)  \eem,~~   \Theta(\t) = \sqrt{i\over \t} {\cal S}\,   \Theta\left(\!-\frac{1}{\t}\right)  = {\cal T} \,\Theta(\t+1)\, ,
\ee
where
\be
{\cal S} = \bem  0&0&1\\ 0&1&0 \\ 1&0&0 \eem , ~~{\cal T} = \bem  e(-\tfrac{1}{8})&0&0\\ 0&0&1 \\ 0&1&0 \eem  . 
\ee
 
To illustrate the relation between vector-valued modular forms and modular forms for a congruence subgroup, consider $\theta(\tau) := \theta_3(2\tau)$. 
This transforms in the following way as a weight $1/2$ modular form for $\Gamma_0(4)$ with a non-trivial multiplier:
\be
\theta(\tau)  = \left(\frac{c}{d}\right) \epsilon_d ~ (c\t+d)^{-\frac{1}{2}} ~  \theta(\gamma\tau)  
\ee
for all $\gamma\in\Gamma_0(4)$, where
\[ 
\epsilon_d:=\left\lbrace\begin{array}{rl} 1~, & d\equiv1\bmod4 \\ i~, & d\equiv3\bmod4 \end{array} \right.
\]
and the Legendre symbol used above is defined as\footnote{$\kappa$ is said to be a quadratic residue modulo $\lambda$ if $\exists~x\in\mathbb{Z}$ such that $x^2\equiv\kappa\bmod\lambda$.}
\[
\left(\frac{\kappa}{\lambda}\right):=\left\lbrace\begin{array}{rl} +1~, & \text{if~}~\kappa\neq0\bmod\lambda\text{ and $\kappa$ is a quadratic residue modulo $\lambda$} \\ -1~, & \text{if~}~\kappa\neq0\bmod\lambda\text{ and $\kappa$ is not a quadratic residue modulo $\lambda$} \\ 0~, & \text{if~}~\kappa\equiv0\bmod\lambda\end{array}\right.~.
\]
Later we will see that these theta functions can be naturally considered as the specialisation at $z=0$ of  the two-variable Jacobi theta functions, defined either as infinite sums or infinite products:
\begin{equation}
\begin{split}
\theta_1(\tau,z) & = -i \sum_{n+{1\over 2}\in \Z} (-1)^{n-{1\over 2}} y^n q^{n^2/2} \\
&=-iq^{1/8} (y^{1/2}-y^{-1/2})\prod_{n=1}^\infty (1-q^n)(1-yq^{n})(1-y^{-1}q^{n})\, ,\\
\theta_2(\tau,z) & = \sum_{n+{1\over 2}\in \Z} y^n q^{n^2/2} \\
&=(y^{1/2}+y^{-1/2})q^{1/8} \prod_{n=1}^\infty (1-q^n)(1+yq^{n})(1+y^{-1}q^{n}) \, ,\\
 \theta_3(\tau,z)  &= \sum_{n\in \Z} y^n q^{n^2/2} = \prod_{n=1}^\infty (1-q^n)(1+yq^{n-1/2})(1+y^{-1}q^{n-1/2})\, , \\
\theta_4(\tau,z) 
&= \sum_{n\in \Z} (-1)^n y^n q^{n^2/2} = \prod_{n=1}^\infty (1-q^n)(1-yq^{n-1/2})(1-y^{-1}q^{n-1/2})~.
\end{split}
\end{equation}
They transform in the following way. 
 
Let 
\be
\Theta(\t,z): = \bem  \theta_1(\tau,z) \\ \theta_2(\tau,z) \\ \theta_3(\tau,z) \\ \theta_4(\tau,z) \eem~, 
\ee
then we have (cf. \S\ref{subsec:shjacobiforms}),
\be
\Theta(\t,z) =  \sqrt{i\over \t} \, e\big(-\frac{z^2}{2\t}\big) \, {\cal S}'\,   \Theta\big(-{1\over \t},{z\over\t}\big)  ={\cal T}' \,\Theta(\t+1,z)\, ,
\ee
where
\be
{\cal S}' = \bem  i &0&0&0\\ 0&0&0&1\\ 0&0&1&0 \\ 0&1&0&0 \eem, ~~ {\cal T}' = \bem  e({-1/8})&0&0&0\\0& e({-1/8})&0&0\\  0&0&0&1 \\ 0& 0&1&0 \eem . 
\ee
 
Another modular form one frequently encounters is the \emph{Dedekind eta function}
\be\label{def:dedeta}
\eta(\t) = q^{1/24} \prod_{n=1}^\infty (1-q^n)~.
\ee
It is a weight $1/2$ modular form with a non-trivial multiplier, satisfying
\be\label{eta_multiplier}
\eta(\t) = \sqrt{i\over \t} \eta\big(-{1\over\t}\big),~~ \eta(\t) = e\big(-{1\over 24}\big) \,\eta(\t+1). 
\ee
It is related to the theta functions by
\be
\eta(\t)^3 = {1\over 2}\th_2(\t)\th_3(\t)\th_4(\t).
\ee
Its 24-th power $\Delta:=\eta^{24}$ is the familiar weight 12 cusp form for the modular group $\SL_2(\ZZ)$.

\subsection{(Skew-)Holomorphic Jacobi forms}
\label{subsec:shjacobiforms}

In this subsection we collect the definitions of (skew-)holomorphic Jacobi forms. These types of objects play a crucial role in moonshine and its connection to physics. This subsection, consisting mostly of definitions, follows \S3.1 of \cite{omjt} very closely.

We first define {elliptic} forms \cite{Dabholkar:2012nd}. For $m$ an integer define the index $m$ {\em elliptic action} of the group $\ZZ^2$ on functions $\phi:\HH\times\CC\to \CC$ by setting
\begin{gather} \label{elliptic}
(\f\lvert_{m} (\l,\m) )(\t,z) := \ex( m\l^2 \t + 2m\l z) \, \f(\t, z+\l \t +\m)
\end{gather}
for $(\l,\m)\in \ZZ^2$. Say that a smooth function $\phi:\HH\times\CC\to\CC$ is an {\em elliptic form} of index $m$ if $z\mapsto\phi(\t,z)$ is holomorphic. 
Denote by $\E_m$ the space of elliptic forms of index $m$.  Observe that any elliptic form $\phi\in\E_m$ admits a {\em theta-decomposition}
\begin{gather}\label{eqn:jac:thtdec}
\phi(\tau,z)=\sum_{r\xmod 2m} h_r(\tau)\theta_{m,r}(\tau,z)~,
\end{gather}
where the theta series are given by $$\theta_{m,r}(\tau,z):=\sum_{\ell=r\xmod 2m}q^{\ell^2/4m}y^\ell~,$$
for some $2m$ smooth functions $h_r:\HH\to \CC$. To see this, note from $\phi(\tau,z)=\phi(\tau,z+1)$ that we have $\phi(\tau,z)=\sum_{\ell \in\ZZ} c_\ell (\tau)y^\ell $ for some $c_\ell :\HH\to\CC$. Then the identity $\phi|_m(1,0)=\phi$ 
implies that $c_r(\tau)q^{-r^2/4m}$ depends only on $r\xmod 2m$. The $2m$ functions $h_r(\tau):=c_r(\tau)q^{-r^2/4m}$ are precisely the theta-coefficients of $\phi$ in the theta-decomposition.

It will be convenient to regard $h_r$ and $\th_{m,r}$ in (\ref{eqn:jac:thtdec}) as defining $2m$-vector-valued functions $h:=(h_r)_{r\xmod 2m}$ and $\th_m:=(\th_{m,r})_{r\xmod 2m}$. 
Then the theta-decomposition (\ref{eqn:jac:thtdec}) may be more succinctly written as $\phi=h^t\th_m$,
where the superscript $t$ denotes matrix transpose.

It follows from the Poisson summation formula that the vector-valued function $\th_{m}=(\th_{m,r})$ has the following behaviour under $\SL_2(\ZZ)$: 
\begin{gather}\label{transf_theta}
\theta_m\left(-\frac1\tau,\frac{z}\tau\right)\frac{1}{\sqrt{\tau}}\ex\left(-\frac{mz^2}{\tau}\right)
=
{\cal S}\theta_m(\tau,z),\quad
\theta_{m}(\tau+1,z)
={\cal T}\theta_m(\tau,z),
\end{gather}
where ${\cal S}=({\cal S}_{rr'})$ and ${\cal T}=({\cal T}_{rr'})$ are unitary matrices defined for a fixed positive integer $m$, given by ${\cal S}_{rr'}:=\frac1{\sqrt{2m}}\ex\left(-\frac{1}{8}-\frac{rr'}{2m}\right)$ and ${\cal T}_{rr'}:=\ex\left(\frac{r^2}{4m}\right)\delta_{r,r'}$.
(Cf. e.g. \S5 of \cite{eichler_zagier}.) This suggests that we obtain elliptic forms $\f=h^t\theta_m\in\E_m$ with good modular transformation properties $\SL_2(\ZZ)$ by requiring suitable conditions on $h$.

To formulate these notions precisely, define the weight $k$ {\em modular}, and {\em skew-modular} actions of $\SL_2(\ZZ)$ on $\E_m$, for $k$ and $m$ integers, by setting 
\begin{gather}\label{def:Jac_skewJac}
\begin{split}
(\f\lvert_{k,m}\gamma)(\t,z) &:= 
\f\left(\frac{a\t+b}{c\t+d},\frac{z}{c\t+d}\right)
\frac1{(c\t+d)^{{k}}} 
\ex\left(- \frac{c mz^2}{c\t+d}\right)
\\
(\f\lvert^\sk_{k,m}\gamma )(\t,z) &:= 
\f\left(\frac{a\t+b}{c\t+d},\frac{z}{c\t+d}\right)
\frac1{(c\bar\t+d)^{{k}}}\frac{c\bar\t+d}{|c\t+d|}
\ex\left(- \frac{c mz^2}{c\t+d}\right)~,
\end{split}
\end{gather} 
for $\f\in\E_m$ and $\gamma=\left(\begin{smallmatrix} a&b\\c&d\end{smallmatrix}\right) \in \SL_2(\ZZ)$. 

Roughly speaking, a Jacobi form of weight $k$ and index $m$ is an elliptic form $\phi(\t,z)$, holomorphic in the $\t$-variable,  which is moreover invariant under  $\lvert_{k,m} \gamma$ for all $\gamma\in \SL_2(\ZZ)$. Note that ${\phi|_{k,m}\left(\begin{smallmatrix}1&1\\0&1\end{smallmatrix}\right)=\phi}$  implies the expansion
\begin{gather}\label{eqn:jac:hol-DFoucffhol}
\f(\t,z) = \sum_{\substack{D,\ell  \in \ZZ\\D=\ell ^2\xmod 4m}} C_\f(D,\ell ) q^{-D/4m}q^{\ell ^2/4m} y^\ell~,
\end{gather}
where $C_\f(D,\ell ) $ depends only on $\ell \xmod 2m$,
corresponding to the theta decomposition 
\begin{gather}\label{eqn:jac:hol-thtcoeffFoucoeff}
h_r(\tau)=\sum_{\substack{D\in\ZZ\\D=r^2\xmod 4m}}C_\phi(D,r)q^{-D/4m}~.
\end{gather}
The invariance under $\SL_2(\ZZ)$ of $\phi=h^t\theta_m$ leads to the modularity of the vector-valued function $h=(h_r)$. In other words, $h=(h_r)$ transforms as a vector-valued modular form and contains precisely the same information as the Jacobi form. To complete the definition, we also need to specify the growth behaviour of $h(\t)$ near the cusp.
We say that $\phi\in\E_m$, invariant under $\lvert_{k,m} \gamma$ for all $\gamma\in \SL_2(\ZZ)$, is a {\em weak holomorphic/holomorphic/cuspidal holomorphic Jacobi form} if the Fourier coefficients satisfy $C_\f(D,r)=0$ unless $-D+ r'^{2}\geq 0$ for all $r'=r \xmod{2m}$,  
$C_\f(D,r)=0$ for $D > 0$, or $C_{\f}(D,r)=0$ for $D \ge0$, respectively. 
We denote the space of weak holomorphic Jacobi forms of weight $k$ and index $m$ by $J^\wk_{k,m}$. Notice that, at odd weight, applying (\ref{def:Jac_skewJac})  to the case $\gamma = - \big(\begin{smallmatrix} 1& 0 \\ 0&1\end{smallmatrix}\big)$ shows that the Jacobi form must be odd under $z\leftrightarrow -z$. It will therefore be convenient to introduce
\be\label{def:thetahat}
\tilde\theta_{m,r} = \theta_{m,r}-\theta_{m,-r}~.
\ee

We now turn to the closely related skew-holomorphic Jacobi forms. 
An elliptic form $\f\in\E_m$ is called 
a {\em weak skew-holomorphic Jacobi form}
if it meets the following conditions. 
First, its theta-coefficients are anti-holomorphic functions on $\HH$; second, it is invariant for the weight $k$ 
skew-modular action (\ref{def:Jac_skewJac}), so that $\f\lvert_{k,m}^\sk\gamma =\f$ for all $\gamma \in \SL_2(\ZZ)$; finally, $\tau \mapsto \f(\t,z)$ remains bounded as $\Im(\tau)\to \infty$ for fixed $z\in \CC$. Thus a weak skew-holomorphic Jacobi form admits a Fourier expansion of the form
\begin{gather}\label{eqn:jac:hol-DFoucffskw}
\f(\t,z) = 
\sum_{\substack{D,\ell \in \ZZ\\  D=\ell ^2  \xmod 4m} }C_\f(D,\ell )\, 
\bar q^{D/4m} q^{\ell ^2/4m} y^\ell ~,
\end{gather}
for some $2m$ functions $D\mapsto C_\f(D,r)$, and we recover its theta-coefficients by writing
\begin{gather}\label{eqn:jac:hol-thtcoeffFoucoeff-skew}
h_r(\tau)=\sum_{\substack{D\in\ZZ\\D=r^2\xmod 4m}}C_\f(D,r)\bar q^{D/4m}~.
\end{gather}

A weak skew-holomorphic Jacobi form $\f$ is called a {\em skew-holomorphic Jacobi form}, or a {\em cuspidal skew-holomorphic Jacobi form}, when the Fourier coefficients satisfy $C_\f(D,r)=0$ for $D < 0$, or $C_{\f}(D,r)=0$ for $D \le0$, respectively.

We will close this subsection with some examples. 
\begin{itemize}
\item 
Define 
\begin{gather}
\begin{split}
\label{phi01}
\phi_{0,1}(\t,z)&=4 \sum_{i=2,3,4}\big(\frac{\th_i(\t,z)}{\th_i(\t,0)}\big)^2~,  \\
\phi_{-2,1} &= -{\th_1(\t,z)^2\over \eta^6(\t)}~. 
\end{split}
\end{gather}
The ring of weak Jacobi forms of even weight is freely generated by $ \phi_{0,1}$ and $ \phi_{-2,1}$ over the ring of modular forms for $\SL_2(\ZZ)$: 
\be
J^{\wk}_{2k,m} = \sum_{j=0}^{m} M_{2k+2j}(\SL_2(\ZZ)) \,   \phi_{-2,1}^j \phi_{0,1}^{m-j}~. 
\ee

The function $  \phi_{0,1}$ plays an important role in Mathieu and umbral moonshine, since 
\( 2 \phi_{0,1}
\) coincides with the $K3$ elliptic genus {{\bf EG}$(\t,z;K3)$}. See \S\ref{sec:EG} for a definition of the elliptic genus.
\item{In \cite{Cheng:2017dlj} it was shown that the modified elliptic genus of the so-called MSW string \cite{Maldacena:1997de} involves a skew-holomorphic Jacobi form with index specified by the charges of the black hole, provided that the moduli is fixed at their  black hole attractor value.  See \S\ref{sec:other_physics} for more discussions on this.}

As a concrete example, consider a single M5 brane wrapping $\PP^2$. 
The modified elliptic genus of the resulting effective string is given by 
${1\over 2 \eta^6(\tau)}\overline{t_{2,2}(\tau, {\bar z\over 2})}$, where $$t_{2,2}(\t,z) = \sum_{r\xmod{4}} \overline{\theta^1_{2,r}(\t)} \theta_{2,r}(\t,z) $$ is a weight 2 skew-holomorphic Jacobi forms. 
In the above we have used the definition 
\begin{equation}
\theta_{m,r}^1(\tau) := {1\over 2\pi i} {\pa\over\pa z}\theta_{m,r}(\tau,z)\lvert_{z=0}~.
\end{equation}
Note that $\theta^1_{2,1}(\t) = -\theta^1_{2,-1}(\t) = \eta^3(\tau)$ coincides with the shadow of the Mathieu moonshine function \eq{second_def_H}. 

\end{itemize}

\subsection{Mock modular forms}
\label{subsec:MMF}

In this section we introduce mock modular forms and the closely related concept of mock Jacobi forms. We follow the treatment of  \S7.1 of \cite{Dabholkar:2012nd} and \S3.2 of \cite{omjt} closely. The subject, initiated by the legendary mathematician Srinivasa Ramanujan, has a fascinating history. We recommend \cite{zagier_mock} for a short account of it.  

Let $w\in \frac12\ZZ$ and let $h$ be a holomorphic function on $\HH$ with at most exponential growth at all cusps.  
We say that $h$ is a (weakly holomorphic) mock modular form of weight $w$ for a discrete subgroup $\Gamma \leq \SL_2(\RR)$ if there is a modular form $g$ of weight $2-w$ such that the sum $\hat h:= h+ g^\ast$ transforms like a holomorphic modular form of weigh $w$ for $\Gamma$. 
Moreover, we say that $g$ is the {\em shadow} of the mock modular form $h$ and $\hat h$ is its {\em completion}.  
In the above we have used the following definition of $g^\ast$. Writing the Fourier expansion of $g$ as $g(\t) = \sum_{n\geq 0} c_g(n) q^n$, then  
\be \label{def:int_shadow}
g^\ast(\t)  := \overline{c_g(0)} {(-\Im(\t))^{1-w}\over w-1} +\sum_{n>0} (-4\p n)^{w-1}\overline{c_g(n)} q^{-n} \Gamma(1-w,4\pi n \Im(\t))~,
\ee
where $\Gamma(1-w,x) = \int_{x}^{\infty} e^{-t} t^{w} dt$ denotes the incomplete gamma function. 
When $c_g(0)=0$, the above coincides with the so-called {\em non-holomorphic weight $w$ Eichler integral} of $g$, given by
\be\label{def:Eichler_integral}
g^\ast (\t):= (-2)^{w-1}\ex(\tfrac{w-1}{4}) \int_{-\bar \tau}^\inf (\t'+\t)^{-w} \overline{g(-\overline{\t'})} {\rm d}\t'~. 
\ee
Note that 
\begin{gather}\label{eqn:jac:mck-partialgstargbar}
-2i\Im(\t)^w\frac{\pa}{\pa \bar \tau} g^\ast(\t) =  
\overline{g(\t)}~, 
\end{gather}
and hence $\hat h$ is annihilated by the weight $w$ Laplacian $\Delta_w := \Im(\t)^{2-w} \pa_\t \Im(\t)^w \pa_{\bar\t}$. Such functions are called harmonic Maass forms, and one can identify $h$ as the (uniquely defined) holomorphic part of the harmonic Maass form $\hat h$. 
Finally, note that from (\ref{def:int_shadow}) it is obvious that the harmonic Maass form $\hat h$  transforms with a multiplier which is the inverse of that of the modular form $g$.

Just as in the case of usual modular forms, one can generalise the above  definition of mock modular forms in various directions, including  incorporating non-trivial multiplier systems and considering vector-valued mock modular forms. 
Next we turn our attention to a specific type of vector-valued mock modular forms, namely those arising from the so-called mock Jacobi forms. For integers $k$ and $m$, we say that 
an elliptic form $\phi\in\E_m$ is a {\em weak mock Jacobi form} of weight $k$ and index $m$ if the following is true. Write the theta-decomposition of $\phi$ as $\phi=\sum_r {h}_r\th_{m,r}$. 
First, $\tau\mapsto \phi(\t,z)$ 
is bounded as $\Im(\tau)\to \infty$ for every fixed $z\in\CC$; second, all the $h_r$ are holomorphic; finally, there exists 
a skew-holomorphic Jacobi form $\s=\sum_r \overline{g_r}\theta_{m,r}\in S^\sk_{3-k,m}$,
such that $\hat\phi:=\sum_r\hat{h}_r\th_{m,r}$ is invariant for the weight $k$ 
modular action $\lvert_{k,m}$ of $\SL_2(\ZZ)$ on $\E_m$ (cf. \eq{def:Jac_skewJac}) with the definition 
\begin{gather}\label{eqn:jac:mck-hhat}
\hat{h}_r(\t):=h_r(\t)+\frac1{\sqrt{2m}}g_r^*(\t)~. 
\end{gather} 
As was discussed in \cite{Zwegers2008} and analysed carefully in \cite{Dabholkar:2012nd,umstar}, meromorphic Jacobi forms -- what one obtains when relaxing the condition on Jacobi forms to allow for poles at torsion points $z\in \QQ + \QQ \tau $ --  naturally give rise to mock Jacobi forms. In particular, all the mock Jacobi forms featured in umbral moonshine can be viewed as arising from meromorphic Jacobi forms.

From a physical point of view, as demonstrated in a series of recent works, the ``mockness'' of these mock modular objects is often related to the non-compactness of relevant spaces in the theory. See, for instance, \cite{VafaWitten1994,Troost:2010ud,Dabholkar:2012nd,Alexandrov:2012au,UMk3}. 
Let us take 2d CFTs with a non-compact target space as an example.
The non-compactness of the target space often leads to a continuous part of the spectrum. In this case the standard CFT arguments might fail. In particular there could be imperfect pairing between the bosonic and fermionic states in the continuous part of the  spectrum and we could end up with a non-holomorphic BPS index, given by the completion of a mock modular object, as a result.  See for instance \cite{Troost:2010ud,Eguchi:2010cb,Eguchi:2004yi, Ashok:2013pya,Ashok:2011cy,Murthy:2013mya} for details for some specific examples, and see the remark at the end of \S\ref{sec:EG} for a more detailed discussion in the context of elliptic genus. 
Another context in which non-compactness appears and leads to a role for mock modular forms is wall-crossing (when approaching the wall, the distance of the bound black hole centers goes to infinity). The BPS counting of the black hole microstates hence depends on the moduli and correspondingly the countour of integration \cite{Cheng2007a}, and the result of the integration is mock modular \cite{Dabholkar:2012nd}. 

Another source of mock modular forms in physics is the characters of supersymmetric infinite algebras, such  as the ${\cal N}=2$ and ${\cal N}=4$ superconformal algebras mentioned in \S \ref{sec:M24moonshine}. Some more examples can be found in for instance \cite{Kac:2014rra} and references therein. 
Interestingly, as we will explain in \S\ref{sec:umK3}, the mockness of the mock modular form  in  (\ref{second_def_H}) can be seen as either arising from CFT with non-compact target space or as a result to the mockness of characters of the ${\cal N}=4$ superconformal algebra. 

We will end this subsection with a few examples. 
\begin{itemize}
\item
Ramanujan wrote down the following simple-looking Eulerian series in his 1920 letter to Hardy \cite{MR2280843},
\begin{align}\notag
\chi_0(q) &=\sum_{n=0}^{\infty} {q^n \over (1-q^{n+1})(1-q^{n+2})\dots (1-q^{2n})}  = 1+ q+ q^2 +2 \,q^3 +\dots~,\\\notag
  \chi_1(q) &=\sum_{n=0}^{\infty} {q^n \over (1-q^{n+1})(1-q^{n+2})\dots (1-q^{2n}) (1-q^{2n+1})}  = 1+2\, q+ 2\,q^2 +3 \,q^3 +\dots~,
\end{align}
as two of the examples of his {\em mock theta functions} (of order 5). 
In fact, they are closely related to mock Jacobi forms. 

Define $I^{3E_8} =\{1,7\}$, $A=\{1,11,19,29\}$, and 
\be\label{eqn:3E8}
H_1^{3E_8}(\t) = q^{-1/120}(2  \chi_0(q)  -4)~, ~~H_7^{3E_8}(\t) = 2  q^{71/120} \chi_1(q) , 
\ee  
then $(H_r^{3E_8})_{r\in I^{3E_8}}$ is a vector-valued mock modular form of weight $1/2$ for the modular group. 
Its shadow is given by the index $30$ theta functions 
\be
g_r ^{3E_8}= 3 \sum_{a\in A} \theta_{30,a\,r}^{1}. 
\ee

Writing $\widehat H_r^{3E_8}(\t) = H_r^{3E_8} +(g_r^{3E_8})^\ast$, we have 
\begin{align}
\bem \widehat H_1^{3E_8}\\\widehat H_7^{3E_8}\eem(\t+1) =\bem \ex(-{1\over 120}) & 0 \\ 0 &\ex(-{49\over 120}) \eem \bem \widehat H_1^{3E_8}\\\widehat H_7^{3E_8}\eem(\t)
\end{align}
and 
\begin{align}
\bem \widehat H_1^{3E_8}\\\widehat H_7^{3E_8}\eem\left(-{1\over \t}\right) =\tau^{1/2} i^{3/2} \bem {1\over 2}\sqrt{1-{1\over \sqrt{5}}} &  {1\over 2}\sqrt{1+{1\over \sqrt{5}}}\\  {1\over 2}\sqrt{1-{1\over \sqrt{5}}} &- {1\over 2}\sqrt{1-{1\over \sqrt{5}}} \eem \bem \widehat H_1^{3E_8}\\\widehat H_7^{3E_8}\eem(\t)
\end{align}

Moreover, $H_i^{3E_8}$ can be viewed as arising from the theta composition of the  mock Jacobi form $\psi^{3E_8}(\t,z):= \sum_{r\in I^{3E_8}} H_r^{3E_8}\sum_{a\in A}\tilde \theta^{3E_8}_{a\,r}$. More specifically, $\hat \psi^{3E_8}(\t,z):=\sum_{r\in I^{3E_8}} \widehat H_r^{3E_8}\sum_{a\in A}\tilde \theta^{3E_8}_{a\,r}$ is non-holomorphic in $\t$ and transforms as a Jacobi form of weight 1 and index 30. 
As the notation suggests, $H_r^{3E_8}$ encodes the graded dimension of the umbral moonshine module underlying the case of umbral moonshine corresponding to Niemeier lattice $N$ with root system $3E_8$, as we will discuss in \S\ref{sec:umbral}.

\item Let $H: \HH \to \CC$ be given by
\be\label{second_def_H}
H(\t) = \frac{-2 E_2(\t) + 48 F_2^{(2)}(\t)}{\eta(\t)^3}=2 q^{-\frac{1}{8}} \left( -1 + 45\, q+ 231\, q^2 + 770\,q^3\dots \right)~,
\ee
where 
$$E_2 = 1- 24\sum_{n=1}^\infty {nq^n \over 1-q^n }$$ is the weight two  Eisenstein series (which is not a modular form) 
and
$$
F_2^{(2)}(\t)= \sum_{\substack{r>s>0\\ r-s=1\, {\rm mod }\; 2 }} (-1)^{r} \,s \,q^{rs/2} = q+q^2-q^3+q^4+\dots ~.
$$
Note that the first few Fourier coefficients of $H/2$ : 45, 231 770, 2277 , 5796,  coincide with dimensions of certain irreducible representations of the sporadic group $M_{24}$! Indeed, in umbral moonshine $H = H^{24A_1}$ plays the role of the graded dimensions of the underlying $M_{24}$-module. See \ref{subsec:Mathieu}.

This function is a mock modular form with shadow $24 \eta^3(\t)$ (and therefore with a multiplier given by the inverse of that of $\eta^3(\t)$).
In other words,
\be\label{completion_H}
\widehat{H}(\t)=H(\t)+24\, (4{i})^{-1/2} \int_{-\bar \t}^{\infty}(z+\t)^{-1/2}\overline{\eta(-\bar z)^3}{\rm d}z,
\ee
transforms as a weight 1/2 modular form for the modular group $\SL_2(\ZZ)$.

Moreover, the two-variable function \(\psi^{24A_1}(\t,z):= H(\t) \tilde{\theta}_{2,1}$ is a mock Jacobi form of weight one and index two.
This mock Jacobi form can be seen as arising from a meromorphic Jacobi form by subtracting its ``polar part''. 
To see this, consider the weight one index two meromorphic Jacobi form
\be\label{eq:psi_EGK3}
\psi(\t,z) := -2i{\theta_1(\t,2z)\eta^3(\t)\over \theta_1^2(\t,z)}\phi_{0,1}(\t,z)= -i {\theta_1(\t,2z)\eta^3(\t)\over \theta_1^2(\t,z)}{{ \bf EG}}(\t,z;K3)~,
\ee
(cf. (\ref{phi01})) which has a simple pole at $z\in \ZZ + \ZZ \t$. 
Then
the following identity holds,
\be\label{polar:M24}
\psi(\t,z) =  \psi^{24A_1}(\t,z) -24\, {\rm Av}^{(2)}\left[{y+1\over y-1}\right]~. 
\ee
In the above ${\rm Av}^{(m)}$ denotes the index-$m$ averaging operator 
 \[{\rm Av}^{(m)}[F(y)] = \sum_{k\in \ZZ} q^{mk^2}y^{2mk}F(q^ky)~,\] 
with the elliptic symmetry ${\rm Av}^{(m)}[F(y)] \lvert_m(\l,\mu) ={\rm Av}^{(m)}[F(y)]  $ for all ${\l,\m\in \ZZ}$, 
and the second term in \eq{polar:M24} can be interpreted as  the canonical ``polar part'' of the meromorphic Jacobi form  $\varphi$. 

\item{Consider the set of binary quadratic forms with discriminant $D$,
\be
{\cal Q}_D := \left\lbrace Q\left(\begin{matrix} X \\ Y \end{matrix}\right)=AX^2+BXY+CY^2\lvert B^2 -4AC = D\right\rbrace~. 
\ee
This has a natural action of $\SL_2(\Z)$, acting as
\begin{equation}
\left(\begin{matrix} X \\ Y \end{matrix}\right)\mapsto\left(\begin{matrix} a & b \\ c & d \end{matrix}\right)\left(\begin{matrix} X \\ Y \end{matrix}\right)~,
\end{equation}
and we are interested in elements in ${\cal Q}_D$ that are not equivalent under the modular group action. 
For instance, an interesting number is the {\em Hurwitz-Kronecker class number} 
\be\label{def:hurwitz}
H(D):= \sum_{Q\in {\cal Q}_D\backslash \SL_2(\ZZ)} {1\over \omega_Q}~,
\ee
where $\omega_Q$ denotes the order of the $\SL_2(\ZZ)$-subgroup that leaves $Q$ invariant. It takes values in $\{1,2,3\}$.
Roughly speaking, this number counts the number of inequivalent quadratic forms with discriminant $D$, and will play an important role in \S\ref{ch:moonshineweight32} and \S\ref{sec:other_physics}.

To each $Q \in {\cal Q}_D$ there is a unique root $\alpha_D$ (satisfying $Q[1,\alpha_D]=0$) in the upper-half plane. 
Clearly, $J(\alpha_Q)$ is independent of which representative of $Q$ one picks in the  $\SL_2(\ZZ)$-orbit and one can similarly define weighted sums like \eq{def:hurwitz}, but now involving values of polynomials of $J$ evaluated at $\alpha_Q$, referred to as the \emph{traces of singular moduli}.
The generating functions of such quantities (summing over $D$ with the grading factor $q^{cD}$ for some $c$) will turn out to have interesting modular properties. 

For instance, the following generating of the Hurwitz-Kronecker class number, extended by $H(0)=-{1\over 12}$,
\be\label{hurwitz_mock}
{\cal H}(\tau) := \sum_{D\leq 0} H(D)q^{|D|} = -{1\over 12}+{1\over 3}q^3 +{1\over 2}q^4+q^7 + q^8 +\dots
\ee
is a mock modular form of weight ${3\over 2}$ for $\Gamma_0(4)$, with shadow given by the theta function $\theta(\t)$ (see \S\ref{subsec:modular_forms}). 
The mock modularity property of this function is realised very early on  in \cite{MR0429750}. 
This form also has the interesting feature that it is bounded at all cusps and hence has slow growth in its Fourier coefficients, corresponding to the subtle growth of the class numbers. 

The physical relevance of this function will be mentioned in \S\ref{sec:other_physics}. 
This is in fact perhaps the first mock modular form which was  found to play a role in physic \cite{MR1285785,Vafa:1994tf}. 
}
\end{itemize}

\section{Conformal field theory in 2 dimensions}
\label{sec:CFT}

This chapter gives a brief summary of some key ingredients of $2$-dimensional conformal field theories (CFTs), and is in no way meant as a complete exposition. CFTs are relevant for moonshine, since in the cases that are known so far the corresponding moonshine modules feature vertex operator algebra (VOA) structures, which capture the structure of the chiral algebra of a 2d CFT. 
Instead of the more formal VOA language, we  opt for the CFT language more familiar to physicists. 
References on the basic knowledge of CFT include \cite{DiFrancescoPhilippe1997Cft/,BlumenhagenRalph2009ItCF,MR2201600,MatthiasRGaberdiel2000Aitc}. See also Professor Xi Yin's TASI lecture notes in this volume \cite{xi_yin_tasi17}.

After the general summary of the basic structure in \S\ref{subsec:CFTGeneral structure}, we quickly review aspects of  (holomorphic) orbifolds that are relevant for moonshine, in particular for the understanding of the modular properties of the moonshine functions. After that we focus on supersymmetric conformal field theories and introduce the so-called elliptic genus, counting BPS states, that will play an important role in the discussion of the physical context of Mathieu and umbral moonshine in \S\ref{chap:moonshineK3}. 

\subsection{General structure}
\label{subsec:CFTGeneral structure}

A conformal field theory is a quantum field theory with conformal symmetry. 
Conformal transformations are coordinate transformations that preserve the conformal flatness of the metric. 
Focusing on Riemannian manifolds of Euclidean signature, a metric is said to be conformally flat if it can be written in the form $ds^2=e^{\omega(x)}\delta_{\mu\nu}dx^\mu dx^\nu$. 
Conformal transformations locally preserve the angles but may deform the lengths arbitrarily, so conformal symmetry is typically associated with the absence of an intrinsic length scale. On the conformal compactification (by adding the point at infinity which is necessary for the special conformal transformation to be well-defined) of $\RR^{n,0}$ for $n\geq3$, all conformal transformations are globally well-defined and form a group isomorphic to $\SO(n+1,1,\RR)$. 
The corresponding local transformations thus form a finite-dimensional Lie algebra isomorphic to $\mathfrak{so}(n+1,1,\RR)$. 
In two dimensions, however, the condition of conformal invariance is equivalent to the Cauchy-Riemann equation and any holomorphic function gives rise to an infinitesimal conformal transformation. Using the generators
\begin{equation}
\label{eq:conf_gen}
\ell_n=-z^{n+1}\partial_z~,~~\bar{\ell}_n=-\bar{z}^{n+1}\partial_{\bar{z}}~
\end{equation}
for $n\in\ZZ$, 
we see that the local conformal transformations form an infinite-dimensional Lie algebra, which contains two commuting copies of the Witt algebra with  commutation relations
\begin{equation}
\label{eq:witt}
\witt:~~[\ell_m,\ell_n]=(m-n)\ell_{m+n}~.
\end{equation}
It is important to emphasise that most of the conformal generators in $2$ dimensions are purely local, i.e. they do not generate globally well-defined transformations. Consider, for example, the Riemann sphere $\RS=\CC\cup\infty$, i.e. the Riemann surface of genus zero. On $\RS$, only $\ell_{0},\ell_{\pm1}$ generate global conformal transformations, which form the M\"obius group $\SO(3,1,\RR)\cong\PSL(2,\CC)$.

The quantisation of a $2$-dimensional CFT is typically done on $\CC$.
The theory on the Riemann sphere determines the theory on any other Riemann surface uniquely, but does not guarantee their consistency, as one must also require crossing symmetry and modular invariance of the torus partition function (see below).
To see how to quantise on $\CC$, we note that $\CC$ with origin removed is conformally equivalent to  a cylinder $S^1\times\RR$. 
Denoting by $y$ and $t$ the coordinates for $S^1$ and the Euclidean time $\RR$, the conformal map $z= e^{t+iy}$ maps the cylinder to $\CC\backslash\{0\}$
and in particular maps the infinite past to the origin. The usual time ordering on the cylinder becomes radial ordering on the plane, and the associated space of states is built on radial slices. 

Anything that resembles a local field $\phi(z,\bar{z})$ is called a field in CFT. If a field depends only on the holomorphic variable $z$ we call it \emph{chiral field} (or \emph{anti-chiral} if it depends only on $\bar{z}$). Upon quantisation, fields become operator-valued distributions that create states in the space of states $\mathcal{H}$, by acting on the vacuum $|0\rangle\in\mathcal{H}$. This is called the \emph{state-field correspondence}, which maps an field $\phi$ to a state
\begin{equation}
\phi\mapsto|\phi\rangle:=\lim_{z,\bar{z}\rightarrow0}\phi(z,\bar{z})|0\rangle~,
\end{equation}
created at the origin on the plane (or past infinity on the cylinder). A crucial property of a CFT is that the above map is bijective; every state corresponds uniquely to a single local operator, whereas for a typical QFT different fields can produce the same asymptotic state. This can be understood through  the fact that under conformal transformation $t\rightarrow-\infty$ is mapped to a single local point on $\RS$.

The product of two  fields inserted at the same point is generically singular. The singularity structure is captured by the so-called \emph{operator product expansion} (OPE)
\begin{equation}
\phi_1(z)\phi_2(z')\sim\sum_{n=0}^\infty D_n(z-z') O_n(z')~,
\end{equation}
where $\sim$ means that we only keep the singular terms. Here $O_n(z)$ are fields of the theory and $D_n(z-z')$ are complex-valued functions with polynomial or logarithmic singularities when $z\to z'$. The non-singular part of $\phi_1(z)\phi_2(z')$ is captured by the \emph{normal-ordered product}, which can be defined as
\begin{equation}
\normord{\phi_1(z)\phi_2(z')}~:=\phi_1(z)\phi_2(z')-\sum_{n=0}^\infty D_n(z-z') O_n(z')~.
\end{equation}
When there are only polynomial singularities in $D_n(z-z')$ we say that the fields $\phi_1$ and $\phi_2$ are local with respect to each other, in the sense that there are no branch cuts and contour integrals are well-defined. 

The conserved current associated with the continuous conformal symmetry of a 2d CFT is the \emph{stress-energy tensor}, and we denote $T(z):=T_{zz}(z)$ and $\bar{T}(\bar{z}):=T_{\bar{z}\bar{z}}(\bar{z})$. Classically, these are the only non-vanishing components. Upon quantisation on a generic Riemann surface this is broken to $\langle T^a_{~a}\rangle=-\frac{c}{12}R$ where $R$ is the Ricci scalar.

Since the treatment of the chiral and anti-chiral parts is identical, we will from now on focus on the former. 
The holomorphicity of the stress-energy tensor $T(z)$ follows from the fact that the associated conserved charges are precisely the generators of infinitesimal holomorphic conformal transformations \eqref{eq:conf_gen}. Specifically, we have the mode expansion 
\begin{equation}
L_n:=\frac{1}{2\pi}\oint T(z) z^{n+1} dz~\Leftrightarrow~~T(z)=\sum_{n\in\ZZ}L_n z^{-n-2}~.
\end{equation}
The modes $L_n$ however, do not generally satisfy the Witt algebra \eqref{eq:witt}. This is because the conformal symmetry is typically  ``softly" broken by quantum effects. The  OPE of $T(z)$ with itself,
\begin{equation}
T(z)T(w)\sim\frac{c/2}{(z-w)^4}+\frac{2T(w)}{(z-w)^2}+\frac{\partial T(w)}{z-w}~, 
\end{equation}
is equivalent via the mode expansions to  the commutation relations
\begin{equation}
[L_m,L_n]=(m-n)L_{m+n}+\frac{c}{12}m(m^2-1)\delta_{m+n,0}~.
\end{equation}
In the above, 
the real constant $c$ is called \emph{central charge}, 
and the new algebra is the \emph{Virasoro algebra} $\vir$, which is a central extension of $\witt$ by the term containing the central charge.
Moreover, the two resulting $\vir$ copies commute, i.e. $[L_m,\bar{L}_n]=0$, and there is a central charge $\bar{c}$ associated with the anti-chiral part, which can in principle be different from $c$. 
The central charge captures important information of a CFT and gives a measure for the number of degrees of freedom, but there can exist multiple different CFTs with the same central charge. 
It is related to a ``soft" breaking of the conformal symmetry because it indicates that the stress-energy tensor, which generates conformal transformations, transforms anomalously under conformal mappings. For instance, for the transformation 
 $z=e^w, w=t+iy$ from the cylinder to the Riemann sphere, we have 
\begin{equation}
T(z)=z^{-2}\left(T_{\text{cyl}}(w)+\frac{c}{24}\right)~, 
\end{equation}
which leads to the following relation between the Hamiltonian $\oint_{\rm{time-slice}}dw ( T_{\text{cyl}}(w)+ \bar {T}_{\text{cyl}}(\bar w))$ and $\oint_{\rm{radial-slice}}{dz\over z}( T(z)+ \bar {T}(\bar z)) $: 
\be\label{hamiltonian}
H= L_0  +  \tilde L_0 - {c+\tilde c \over 24}. 
\ee
Similarly, we have the momentum (or {\em spin})
\be\label{momentum}
P= L_0  -  \tilde L_0 - {c-\tilde c \over 24}. 
\ee

As a result, the eigenvalues of $L_0$ plays the role of the chiral part of the energy, and the central charge gives rise to non-vanishing ground state energy.
The eigenvalue $h$ under $L_0$ of an eigenstate $|h\rangle\in\mathcal{H}$, i.e. $L_0|h\rangle=h|h\rangle$, is called the \emph{conformal weight} of $|h\rangle$. If, moreover, $L_n|h\rangle=0$ for all $n>0$, then $|h\rangle$ is called a \emph{Virasoro primary state} and the corresponding field called a primary field. This terminology also extends to the corresponding fields $\phi_h$ via the state-field correspondence. A state of the form $L_{-k_1}L_{-k_2}\cdots L_{-k_n}|h\rangle$ ($k_i>0$) is called a \emph{Virasoro descendant} of $|h\rangle$.  If $|h\rangle$ is a primary state, then along with all of its descendants they form a so-called \emph{Verma module} for  $\vir$. 
The primary state $|h\rangle$ is then called the \emph{highest-weight state} of the module, since it has the lowest (somewhat confusingly) conformal weight among all of its descendants. 

Since the states organise themselves into Virasoro representations, one can decompose the space of states of a CFT into a direct sum of $\vir$ and $\overline{\vir}$ modules. In general, focussing on the chiral part, one can have an enlarged symmetry algebra that contains $\vir$. This is called the \emph{chiral algebra} of the CFT, and is denoted here by $\mathcal{V}$.
We are mainly interested in \emph{rational conformal field theories} (RCFTs), which contain a finite number of such modules; let $\Phi(\mathcal{V})$, $\overline{\Phi}(\overline{\mathcal{V}})$ denote the sets of these (chiral and anti-chiral respectively). The space of states can then be written as
\begin{equation}
\label{eq:split_modules_hilbert_space}
\mathcal{H}~=\bigoplus_{M\in\Phi(\mathcal{V}),\overline{N}\in\overline{\Phi}(\overline{\mathcal{V}})}\mathcal{Z}_{M,\overline{N}}\left(M\otimes\overline{N}\right)~.
\end{equation}
The states in such RCFTs get organized in $\mathcal{V}$-modules, whose highest-weight states correspond to \emph{chiral primaries}, which are not only Virasoro primaries, but also primaries with respect to $\mathcal{V}$.  The chiral descendants are generated by acting with $\mathcal{V}$ on the chiral primaries. This means that if $A(z)\in\mathcal{V}$, then $A(z)M\subset M$ for any $M\in\Phi(\mathcal{V})$.

The subspace $V\subset\mathcal{H}$, corresponding to the chiral algebra $\mathcal{V}$ via the state-field correspondence, always forms an irreducible $\mathcal{V}$-module $V:=M_1\in\Phi(\mathcal{V})$, which contains the vacuum and all states corresponding to the Virasoro primaries generating $\mathcal{V}$, also commonly called \emph{currents}, along with their chiral descendants. Modular invariance requires that the eigenvalues of $L_0-\overline{L}_0$ are integers, which also means that all states in any $M\in\Phi(\mathcal{V})$ have the same weight up to an integer. Specifically, all states in the vacuum module $V$ should have integer weights. However, by dropping modular invariance as an initial requirement, the chiral algebra can possibly contain currents of half-integer weights (fermionic currents), or any rational weight (parafermions). The price to pay is that these currents have non-local OPEs (in the sense discussed previously), with the corresponding branch cuts leading to the introduction of various sectors (for fermionic currents these would be the Ramond and Neveu-Schwarz sectors). The modular invariant theory can then be constructed by a suitable projection.

The chiral algebras themselves are the central objects in the theory of Vertex Operator Algebras (VOAs), where they are discussed in an axiomatic manner. 
In the  context of moonshine, an important property of a chiral algebra is the finite group part of $\Aut(\mathcal{V})$.  The most famous example is the Monster CFT $V^\natural$, which is a VOA with $\Aut(V^\natural)=\MM$, the Monster group. Furthermore, $V^\natural$ is an example of a \emph{holomorphic VOA}, i.e. a VOA that has a unique irreducible $\mathcal{V}$-module, namely the space $V$.

In 2d CFT one is interested in calculating correlation functions of fields, inserted at specific points on a Riemann surface $\Sigma$. 
These can be cast in terms of chiral quantities called \emph{chiral blocks}. Writing $\Sigma=\Sigma_{g,n}$ with genus $g$ and $n$ marked points $p_1,\dots,p_n$,   a chiral block is a multilinear map from $M_1 \otimes\cdots\otimes M_n$ to a meromorphic function. This notation means that a field in $M_i\in\Phi(\mathcal{V})$ is inserted at the point $p_i$. 
In the case of RCFT, they can often be  obtained as solutions to certain differential equations  \cite{FelderGiovanni1996TKeo,FuchsJürgen1992ALaa,KohnoToshitake2002Cfta}. 
The chiral blocks form representations of the \emph{mapping class group} $\Gamma_{g,n}$, which captures the discrete (and almost always infinite) symmetries of $\Sigma_{g,n}$. 
It can be defined by the quotient $\Gamma_{g,n}\cong\Aut(\Sigma_{g,n})/\Aut_0(\Sigma_{g,n})$, where $\Aut_0(\Sigma_{g,n})$ is the component of $\Aut(\Sigma_{g,n})$ that is connected to the identity. Hence, $\Gamma_{g,n}$ maps between equivalent Riemann surfaces $\Sigma_{g,n}$, which only differ by a discrete automorphism. 
As a result, the \emph{moduli space} $\mathcal{M}_{g,n}$, which parametrises the conformally inequivalent Riemann surfaces, has naturally the following quotient form,
\begin{equation}
\label{eq:moduli_space}
\mathcal{M}_{g,n}=\mathcal{T}_{g,n}/\Gamma_{g,n}~,
\end{equation} where $\mathcal{T}_{g,n}$ is the so-called \emph{Teichm\"uller space}. 

Chiral blocks have in general non-trivial monodromy as functions of the moduli space $\mathcal{M}_{g,n}$ (see for example \cite{MR2201600} for more details).
Chiral blocks will thus generally be multi-valued functions on $\mathcal{M}_{g,n}$, and in order to make them well-defined one should lift them to $\mathcal{T}_{g,n}$. As a result, they will then carry a representation of the mapping class group $\Gamma_{g,n}$.
This is one way to understand the origin of the modular properties of torus blocks, and in particular moonshine modules. 

To explain this, let us now focus on the case of $\Sigma_{1,1}$, i.e. tori with a single marked point. As explained in \S\ref{subsec:modular_forms}, a torus can be described up to a scale by $\CC/\Lambda_\t$, where $\Lambda_\t$ is the lattice in $\CC$ generated by the vectors 1 and $\t\in \HH$. An $\SL_2(\Z)$ transformation leaves the lattice invariant and as a result the mapping class group  $\Gamma_{1,0} = \PSL_2(\Z)$ is given by the part of  $\SL_2(\Z)$ that acts non-trivially on the Teichm\"uller space  $\HH$. As any point is equivalent to any other point on a torus due to its translation symmetries, we also have $\Gamma_{1,0}=\Gamma_{1,1}$ and ${\cal M}_{1,0} = {\cal M}_{1,1}$.

Chiral blocks on $\Sigma_{1,1}$, when lifted to $\mathcal{T}_{1,1}$, will consequently be functions of the modular parameter $\tau$. 
For RCFTs, they form a space of finite dimensions, and the dimension is given by the number of irreducible modules in $\Phi(\mathcal{V})$. They admit a natural basis given by the \emph{graded dimensions}, or \emph{characters}, of the irreducible modules  $M\in\Phi(\mathcal{V})$
\begin{equation}
\label{eq:character}
{\it ch}_M(\tau)=\tr_{M} q^{L_0-c/24} ~, 
\end{equation}
where $q=e^{2\pi i\tau}$ as before. 
As discussed previously, the characters furnish  a representation of $\Gamma_{1,1}=\PSL(2,\ZZ)$, so that the $ch_M(\tau)$ are components of a weakly holomorphic vector-valued modular form for $\PSL(2,\ZZ)$. In other words, they mix  with each other under the action of the modular group and the way they mix determines their OPE via the Verlinde formula. The modularity of characters of RCFTs is rigorously shown in the context of VOAs by Zhu's Theorem \cite{Zhu_ModInv}.

The \emph{partition function} of a 2d CFT is defined as the $0$-point correlation function on the torus, which encodes the spectrum of the theory. In the operator formalism, a torus with modular parameter $\tau=\tau_1+i\tau_2$ can be obtained from the Riemann sphere by first conformally mapping it to the cylinder $S^1\times\RR$, and then imposing periodic boundary conditions on the Euclidean time direction $\RR$. The Hamiltonian and momentum operators $H,P$ then propagate states along both cycles of the torus, so the spectrum is embodied in the trace of the corresponding evolution operator over the space of states,
\begin{equation}
\label{eq:part_def}
Z(\t,\bar \t) := \tr_\mathcal{H}  e^{2\pi  i \t_1 P -2\pi\t_2  H }
~.
\end{equation}
Using \eq{hamiltonian}-\eq{momentum}, we can rewrite it as
\begin{equation}
Z(\tau,\bar{\tau})=\tr_{\mathcal{H}}q^{L_0-\frac{c}{24}}\bar{q}^{\bar{L}_0-\frac{\bar{c}}{24}}~,
\end{equation}
making it manifest that it is a generating function of the multiplicities of states at given chiral and anti-chiral conformal weights in $\mathcal{H}$. From \eqref{eq:split_modules_hilbert_space} we see that it has the following decomposition in terms of chiral blocks
\begin{equation}
Z(\tau,\bar{\tau})=\sum_{M\in\Phi(\mathcal{V}),\overline{N}\in\overline{\Phi}(\overline{\mathcal{V}})}\mathcal{Z}_{M,\overline{N}}~{\it ch}_{M}(\tau)
\overline{ {\it ch}_{\bar N}(\tau)}~.
\end{equation}
The partition function \eqref{eq:part_def} can also be computed using the path integral formalism when a Lagrangian description of the CFT is available.
In this language, we have $Z=\int D\phi~e^{-S[\phi]}$, with the fields having appropriate boundary conditions on the two cycles of the torus.
Also from this point of view, it is clear that the partition function should be modular invariant.
This invariance imposes severe constraints on the spectrum of 2d CFTs.
For instance, modular invariance was used to classify  supersymmetric minimal models and further extensions. See  \cite{Cappelli:1987xt,MR1407165} and references therein for some of these results.
In the context of moonshine, we are mainly interested in the chiral CFT, where the modular properties are not as stringent.

\subsection{Orbifolds}
\label{sec:orbifolds}

A special class of CFTs which is of particular interest for moonshine is the so-called  \emph{orbifold CFTs} \cite{Dijkgraaf:1989hb}.
The orbifold construction essentially entail ``gauging" a discrete symmetry group $G$ of the chiral algebra $\cal{V}$. 
More precisely, it builds a theory whose chiral algebra contains the $G$-invariant subalgebra $\mathcal{V}^G$ of $\mathcal{V}$, 
by retaining the $G$-invariant states of the original theory and introducing new ``$g$-twisted" sector states, for every $g\in G$.

There are two important ways orbifold considerations enter the study of moonshine. First, 
we will see in \S\ref{cha:weight0} explicit constructions of moonshine chiral CFTs obtained by $\ZZ_2$-orbifolds.  Second, the  partition functions twined by the finite group symmetries provide the necessary information about the group actions on the moonshine CFT and constitute the modular objects playing a central role in moonshine.  Generalising this to the twisted sectors leads to the so-called generalised moonshine, which we will mention in the next part of the lecture.

\myparagraph{Orbifold chiral algebra.}
Here we are mainly interested in orbifolds of chiral RCFTs (rational VOAs). 
We are interested in automorphisms of the operator algebra.  If such an automorphism acts trivially on the operator algebra, i.e. without permuting the modules $M_i^e$, then it is said to be \emph{inner}. In particular it preserves the chiral algebra of the chiral CFT.  
Let $\mathcal{V}$ denote the chiral algebra, $G\subseteq \Aut(\mathcal{V})$ a finite symmetry group, and $M^e_1,\dots,M^e_n$ its irreducible $\mathcal{V}$-modules. Here $e\in G$ denotes the identity element which will later be generalised to arbitrary $g\in G$. 
In particular, we have $M_1^e=V$, the vacuum module corresponding to $\mathcal{V}$. 

Given such a symmetry, the chiral algebra is decomposed in $G$-representations as
\begin{equation}
\label{eq:decomp_V}
\mathcal{V}=\bigoplus_a ~\rho_a \otimes \mathcal{V}_a~,
\end{equation}
where the corresponding spaces $V_a$ contain states that transform under the irreducible representations $\rho_a$ of $G$, and $a$ runs over all of them.  The $G$-invariant subalgebra 
\begin{equation}
\mathcal{V}^G:=\left\lbrace\phi\in\mathcal{V}~|~h\phi=\phi~~\forall~h\in G\right\rbrace~,
\end{equation}
corresponding to the trivial representation $\rho_0$ of $G$, 
is called the \emph{orbifold chiral algebra} in this setup.
Note that  
while $V$ is irreducible as a $\mathcal{V}$-module, it is generically reducible as a $\mathcal{V}^G$-module, as shown in \eqref{eq:decomp_V}. We instead identify the corresponding space $V_a$, corresponding to $ \mathcal{V}_a$,  as the irreducible $\mathcal{V}^G$-modules relevant for the orbifold CFT.

An analogous statement holds for the rest of the $\mathcal{V}$-modules, and we have decompositions
\begin{equation}
M_i^e=\bigoplus_a~\rr_a\otimes M_{i,a}^e~.
\end{equation}
An important subtlety is that $\rr$ now runs over all irreducible \emph{projective representations} of $G$.
Projective representations generalise the usual notion of representations introduced in \S\ref{subsec:rep}, by allowing them to respect the group operation up to a phase,
\begin{equation}
\label{cocycles}
\rr(h_1h_2)=c_e(h_1,h_2)\rr(h_1)\rr(h_2)~,
\end{equation}
where $c_e(h_1,h_2)$ is a $\UU(1)$-valued 2-cocycle, representing a class in the group cohomology $H^2(G,\UU(1))$ of $G$. 
 This type of behaviour is allowed in CFT because such a phase cancels when  the chiral and the anti-chiral contributions are combined and hence is not in conflict with the modular invariance of the final theory.
See \cite{Tania-LuminiţaCostache2009Oipr} for a nice survey on projective representations of finite groups. Also note that the $G$-invariance of the vacuum  implies that the vacuum module $V$ carries true representations in the decomposition $V=\bigoplus_a \rho_a\otimes V_a$.

\myparagraph{Twinings.}
For each $h\in G$, acting as an inner automorphism of the operator algebra, we define the \emph{twined characters}
\begin{equation}
\label{eq:twined_char}
{\it ch}_i\left(\mbox{\fontsize{9}{10}\selectfont $h\,$}\underset{e}{\Box}~;\t\right) :=\tr_{M^e_i}\left[h~q^{L_0-\frac{c}{24}}\right]~.
\end{equation}
Note that the the special case $h=e$ simply gives the usual character or \emph{graded-dimensions}, of $M_i^e$.
In terms of the decomposition into irreducible $\mathcal{V}^G$-modules, the twined characters are expressed as
\begin{equation}
\label{eq:twined_chars}
{\it ch}_i\left(\mbox{\fontsize{9}{10}\selectfont $h\,$}\underset{e}{\Box}~;\t\right)=\sum_a\chi_{a}(h)~ ch\left(M_{i,a}^e\right)~,
\end{equation}
where $\chi_a$ are projective characters of $G$, and ${\it ch}\left(M_{i,a}^e\right)$ are the graded dimensions of $M_{i,a}^e$. Using the orthogonality of the projective representations analogous to \eqref{eq:orthog_char} one can obtain the character ${\it ch}_i\left(\mbox{\fontsize{9}{10}\selectfont $h\,$}\underset{e}{\Box}~;\t\right)$ from the $2$-cycle and the character of the projective representation $\rho_a$. 

In a similar fashion, we define the twined partition function as
\begin{equation}
\label{eq:twined_part_def}
Z\left(\mbox{\fontsize{9}{10}\selectfont $h\,$}\underset{e}{\Box}~;\t,\bar{\t}\right):=\tr_{\mathcal{H}} \left[h~q^{L_0-\frac{c}{24}}\bar{q}^{\bar{L}_0-\frac{\bar{c}}{24}} \right]~.
\end{equation}
In the path integral language, the twined partition function is obtained by imposing $h$-twisted boundary condition for the fields on the cycle of the torus which is identified with the ``temporal'' circle, while the boundary condition along the spatial circle remains unchanged, i.e. $\phi(z+\tau)=h\cdot\phi(z)$ and $\phi(z+1)=\phi(z)$. 
From this point of view, it is clear that $Z\left(\mbox{\fontsize{9}{10}\selectfont $h\,$}\underset{e}{\Box}~;\t,\bar{\t}\right)$ should be invariant under a subgroup of $\SL_2(\ZZ)$ that preserves the $h$-twisted boundary condition ($\SL_2(\ZZ)$ transforms the boundary conditions on the two independent cycles of the torus as in \eq{eq:orb_b.c.} below).

\myparagraph{Twisted sectors.}
Provided that $\mathcal{V}$ is sufficiently nice, in the sense that it satisfies the so-called $C_2$-cofiniteness condition (see \cite{MR2201600} for the definition), then for any $g\in\text{Inn}(\mathcal{V})$, the inner automorphisms of $\mathcal{V}$, one can define an irreducible \emph{$g$-twisted $\mathcal{V}$-module} $M^g_i$ for each $i=1,\cdots,n$ \cite{Dong2000}. 
In an orbifold theory these modules make up the $g$-twisted sector of the theory. 
Clearly, $G$ is no longer a symmetry group for these modules; only the centraliser subgroup $C_G(g)$ (cf. \S\ref{subsec:Groups}) remains as a symmetry of the $g$-twisted sector. As a result, for any commuting pair $g,h\in G$, we can analogously define the \emph{twisted-twined characters}
\begin{equation}
\label{eq:twisted-twined_char}
{\it ch}_i\left(\mbox{\fontsize{9}{10}\selectfont $h\,$}\underset{g}{\Box}~;\t\right):=\tr_{M^g_i}\left[h~q^{L_0-\frac{c}{24}}\right]~,
\end{equation}
i.e. the twined characters in the twisted sectors (of which \eqref{eq:twined_char} is a special case). In the path integral language, they are obtained by additionally imposing $g$-boundary conditions for the spacial cycle of the torus, i.e. we have $\phi(z+1)=g\cdot \phi(z)$ as well as $\phi(z+\tau)=h\cdot \phi(z)$. They similarly admit the decomposition
\begin{equation}
M_i^g=\bigoplus_a~\rr_a\otimes M_{i,a}^g~,
\end{equation}
where the sum now runs over all irreducible projective representations of $C_G(g)$. Accordingly, an obvious generalisation of  \eqref{eq:twined_chars}, obtained by replacing $e$ with $g$ and $G$ with $C_G(g)$, also holds for the twisted sectors.

We have already mentioned that the twined partition functions enjoy modular properties. Similarly the characters \eqref{eq:twisted-twined_char} form vector-valued modular forms for some congruence subgroup with certain multiplier systems. 
This can be understood via the $\SL_2(\Z)$-action on the boundary conditions: 
under modular transformations $(\begin{smallmatrix} a& b \\c&d\end{smallmatrix})\in\SL(2,\ZZ)$ on the torus, the boundary conditions $(g,h)$ on the two cycles change as
\begin{equation}
\label{eq:orb_b.c.}
(g,h)\mapsto\left(h^cg^d,h^ag^b\right)~.
\end{equation}
As a result, the twisted-twined characters transform as
\begin{equation}
\label{eq:tw_tw_char}
{\it ch}_i\left(\mbox{\fontsize{9}{10}\selectfont $h\,$}\underset{g}{\Box}~;\frac{a\tau+b}{c\tau+d}\right) =\sum_{j=1}^n\psi(\gamma,g,h)_{ij}~ {\it ch}_i\left(\mbox{\fontsize{9}{10}\selectfont $h^ag^b\,$}\underset{h^cg^d}{\Box}~;\t\right)~,
\end{equation}
where $\psi(\gamma,g,h)$ is an $n\times n$ matrix with scalar entries.
The special case of \emph{holomorphic VOAs} , i.e. those that contain only a single irreducible (untwisted) $\mathcal{V}$-module $\mathcal{H}$, is the easiest to describe. In this case, the chiral partition function coincides with the character of the chiral algebra and \eq{eq:tw_tw_char} becomes  
 \cite{BantayP.1990OaHa}
\begin{equation}\label{eqn:modular_boundary_condition}
\begin{split}
&Z\left(\mbox{\fontsize{9}{10}\selectfont $h\,$}\underset{g}{\Box}~;\t+1\right)=c_g(g,h)Z\left(\mbox{\fontsize{9}{10}\selectfont $gh\,$}\underset{g}{\Box}~;\t\right)~, \\
&Z\left(\mbox{\fontsize{9}{10}\selectfont $h\,$}\underset{g}{\Box}~;-{1\over\t}\right)=\overline{c_h\left(g,g^{-1}\right)}Z\left(\mbox{\fontsize{9}{10}\selectfont $g^{-1}$}\underset{h}{\Box}~;\t\right)~,
\end{split}
\end{equation}
where now the phases are given by a $2$-cocycle representing a class in $H^2(C_G(g),U(1))$ as in \eqref{cocycles}. Moreover, all the phases for all $g$ should descend from a $3$-cocycle representing a class in $H^3(G,U(1))$\cite{Roche:1990hs,BantayP.1990OaHa}.

In a non-chiral CFT, the spectrum consists of the $G$-invariant parts of all the twisted sectors, leading to the following expression for the partition function 
\begin{equation}\label{eqn:orb_PF}
Z(\tau,\overline{\tau})=\frac{1}{|G|}\sum_{gh=hg}Z\left(\mbox{\fontsize{9}{10}\selectfont $h\,$}\underset{g}{\Box}~;\t,\bar{\t}\right)\epsilon(g,h)
\end{equation}
where $\epsilon(g,h)$ is a phase called the {\em discrete torsion}, which is just $1$ in the simplest cases of orbifold constructions. 
As usual, the above partition function is modular invariant in a consistent orbifold CFT.

\subsection{Elliptic genus}
\label{sec:EG}

In the previous subsections we have discussed conformal theories in general. 
In the context of string theory and in this lecture, we often encounter  2d CFTs with supersymmetries. 
In this subsection, we will introduce introducing some necessary background on superconformal algebras and their representations, and in particular explain what an elliptic genus is, first from a physics point of view and then from a geometric point of view. 

With supersymmetries, the presence of fermions leads to many new features, stemming from the fact that there is now an extra $\Z_2$ grading on the Hilbert space: $V= V_0 \oplus V_1$. (In the context of moonshine, this leads to supermodules of finite groups, cf. \eq{def:supermodule}. )
For instance, in the context of type II superstrings compactified on Calabi-Yau manifolds, the relevant ``internal" CFT is a non-linear sigma model with ${\cal N}=2$ supersymmetry. 
The Calabi-Yau structure of the target space guarantees that the theory has the ${\cal N}=2$ extension of Virasoro symmetry, given by the so-called ${\cal N}=2$ superconformal algebra~(SCA).
In particular, superstrings on $K3$ manifolds and the corresponding elliptic genus will play an important role in \S\ref{chap:moonshineK3}.

The terminology ``${\cal N}=2$" refers to the fact that we include 2 fermionic currents in the algebra on top of the bosonic energy-momentum tensor $T(z)$. Furthermore, there's now an extra automorphism, called the R-symmetry, that rotates different  fermionic currents onto each other.

We denote the two fermionic currents by $G_+(z)$ and $G_-(z)$ and the $U(1)$ R-symmetry current rotating the two by $J(z)$. 
The algebra reads
\bea\notag
\lbrack L_m,L_n\rbrack&=& (m-n) L_{m+n} + \frac{c}{12} m (m^2-1) \,\d_{m+n,0}\\ \notag
\lbrack J_m,J_n \rbrack&=&\frac{c}{3} m \,\d_{m+n,0}\\ \notag
\lbrack L_n,J_m \rbrack &=& -m \,J_{m+n}\\ \label{n_2_superconformal_algebra}
\lbrack L_n,G_r^{\pm} \rbrack &=& (\frac{n}{2} - r)\, G_{r+n}^\pm\\ \notag
\lbrack J_n,G_r^{\pm} \rbrack &=&\pm G_{r+n}^{\pm} \\ \notag
\{G^{+}_{r},G^-_{s}\} &=& 2 L_{r+s} + (r-s) J_{r+s} +\frac{c}{3}\, (r^2 -\frac{1}{4})\, \d_{r+s,0}\;,
\eea
and all other (anti-)commutators are zero. 
As before we have two possible boundary conditions for the fermions
\be
\begin{cases}
2r =  0 \text{  mod   }2 &\text{for R sector} \\ &\\ 
2r = 1 \text{  mod   }2 &\text{for NS sector}\;.
\end{cases} 
\ee

Two comments about this algebra are in order here.
First, we have now two generators, $L_0$ and $J_0$, of the Cartan subalgebra. As a result, the representations will now be graded by two ``quantum numbers", given by the eigenvalues of the $L_0$ and $J_0$ of the highest weight vector. 
The second new feature is that there is a non-trivial inner automorphism of the algebra, which means that the algebra remains the same under the following redefinition
\begin{align} \notag
L_n &\mapsto L_n + \eta J_n + m\, \eta^2 \,\d_{n,0}\\[4pt]
{\bf SF}_\eta :\hspace{5pt} J_n&\mapsto J_n + 2m\,  \eta \, \d_{n,0}\\[4pt] \notag
G_r^\pm &\mapsto G_{r\pm\eta}^\pm\;
\end{align}
with $\eta \in \Z$. This automorphism is called \emph{spectral flow}, and in the above we have written  $m:= c/6$. 
If instead we choose $\eta\in \Z+1/2$ we exchange the Ramond and the Neveu-Schwarz algebra. 
Note that the only operator (up to trivial rescaling and the addition of central terms, of course) invariant under such a transformation is $4mL_0 -J_0^2$.
Recall also that NS sector states give spacetime fermions and Ramond sector states give spacetime bosons. Hence the spectral flow operator has an intimate relation to spacetime supersymmetries.

\myparagraph{Ramond ground states and the Witten index.}
In what follows we will focus on the Ramond algebra and define the Ramond ground states of ${\cal N}=2$ SCFT.  
As usual, we require the ground states to be annihilated by all the positive modes:
$$
L_n \lvert \f \rangle=J_m \lvert \f \rangle = G^\pm_r \lvert \f \rangle = 0 \quad\text{for all} \quad m,n,r > 0\;.
$$
Moreover, they have to annihilated by the zero modes of the fermionic currents 
$$
G^\pm_0\lvert \f \rangle=0\;.
$$
This condition fixes their $L_0$-eigenvalue to be
$$
\frac{1}{2} \{G^+_0,G^-_0\}  \lvert \f \rangle= \left(L_0 -\frac{c}{24}\right) \lvert \f \rangle = 0 \;.
$$

Let's ignore the right-moving part of the spectrum for a moment and consider a chiral Hilbert space $V$. We define its Witten index as
$$
\text{WI}(\t,V) = \Tr_V \big((-1)^{ J_0}q^{ L_0-\frac{ c}{24}} \big)\; . 
$$
If a state $\lvert \psi \rangle$ is not annihilated by $G^+_0$, then the states  $\lvert \psi \rangle$ and $G^+_0\lvert \psi \rangle$ together contribute 0 to $\text{WI}(\t,V)$ since $[L_0,G^+_0]=0$ while $[J_0,G^+_0]=G^+_0$. The same argument holds for $G^-_0$ and we conclude that only Ramond ground states can contribute to the Witten index. As a result, the Witten index $ \text{WI}: \{{\cal N}=2 \;\text{SCFT}\} \to \Z$ is independent of $\t$ and counts (with signs) the number of Ramond ground states in $V$. 

Notice moreover that the Witten index for ${\cal N}=2$ SCFT acquires an interpretation as computing the graded dimension of the cohomology of the $G^+_0$ operator, satisfying $( G_0^+)^2=0$. For $\{ G^+_0,( G^+_0)^\dag\}  =\{ G^+_0, G^-_0\}   = L_0 -\frac{c}{24}$, the Ramond ground states have the interpretation as the harmonic representative in the cohomology. This fact underlies the rigidity property of the Witten index and the elliptic genus which we will define now. 

The same analysis can be trivially extended when one has a non-chiral theory with both left- and right-moving degrees of freedom: the Witten index 
$$
\text{WI}(\t,\bar \t,V) = \Tr_V \big((-1)^{\til J_0+J_0}\bar q^{\til L_0-\frac{\til c}{24}} q^{ L_0-\frac{ c}{24}} \big)\; 
$$
counts states that are Ramond ground states for both the left- and the right-moving copy of ${\cal N}=2$ SCA.

\myparagraph{The ${\cal N}=2$ elliptic genus.} 
It is fine to be able to compute the graded dimension of a cohomology, but we can go further and compute more interesting properties of this vector space. For instance, we have learned that the representations of ${\cal N}=2$ SCA are labelled by two quantum numbers corresponding to the Cartan generators $L_0$ and $J_0$. It will hence be natural to consider the following quantity which computes the dimension of $\til G_0^+$ cohomology graded by the left-moving quantum numbers $L_0, J_0$. 

The elliptic genus of a ${\cal N}=(2,2)$ SCFT is defined as the following Hilbert space trace 
\be\label{elliptic_genus}
 {\bf EG}(\t,z) = \Tr_{{ \cal H}_{\text {RR}}}\Big( (-1)^{J_0+\bar J_0} y^{J_0} q^{L_0-c/24}\bar q^{\bar L_0-c/24}  \Big) \ ,\quad y= e^{2\p i z} \;,
\ee
where   ${ \cal H}_{\text {RR}}$ denotes the Hilbert space of states  that  are in the Ramond sector of the ${\cal N}=2$ SCA both for the left- and right-moving copy of the algebra. From the same argument as that for the Witten index, this quantity will be independent on $\bar q$ and will hence be holomorphic as a function of both $\tau$ and $z$.  

Note that the elliptic genus can be seen as something between  the partition function and the Witten index. 
While the former counts all states and the latter counts only RR ground states, the elliptic genus counts states that are Ramond ground state on the one side and unconstrained on the other side. It contains a lot more information but still has the rigidity property of the Witten index which makes it possible to compute for many SCFTs, and as such it offers  a good balance between information content and computability. 
  
When the theory has a finite group symmetry $G$ which commutes with the superconformal symmetries, one can define the ellpitic genus twined by $g\in G$ as 
\be\label{def:twined_elliptic_genus}
 {\bf EG}_g(\t,z) = \Tr_{{ \cal H}_{\text {RR}}}\Big(g\, (-1)^{J_0+\bar J_0} y^{J_0} q^{L_0-c/24}\bar q^{\bar L_0-c/24}  \Big) .
\ee
These objects will play an important role in the discussions in Part II and III.

\myparagraph{Modular properties.}
As in the case of partition functions (cf. \S\ref{subsec:CFTGeneral structure}), a path integral interpretation of the elliptic genus suggests it has nice transformation property under the torus mapping class group. 
Moreover, the inner automorphism of the algebra (the spectral flow symmetry) implies that the graded dimension of a $L_0$-, $J_0$- eigenspace should only depends on its eigenvalue under the   eigenvalue of the combined operator $4mL_0 -J_0^2$ and the charge of $J_0$ mod $2m$ where $m=c/6$. 
Hence, the Fourier expansion of the elliptic genus should take the form
$$
{\bf EG}(\t,z) = \sum_{n,\ell} q^n y^\ell \,c(4mn-\ell^2,\ell) \;.
$$
where $c(D,\ell)$ only depends on $D$ and $\ell \xmod 2m$
(cf. \eq{eqn:jac:hol-DFoucffhol}). 
From these facts one can deduce that the elliptic genus of an ${\cal N}=(2,2)$ SCFT with central charge $c=6m$  is a weak Jacobi form of  weight zero and index $m$. Similarly, following the same argument and that in \S\ref{sec:orbifolds}, the twined elliptic genera are also weak Jacobi form of  weight zero and the same index, but with the modular group $\SL_2(\ZZ)$ in \eq{def:Jac_skewJac} replaced by a certain subgroup which depends on the twining symmetry $g$.

\myparagraph{The geometric elliptic genus.} 
For a compact complex manifold $M$ with dim$ _\C M=d_0$, we can define its elliptic genus  as the character-valued Euler characteristic of the infinite-dimensional formal vector bundle 
\cite{Ochanine,Witten1987,Landweber_book, WittenInt.J.Mod.Phys.A9:4783-48001994, KawaiNucl.Phys.B414:191-2121994}
$$
{\bf E}_{q,y} 
=
y^{d/2}{\textstyle \bigwedge}{ }_{-y^{-1}} T_M^\ast
\textstyle{\bigotimes}_{n\geq 1} \textstyle\bigwedge{ }_{-y^{-1}q^n} T_M^\ast\bigotimes_{n\geq 1} \textstyle\bigwedge{ }_{-yq^n} 
T_M \bigotimes_{n\geq 0} S_{q^n} (T_M\oplus  T_M^\ast),
$$
where $T_M$ and \(T_M^\ast\) are the holomorphic tangent bundle and its dual, and we adopt the notation
\[ 
\textstyle\bigwedge{ }_q V = 1 + q V + q^2 \textstyle \bigwedge^2 V + \dots,\quad {\rm{and}}\quad S_q V = 1 + q V +q^2 S^2V \dots,
\]
with $S^kV$ denoting the $k$-th symmetric power of $V$. In other words, we have 
\be\label{def_eg}
{\bf EG}(\t,z;M) = \int_M ch({\bf E}_{q,y}) {\rm Td}(M).
\ee
From the above definition we see that this ``stringy'' topological quantity reduces to the familiar ones: the Euler number, the signature, and the $\hat{A}$ genus of $M$, when  we specialise $z$ to $z=0,\t/2,(\t+1)/2$, respectively.

When $M$ has vanishing first Chern class, in particular when $M$ is a Calabi--Yau manifold, its elliptic genus ${\bf EG}(\t,z;M)$ can be shown to be  a weak Jacobi form of weight zero and index $d_0/2$ \cite{KawaiNucl.Phys.B414:191-2121994}. 
Note that the supersymmetric sigma model on a Calabi--Yau manifold flows to a superconformal SCFT in the infrared. The elliptic genus of this ${\cal N}=(2,2)$ SCFT, defined as in \eq{elliptic_genus}, then coincides with the geometric elliptic genus defined in (\ref{def_eg}) of the Calabi--Yau manifold. 

\myparagraph{Examples: $K3$ and $T^4$.}
There are two topologically distinct Calabi-Yau two-folds: $K3$ and $T^4$. Since both are equipped with a hyper-K\"ahler structure, extending the K\"ahler structure of generic Calabi--Yau manifolds, the superconformal symmetry is enhanced from $\mathcal{N}=(2,2)$ to $\mathcal{N}=(4,4)$.
From the above argument, we expect their elliptic genus to be weight zero weak Jacobi forms with index 1. 
Coincidentally, the space of such a form is one-dimensional and is spanned by $\phi_{0,1}(\t,z)$ (cf. \eq{phi01}), and hence we only need one topological invariant of the Calabi-Yau two-folds to fix the whole elliptic genus.
From
\[{\bf EG}(\t,z=0;T^4) =\chi(T^4) = 0 \quad,\quad {\bf EG}(\t,z=0;K3) =\chi(K3) = 24  
\]
and 
\[\phi_{0,1}(\t,z=0) = 12\]
we obtain
\[
{\bf EG}(\t,z;T^4)=0\quad,\quad {\cal Z}(\t,z;K3) =2  \phi_{0,1}(\t,z)\;.
\]
This clearly demonstrates the power of modularity in gaining extremely non-trivial information about the spectrum of a SCFT.
  
\myparagraph{Remark.} 
The argument for the holomorphicity of the elliptic genus fails in an interesting way for theories whose spectrum contains a continuous part. Due to the possible spectral asymmetry (i.e. non-perfect pairing between bosonic and fermionic states), the elliptic genus, when defined as a trace/integral over the full Hilbert space with continuous spectrum included, of such a theory could develop a non-trivial $\bar q$-dependence. 
For such an object the usual path-integral intuition still holds and the resulting non-holomorphic function transforms as a Jacobi form. Restricting to the discrete part of the spectrum, the analogous trace will be holomorphic but will no longer be modular. In particular, it will be a mock Jacobi form. 
As a result, in this context the holomorphic part of the elliptic genus is a well-defined notion both from a physical and mathematical point of view. From the physics perspective, the holomorphic part corresponds to the contribution from the discrete part of the spectrum \cite{Hanany:2002ev,Troost:2010ud,Eguchi:2010cb,Ashok:2011cy}. From the mathematical point of view, the holomorphic part corresponds to the holomorphic part of the harmonic Maass form \cite{MR2097357}.
We will encounter such a situation in \S\ref{sec:umK3}.

\newpage

\part{Moonshine}

In this part we  describe classical and recent cases of moonshine connections, focussing on the mathematical statements and postponing the physics till the next part. 
We  organise the different cases in terms of the weights of the modular objects involved, and we will see that in different weights the moonshine relations have  features that are different in interesting ways.

\section{Moonshine at weight zero}
\label{cha:weight0}

Here we review the two moonshine connections, monstrous and Conway moonshine, that occur at weight zero. They are the moonshine cases that are best understood at the moment, in terms of the specification of the modular objects, the origin of the symmetries, and their physical context.

\subsection{Monstrous moonshine}
\label{sec:mm}

Monstrous moonshine is arguably one of the most fascinating chapters of mathematics in the last century, where finite groups and modular objects were first noticed to be related via physical structures. As the theory of moonshine further develops, we believe that monstrous moonshine will remain the most distinguished example of moonshine phenomenon from various points of view. In this section we briefly describe the features of monstrous moonshine, and we refer to \cite{MR2201600,mason2014vertex,MR3375653} and references therein for other excellent reviews of this beautiful story,  in particular the historical aspects of it.

The term \emph{monstrous moonshine}, coined in \cite{conway_norton}, refers to the unexpected connection between the representation theory of the  Monster group $\mathbb M$ and the  modular form
\begin{equation}\label{Jcoeff}
J(\tau)=\sum_{n\geq-1}a_n \, q^n=q^{-1}+196884q+21493760q^2+864299970q^3+\cdots~,
\end{equation}
which we encountered in \eq{eq:def_J}. 
The development of monstrous moonshine was initiated with 
the key observation, due to McKay,  that the coefficient $196884$ in the $q$-expansion of $J$ can be decomposed as $196884=1+196883$, 
where the summands are the dimensions of the two smallest irreducible represenations of $\mathbb{M}$. Similar decompositions were observed for the next few coefficients by Thompson in \cite{Tho_NmrlgyMonsEllModFn}:
\begin{equation}
\label{eq:J_decomp}
\begin{split}
1&=1 \\
196884&=196883+1 \\
21493760&=21296876+196883+1 \\
864299970&=842609326+21296876+2\cdot196883+2\cdot1 \\
&\cdots 
\end{split}
\end{equation}
where $1$, 196883, 21296876, and 842609326 are dimensions of certain irreducible representations of $\mathbb M$.  
The observation led to the conjecture of the existence of an infinite-dimensional $\ZZ$-graded Monster module,
\begin{equation}
V=\bigoplus_{n\geq-1}V_n~,
\end{equation}
such that $\dim V_n=a_n$ for all $n\geq-1$. 
In other words, $J$ acquires the interpretation as the graded dimension of $V$ 
\begin{equation}
\label{eq:J_graded_trace}
J(\tau)=\sum_{n\geq-1}\dim V_n q^n.
\end{equation}
Notice that $V_0$ is empty, corresponding to the vanishing constant coefficient of $J$.

\myparagraph{The conjecture.}
This conjecture as stated above is not interesting since one could take each $V_n$ to contain $c(n)$ copies of the trivial representation of $\mathbb{M}$ to make \eq{eq:J_graded_trace} true, given the fact that all $a_n$ are non-negative integers.
To access the information on the $\mathbb M$-action, Thompson also proposed in \cite{Tho_FinGpsModFns} to look at the graded characters of $V$, the so-called \emph{McKay-Thompson series}  defined by 
\begin{equation}
T_g(\tau)\coloneqq\sum_{n\geq-1}\tr_{V_n}(g)q^n~,
\end{equation}
for each element $g\in\mathbb{M}$ (with $T_e=J$). Note that the $q$-series $T_g(\tau)$ must also have vanishing constant term.  
As is clear from the definition, the $T_g$ are class functions, i.e. $T_g=T_{hgh^{-1}}$. As a result, there are at most $194$ distinct $T_g$ as $\MM$ has 194 conjugacy classes. In fact, it turns out that $T_g$ only gives rise to 171 distinct functions. 
The main point of monstrous moonshine lies in the fact that these graded trace functions also exhibit modular properties and are moreover the unique Hauptmoduls with no constant terms (cf. \S\ref{subsec:modular_forms}), as stated in the following astonishing conjecture made by Conway and Norton \cite{conway_norton}:
{\conj \label{conj:conwa_norton}  {\bf(Monstrous Moonshine Conjecture)}\\
For each $g\in \MM$ the  McKay-Thompson series $T_g$ coincides the unique Hauptmodul $J_{\Gamma_g}$ with expansion $q^{-1}+O(q)$ near $\t\to i\infty$, for some genus zero subgroup $\Gamma_g\leq\SL_2(\RR)$. 
Furthermore, each $\Gamma_g$ contains $\Gamma_0(N)$ as a normal subgroup, for some $N$ dividing the quantity $|g|\gcd(24,|g|)$.
}
~\\~\\
Given the importance of this conjecture, we will pause to make a few comments. 
Note that  $\Gamma_g$ is often not a subgroup of $\SL_2(\ZZ)$; only for some $g$ we have $\Gamma_g= \Gamma_0(N)$ (cf. \eq{def:cong_subgr}), for some $N$ satisfying the conditions mentioned above. In general, $\Gamma_g$ is a normaliser of $\Gamma_0(N)$ in $\SL_2(\RR)$, which in general involves the so-called Atkin-Lehner involutions. 
For later purpose we will be particular interested in the groups of the form
\be\label{eqn:groupAL}
\Gamma^{N+K} :=
	\left\{
		\frac1{\sqrt{n}}\begin{pmatrix}
		an&b\\cN&dn
		\end{pmatrix}
		\mid
		adn-bcN/n=1,\;
		 n \in K
	\right\},
\ee
where $K<{\rm Ex}_N$ is a subgroup of the group of exact divisors of $N$. We say that $e$ is an exact divisor of $N$ if $e|N$ and  $(f,{N\over f})=1$, and they form a group with 
multiplication $f\ast f'  = {ff' \over (f,f')^2}$.
An especially simple case is when $N$ is a prime number $p$, and the full normaliser (corresponding to $K=\{1,p\}$) is given by
\be
\Gamma_0(p)+:= \Big\langle \Gamma_0(p), {1\over \sqrt{p}} \bem 0&-1\\ p& 0\eem \Big\rangle ~. 
\ee
A harbinger of monstrous moonshine, predating the observation by McKay, is the following observation made by Ogg \cite{ogg1974automorphismes}. 
He noted that $\Gamma_0(p)+$ defines a genus zero quotient on the upper-half plane if and only if 
\be
p\in \{ 2, 3, 5, 7, 11, 13, 17, 19, 23, 29, 31, 41, 47, 59, 71\}, 
\ee
and this is precisely the set of primes dividing the order of the Monster group, and subsequently offered a bottle of Jack Daniel's to anyone who can explain the coincidence \cite{ogg1974automorphismes}. 
Monstrous moonshine sheds light on this mysterious coincidence through the fact that the Hauptmoduls of all the genus zero $\Gamma_0(p)+$ feature in moonshine as the McKay-Thompson series $T_g$ for a $g\in \MM$ of order $p$. 
In the case $\Gamma_g\not\subset \SL_2(\ZZ)$ the modularity of CFT does not help to explain the appearance of modularity for  $\Gamma_g$, since in CFT modularity arises from the mapping class group of the torus (cf. \S\ref{sec:CFT}). 
The crucial genus zero property of monstrous moonshine  received a useful paraphrasing \cite{DunFre_RSMG} as the property that these Hauptmoduls can be obtained (up to a constant) as a {\em Rademacher sum}, a regularised sum over the images of the polar term (in this case $q^{-1}$) under the action of the appropriate subgroup of $\SL_2(\RR)$ (in this case $\Gamma_g$). This {\em Rademacher summability} property subsequently played a key role in the discovery of umbral moonshine (cf. \S\ref{sec:umbral}).  Recently, the genus zero property is explained by noting that $\Gamma_g$ plays the role of the stringy symmetry group in the string realisation of the Monster theory and by requiring a physical analyticity condition on the supersymmetric index of the theory. See \S\ref{sec:physics_monstrous} for more details.

\myparagraph{The moonshine module.}
This conjecture was verified numerically by Atkin, Fong and Smith (cf.~\cite{fong1980characters,smith1985head}), following the idea of Thompson (see \cite{MR2201600} for references). To be more specific, they showed that, for each $n\geq -1$, the $q^n$-coefficient of the Hauptmoduls specified in \cite{conway_norton} coincide with the characters of a certain virtual representation of $\MM$.
A  constructive verification was later obtained by Frenkel, Lepowsky and Meurman \cite{FLMPNAS,FLMBerk}, with the explicit construction of a Monster module $V=V^\natural$. This module has the structure of holomorphic VOA, i.e. a VOA with a single irreducible $V^\natural$-module, namely itself. 

The starting point for constructing $V^\natural$ is $24$ chiral bosons $X^i(z)$, compactified on the $24$ dimensional torus $\RR^{24}/\Lambda$ defined by the Leech lattice $\Lambda$. 
This results into a VOA $V(\Lambda)$ with central charge $c=24$, leading to a partition function whose $q$-expansion starts with 
\( Z_{V(\Lambda)}(\tau)=q^{-1}+\dots\).  
This, together with the modular invariance, fixes the function to be the same as $J(\tau)$ up to an additive constant. 
At the same time, we know what this constant has to be since the Leech lattice has no root vectors and hence $\Theta_{\Lambda}(\t) =\sum_{v\in \Lambda} q^{\langle v,v\rangle/2}  = 1+ O(q^2)$, leading to 
\be
Z_{V(\Lambda)}(\tau)= {\Theta_{\Lambda}(\t)\over \eta^{24}(\t)} = J(\t)+24~. 
\ee
In other words, thanks to the root-free property of the Leech lattice, the lattice vertex operators of the form $e^{ik\cdot\Phi}$ all have weight larger than one, and the only remaining weight one primaries are the $24$ fields $\partial X^i$. 

In order to have an exact matching with $J$ we would like to remove these primaries, which can be achieved by a simple  $\ZZ_2$ orbifold of $V(\Lambda)$, acting  as $X^i\rightarrow-X^i $, which corresponds to the $\lbrace{\rm id},-{\rm id}\rbrace\cong\ZZ_2$ symmetry of $\Lambda$, contained in $\Aut(\L)\cong\Co_0$.
Indeed, one can easily compute the partition function of the orbifolded theory explicitly as follows. 
Note that the $\ZZ_2$-twined partition function of 24 chiral bosons is given by
\be
Z\left(\mbox{\fontsize{9}{10}\selectfont $-$}\underset{+}{\Box}~;\t\right)= {1\over q\prod_{n>0}(1+q^{n})^{24}} = \left(2 \eta(\t)\over \theta_2(\t) \right)^{12}. 
\ee
The orbifold entails that $V^\natural$ is the direct sum of the $\ZZ_2$-invariant projections of the untwisted and twisted sectors respectively (cf.  \eq{eqn:orb_PF}). From \eq{eqn:modular_boundary_condition} and \eq{eqn:transf_theta} we have 
\begin{gather}
\begin{split}
Z_{V^\natural}(\tau)& ={1\over 2}\left( J(\t)+24 + \left(2 \eta(\t)\over \theta_2(\t) \right)^{12} -  \left(2 \eta(\t)\over \theta_3(\t) \right)^{12} +\left(2 \eta(\t)\over \theta_4(\t) \right)^{12} \right) \\& = J(\t). 
\end{split}
\end{gather}

It remains to see that Aut$(V^\natural)$ is the Monster. Note that Aut$V(\Lambda)$ has a continuous piece $T$ which is a 24-dimensional torus corresponding to the translation symmetry of the chiral bosons and to the 24 weight-one primary fields $\pa X^i$. The total symmetry is captured by the (non-split) short exact sequence 
\be
1 \rightarrow T \rightarrow \rm{Aut}V(\Lambda) \rightarrow \Co_0 \rightarrow 1~. 
\ee 
The $\ZZ_2$-orbifold breaks  the automorphism group to its discrete part $2^{24}.\Co_0$, which preserves the decomposition $V^\natural=V^\natural_+\oplus V^\natural_-$ and is suggestively similar to  a certain maximal subgroup $2^{1+24}.\Co_1$ of $\MM$. 
It is clear from the contribution to the weight two (and similarly for weight three, four, $\dots$) states in $V^\natural$ from $V^\natural_+$ and $V^\natural_-$ that the Monster must mix them and hence cannot preserve the (un)twisted sector individually. 
Note that the 196884-dimensionl space of weight two states of $V^\natural$ has the structure of a commutative and non-associative algebra (as is true for any VOA), and can be shown to be precisely the {\em Griess algebra} constructed in 1980 and used to construct the Monster group itself \cite{Gri_FG}.
From this and the VOA structure of $V^\natural$ one can show that Aut$(V^\natural)$ is indeed the Monster, and can be obtained by adjoining a certain order two symmetry mixing  $V^\natural_\pm$ to the discrete part of Aut$(V(\Lambda))$.

\myparagraph{The proof of monstrous moonshine. }
To prove that the $V^\natural$ constructed by Frenkel, Lepowsky and Meurman indeed ``does the job'', one needs to show that 
\be
T_g^{V^\natural}(\tau):= \tr_{V^\natural} g \, q^{L_0-c/24} 
\ee
coincides with the corresponding Hauptmodul $J_{\Gamma_g}$ specified in \cite{conway_norton}. It was known that the coefficients of Hauptmoduls satisfy certain recursive formulas and one can determine all coefficients from just a handful of them. 
In the simplest case the recursive formulas are encoded in the remarkable identity
\begin{equation}
\label{eq:replicability_J}
p^{-1}\prod_{m>0\atop n\in\ZZ}\left(1-p^mq^n\right)^{a_{mn}}=J(\rho)-J(\tau)~,
\end{equation}
independently discovered by Zagier, Borcherds and others. Here $p=e^{2\pi i\rho}$, and 
$a_i$ denotes  the $q^i$ coefficients in the $q$-expansion of $J$ (cf. (\ref{Jcoeff})). This identity results in infinitely many relations between $a_i$, which enables one to completely fix all the coefficients from just $a_1,a_2,a_3,a_5$. Clearly, the proof can be achieved if one can show the existence of the same type of identities among the coefficients of $T_g^{V^\natural}(\tau)$, and just explicitly compare the handful of coefficients that are necessary to fix the whole functions on both sides.

This is precisely what Borcherds did, and he obtained the replication formulas by introducing the notion of a \emph{generalised Kac-Moody algebra}, which can be viewed as a generalisation of Kac--Moody algebras that allows for imaginary simple roots. Subsequently, he constructed a generalised Kac-Moody algebra (also called ``Borcherds- Kac-Moody algebra'') $\mathfrak{m}$, called the Monster Lie algebra. Roughly speaking, the construction was achieved by studying the cohomology of a BRST-like operator, which acts on $V^\natural\times\Gamma^{1,1}$, where $\Gamma^{1,1}$ is the unique unimodular lattice of signature $(1,1)$. This construction has a  natural interpretation in string theory of considering second quantised strings in the background of $V^\natural$. See \S\ref{sec:physics_monstrous} for more details. 

Borcherds managed to derive the replication formulas \eqref{eq:replicability_J} as the \emph{denominator identities} of the Monster Lie algebra $\mathfrak{m}$ that he attached to $V^\natural$. As in usual Kac--Moody algebras, the denominator identity results from applying the Weyl-Kac character formula of a Lie algebra to the trivial representation, and in this case relates an infinite sum to an infinite product, precisely the structure we see in \eq{eq:replicability_J}. 
Moreover, by considering the $\MM$-action on $V^\natural$ one can also obtain from $\mathfrak{m}$ the analogous identity
\begin{equation}
\label{eq:replicability_T}
p^{-1}\exp\left[-\sum_{k>0}\sum_{m>0\atop n\in\ZZ} a_{mn}^{g^k}\frac{p^{mk}q^{nk}}{k}\right]=J_{\Gamma_g}(z)-J_{\Gamma_g}(\tau)~,
\end{equation}
satisfied by the other Hauptmoduls, where
$a_i^g$ are the $q$-expansion coefficients of $J_{\Gamma_g}$. Combining the above components then proves the validity of $V^\natural$ as the module of monstrous moonshine.

\myparagraph{Generalised monstrous moonshine.}
In \cite{generalized_moonshine} Norton proposed a generalisation of monstrous moonshine under the name of \emph{generalised monstrous moonshine}. He suggested that there is a rule to assign to each element $g\in\MM$ a graded projective representation $V(g)=\bigoplus_{n\in\QQ}V(g)_n$ of the centralizer group $C_\MM(g)$, and to each pair $(g,h)$ of commuting elements of $\MM$ a holomorphic function $T_{(g,h)}$ on the upper half-plane $\HH$, which satisfies the following conditions:
\begin{enumerate}[label=(\roman*)]
\item 
$~T_{\left(g^ah^c,g^bh^d\right)}(\tau)=\gamma ~T_{(g,h)}\left(\frac{a\tau+b}{c\tau+d}\right)~~~~$ with $~~\begin{pmatrix} a &b \\ c&d\end{pmatrix}\in\SL_2(\ZZ)$ and $\gamma$ being a $24$th root of unity.
\item $~T_{\left(g,h\right)}(\tau)=T_{\left(k^{-1}gk,k^{-1}hk\right)}(\tau)~~~~$ with $k\in\MM$.
\item There is a lift $\tilde{h}$ of $h$ to a linear transformation on $V(g)$ such that
\begin{equation}
T_{(g,h)}(\tau)=\sum_{n\in\QQ}\tr_{V(g)_n}\left(\tilde{h}~q^{n-1}\right)~.
\end{equation}
\item \label{haupt} $T_{(g,h)}(\tau)$ is either a constant or a Hauptmodul for some genus-zero congruence subgroup of $\SL_2(\ZZ)$.
\item $T_{(e,h)}(\tau)$ coincide with $T_h(\tau)$, the McKay-Thompson series attached to $h\in\MM$ by monstrous moonshine.
\end{enumerate}

As we can see from the discussion in \S \ref{sec:orbifolds}, all of these properties, apart from \ref{haupt}, can be understood in the framework of holomorpic orbifolds \cite{DixonL.1988Batb}, applied to $V^\natural$. 
In particular, the function $T_{(g,h)}$ can be thought of  the $h$-twined character of the twisted module $V^\natural_g$. 
The proof of generalised monstrous moonshine was carried out recently in \cite{carnahan2012generalized}, where a generalised Kac--Moody algeba ${\mathfrak m}_g$, generalising the monster Lie algebra ${\mathfrak m}$, is constructed for all $g\in \MM$.

\subsection{Conway moonshine}
\label{sec:conway_moonshine}

Conway moonshine establishes the relation between  $\Co_0$, related to Conway's sporadic group $\Co_1$ by $\Co_1\cong\text{Co}_0/\lbrace\pm\text{Id}\rbrace$, and Hauptmoduls of certain genus zero subgroups of $\SL_2(\RR)$.

Recall from  \S\ref{sec:sporadic_lattices} that $\Co_0$ is isomoprhic to the automorphism group of the Leech lattice $\Lambda$. In this context, hints of Conway moonshine had already appeared in the original montrous moonshine paper \cite{conway_norton}, where the authors assigned genus zero groups $\Gamma_g<\SL_2(\mathbb{R})$ to elements $g\in\text{Co}_0$: let  $\left\lbrace\lambda_i,\lambda_i^{-1}\right\rbrace_{i=1}^{12}$ be the 24 eigenvalues of the natural $g$-action  on the Leech lattice $\Lambda\otimes_{\mathbb{Z}}\mathbb{C}$ (embedded in a complex vector space), then 
$\Gamma_g$ is given by the  invariance groups of the holomorphic function
\begin{equation}
t_g(\t):=q^{-1}\prod_{n>0}\prod_{i=1}^{12}\left(1-\lambda_iq^{2n-1}\right)\left(1-\lambda_i^{-1}q^{2n-1}\right)=q^{-1}-\chi_g+\mathcal{O}(q)~. 
\end{equation}
Note that $\chi_g = \sum_i (\lambda_i+\lambda_i^{-1})$ is generically non-vanishing, and $t_g$ has non-zero constant terms unlike the monstrous moonshine functions discussed in the previous subsection.

Conway moonshine, on the other hand, introduces a Conway module $V^{s\natural}$ whose McKay--Thompson series coincide with Hauptmoduls with vanishing constant terms. 
It was developed in \cite{DuncanJohnF.2007SfCl, DuncanJohnf.r.2015TMMF} (see also \cite{MR3465528}, \cite{Cheng:2014owa} and \cite{Harrison:2018joy} for nice summaries of the construction). 
The Conway module $V^{s\natural}$ is the unique, up to isomorphisms, super-VOA (SVOA) with $c_{V^{s\natural}}=12$ and $\mathcal{N}=1$ superconformal structure, which has no states with weight $1/2$. It can be constructed as a $\ZZ_2$ orbifold of the theory with eight bosons on the $E_8$ torus together with their fermionic superpartners.  Alternatively, it can  be constructed as a $\ZZ_2$ orbifold, acting as $k_a \to -k_a$, of $24$ free chiral fermions $k_a$, $a=1,2,\dots,24$. 
This is to be compared with the monstrous moonshine module $V^\natural$, where the corresponding Monster module $V^\natural$ is built as a $\ZZ_2$ orbifold of the Leech lattice VOA ($24$ chiral bosons compactified on $\RR^{24}/\Lambda$), resulting in theory with  $c_{V^\natural}=24$ and no states of weight $1$.  
It turns out that $V^{s\natural}$ has an interesting relation to stringy symmetries of K3 surfaces (see \S\ref{subsec:conwayK3}).  In what follows we will give more details on the Conway module $V^{s\natural}$.

Consider 24 real chiral fermions $k_a$ and the corresponding complex fermions  
\begin{equation}
\psi^\pm_j=\frac{1}{\sqrt{2}}\left(k_{2j-1}\pm ik_{2j}\right)~,~~~j=1,\dots,12~, 
\end{equation}
with the following non-vanishing OPEs and stress-energy tensor
\begin{equation}
\psi^\pm_i(z)\psi^\mp_j(w)\sim\frac{\delta_{ij}}{z-w}~~,~~~~~L=-\frac{1}{2}\sum_{i=1}^{12}\normord{\psi_i^+\partial\psi_i^- + \psi_i^-\partial\psi_i^+}~.
\end{equation}
Denote by $\mathfrak{a}$ the 24-dimensional vector space spanned by the fermions. Since fermions allow for both periodic and anti-periodic boundary conditions, there exist two sectors in the theory. The  antiperiodic (Neveu-Schwartz) sector contains a single ground state $|0\rangle$ and excitations of half-integer weight, while the periodic (Ramond) sector contains integral-weight excitations and has $2^{12}$ degenerate ground states. The degeneracy is due to the Clifford algebra satisfied by the zero modes, 
\begin{equation}
\left\lbrace\psi^\pm_{i,0},\psi^\pm_{j,0}\right\rbrace=0~,~~~\left\lbrace\psi^\pm_{i,0},\psi^\mp_{j,0}\right\rbrace=\delta_{ij}~,
\end{equation}
which moreover commute with $L_0$. 
As a result, one can build the Ramond ground states by acting with $\psi^-_{i,0}$ on a ground state $|s\rangle$  satisfying $\psi^+_{i,0}|s\rangle=0$. 
Namely, the Ramond ground states are given by the mononomials
\begin{equation}
\label{eq:conway_mononomials}
\psi^-_{i_1,0}\cdots\psi^-_{i_k,0}|s\rangle~,
\end{equation}
which form a spinor in twenty-four dimensions with Euclidean signature.

Next we want to construct an action of $\text{Co}_0$ on the states described above. 
To do so, recall that $\text{Co}_0\cong\Aut(\Lambda)$, so the Conway group is isomorphic to a subgroup of $\SO(24)$ and we can make the natural identification $\mathfrak{a}=\Lambda\otimes_\mathbb{Z}\mathbb{C}$, i.e. let fermions ``live" on the Leech lattice. 
Then consider  a group element $g\in\text{Co}_0$ with complex eigenvalues $\lambda_i^{\pm1}$, and choose the basis of  $\mathfrak{a}$  such that the fermions $\psi^\pm_i$ are acted upon as eigenvectors:
\begin{equation}
\label{eq:conway_action_fermions}
g\psi^\pm_i=\lambda_i^{\pm1}\psi^\pm_i~,~~~\lambda_i\equiv e^{2\pi ia_i}~,~~~i=1,\dots,12~.
\end{equation}
Moreover, since the ground states in the Ramond sector form a representation of the Clifford algebra associated to $\mathfrak{a}$, we should lift $G<SO(24)$ to a subgroup $\widehat{G}<\text{Spin}(\mathfrak{a})$.
An element $x\in\text{Spin}(\mathfrak{a})$ has the property $xux^{-1}\in\mathfrak{a}$ for $u\in\mathfrak{a}$. We define the \emph{lift} $\hat{g}\in\widehat{G}<\text{Spin}(\mathfrak{a})$ of $g\in G<\SO(\mathfrak{a})$ by requiring that it results in the same action as $g$ when acting on $\mathfrak{a}$,
\begin{equation}
\label{lift_def}
\hat{g}(u):=\hat{g}u\hat{g}^{-1}=gu
~,~~~\forall~u\in\mathfrak{a}~.
\end{equation}
The map $u\mapsto\hat{g}(u)$ is a linear transformation on $\mathfrak{a}$ belonging to $\SO(\mathfrak{a})$, so $\hat{g}\mapsto\hat{g}(\cdot)$ defines a map $\text{Spin}(\mathfrak{a})\rightarrow\SO(\mathfrak{a})$ with kernel $\lbrace\pm\mathbf{1}\rbrace$, i.e. $\text{Spin}(\mathfrak{a})$ is a double cover of $\SO(\mathfrak{a})$. It turns out that for $G\cong\text{Co}_0$ there exists a unique lift $\widehat{G}\cong\Co_0$ (see \cite{DuncanJohnf.r.2015TMMF} for more details).

While the NS ground state $|0\rangle$ is invariant under $\Co_0$, the group action on the $2^{12}$ Ramond ground states turns out to be
\begin{equation}
\hat{g}|s\rangle=\prod_{i=1}^{12}e^{\pi i a_i}|s\rangle=\nu|s\rangle~,~~~\nu\equiv\prod_{i=1}^{12}\nu_i~,~~~\nu_i\equiv e^{\pi ia_i}=\lambda_i^{1/2}~,
\end{equation}
where $|s\rangle$ is the ground state described in \eqref{eq:conway_mononomials}.
Notice that a priori there is a sign ambiguity for $\nu_i$, since it is the square root of $\lambda_i$. But actually the choice of sign is unique since the lift of $\text{Co}_0$ is unique.  There is a further ambiguity in the definition of the $g$-action on the fermions, in that we can swap the complex eigenvalues. This translates into setting $\lambda_i\leftrightarrow-\lambda_i^{-1}$ in \eqref{eq:conway_action_fermions}, and is referred to as a choice of polarisation.

The last step is to consider a $\ZZ_2=\lbrace1,\mathfrak{z}\rbrace$ orbifold of the theory described so far, acting as $\mathfrak{z}\psi^\pm_i=-\psi^\pm_i$ on the fermions. In other words, it acts as $(-1)^F$ where $F$ is the fermion number. Supposing that it acts trivially on both ground states $|0\rangle$ and $|s\rangle$\cite{DuncanJohnf.r.2015TMMF}, it splits the two sectors into even/odd eigenspaces,
\begin{equation}
\text{NS}=\text{NS}^0\oplus\text{NS}^1~,~~\text{R}=\text{R}^0\oplus\text{R}^1~,
\end{equation}
where the eigenvalues of $\text{NS}^j/\text{R}^j$ are given by $(-1)^j$. From this point on, one can construct two closely related SVOAs. 
A useful description is by exploiting the fact that $\text{NS}^0$ forms a (bosonic) VOA on its own, that of the lattice $D_{12}$. Equivalently, it is the VOA associated to the affine Kac-Moody algebra $\widehat{\mathfrak{so}}(24)_1$, at level $1$. The latter has four irreducible integrable modules, namely the vacuum module $A\cong\text{NS}^0$, the vector module $V$, the spinor module $S$ and the conjugate spinor module $C$. By extending the $D_{12}$ VOA $\text{NS}^0$ by either of the spinor modules, one arrives at two SVOAs:
\begin{equation}
\label{eq:Vf,Vs}
\begin{split}
&V^{f\natural}=\text{NS}^0\oplus S\cong\text{NS}^0\oplus \text{R}^0 \\
&V^{s\natural}=\text{NS}^0\oplus C\cong\text{NS}^0\oplus \text{R}^1~. 
\end{split}
\end{equation}
From the orbifold point of view we have the identifications $S\cong\text{R}^0$ and $C\cong\text{R}^1$ in our notation.

The two SVOAs $V^{f\natural}$ and $V^{s\natural}$ are isomorphic as SVOAs, and are uniquely characterized by their central charge $c_{V^{f\natural}}=c_{V^{s\natural}}=12$ and the absence of weight $1/2$ states. In \cite{DuncanJohnF.2007SfCl} is was shown that the $\mathcal{N}=1$ supercurrent of $V^{f\natural}$ is fixed by a subgroup of $\text{Spin}(24)$ isomorphic to $\Co_0$, which is identified by the group $\widehat{G}$ in the notation above. 
In particular, note that $Z(\text{Spin}(24))\cong\ZZ_2\times\ZZ_2$ where the first $\ZZ_2$ factor can be identified with the kernel of $\text{Spin}(24) \to SO(24)$ and the latter with the center of $SO(24)$. The centre $Z(\widehat{G})\cong\ZZ_2$ can be identified with the second $\ZZ_2$ in Z(\text{Spin}(24)), and has the same action as the $\ZZ_2$ in the orbifold construction.
As a result, it follows immediately from \eqref{eq:Vf,Vs} that $\Co_0$ does not act faithfully on $V^{f\natural}$, since the latter is invariant under the action of the centre $Z(\widehat{G})$.
Instead, $V^{f\natural}$ carries a faithful action of the quotient group $\widehat{G}/\ZZ_2\cong\Co_1$. On the other hand, $\Co_0$ acts faithfully on $V^{s\natural}$, and  this is ultimately the reason why we consider $V^{s\natural}$ instead of $V^{f\natural}$ in what follows. Another notable difference between $V^{f\natural}$ and $V^{s\natural}$ is that the $\mathcal{N}=1$ supercurrent in $V^{f\natural}$ fixed by $\Co_0$ is not contained in $V^{s\natural}$, but rather in $V^{s\natural}_{\text{tw}}$ (inside the $\text{R}^0$ part).
A ``canonically twisted'' (or Ramond sector) module for $V^{s\natural}$ can also be constructed as
\begin{equation}
V^{s\natural}_{\text{tw}}=\text{NS}^1\oplus\text{R}^0~,
\end{equation}
which is twisted with respect to the $Z(\widehat{G})$ symmetry. The action of $\Co_0$ on this twisted module is also faithful. 

In order to formulate the Conway moonshine statement, first define the functions
\begin{equation}
\label{eta_g_and_C_g}
\begin{split}
&\eta_{\pm g}(\tau):=q\prod_{n=1}^{\infty}\prod_{i=1}^{12}\left(1\mp\lambda_i^{-1}q^n\right)\left(1\mp\lambda_iq^n\right) \\
&C_{\pm g}:=\nu\prod_{i=1}^{12}\left(1\mp\lambda_i^{-1}\right)=\prod_{i=1}^{12}\left(\nu_i\mp\nu_i^{-1}\right)~.
\end{split}
\end{equation}
The twined partition functions of Conway moonshine are then given by
\begin{equation}
\begin{split}
T^s_g(\tau):=\text{str}_{V^{s\natural}}\left[\hat{g}q^{L_0-1/2}\right]=\text{tr}_{V^{s\natural}}\left[\mathfrak{z}\hat{g}q^{L_0-1/2}\right]=\frac{\eta_g(\tau/2)}{\eta_g(\tau)}+\chi_g \\
T^s_{g,\text{tw}}(\tau):=\text{str}_{V^{s\natural}_{\text{tw}}}\left[\hat{g}q^{L_0-1/2}\right]=\text{tr}_{V^{s\natural}_{\text{tw}}}\left[\mathfrak{z}\hat{g}q^{L_0-1/2}\right]=C_g\eta_g(\tau)-\chi_g~,
\end{split}
\end{equation}
where the super-gradings can be defined by inserting $(-1)^F$ into the trace, whose action coincides with that of $\mathfrak{z}$ (recall that we identified $\ZZ_2=\lbrace1,\mathfrak{z}\rbrace$ with the centre of $\Co_0$). The main theorem of Conway moonshine states \cite{DuncanJohnf.r.2015TMMF}:

{\thm  \label{thm:conway_moonshine} The function $T_g^s(2\tau)=q^{-1}+O(q)$ is a Hauptmodul for a genus zero group $\Gamma_g<\SL_2(\RR)$ that contains some $\Gamma_0(N)$, for every $g\in\Co_0$. If $g$ has a fixed point in its action on $\Lambda$, then $T^s_{g,\text{tw}}(\tau)$ is equal to the constant $-\chi_g$. Furthermore, if $g$ has no such fixed point, then $T^s_{g,\text{tw}}(\tau)$ is also a Hauptmodul for a genus zero subgroup of $\SL_2(\mathbb{R})$.}\\

In \S\ref{subsec:conwayK3} we will see a prominent role played by Conway moonshine in relation to the symmetries of $K3$ CFTs.

\section{Moonshine at weight one-half}
\label{sec:Moonshine at weight one-half}

Somewhat unexpectedly, a wave of moonshine development started in 2010 which led to the discovery of many more examples of moonshine connections. 
The modern examples share some similarities, but also display important differences with the classical moonshine examples discussed in \S\ref{cha:weight0}. 
The modular objects in these examples are typically mock modular forms, an important and natural generalisation of modular forms introduced in \S\ref{subsec:MMF}, which furthermore have non-vanishing weights. The first and very fruitful arena that was explored is that of weight 1/2 mock modular forms. In this section we will describe interesting  examples of moonshine relating finite groups and weight 1/2 mock modular forms.

\subsection{Mathieu moonshine}
\label{subsec:Mathieu}

The first example of the new type of moonshine, {\em Mathieu moonshine}, was initiated with certain observations about the weight $1/2$ mock modular form $H$ introduced in (\ref{second_def_H}), in an analogous fashion as how observations about the classical $J$ function initiated the development of monstrous moonshine. In \cite{Eguchi2010} it was pointed out that the first few Fourier coefficients of $H$ coincide with twice the dimensions of certain irreducible representations of the largest sporadic group $M_{24}$. 
Moreover, this observation was made in an interesting physical context which we will mention briefly below, and will explain in more detail in \S\ref{chap:moonshineK3}. 

By now we have understood  that, from many different points of view,  Mathieu moonshine should really be thought of as a component of umbral moonshine, which we will review in the next subsection. 
However, in many ways Mathieu moonshine stands out among the other cases of umbral moonshine, not just historically but also in terms of its direct relation to the $K3$ elliptic genus. As a result, we will devote a separate subsection to Mathieu moonshine before discussing umbral moonshine.

Recall that the mock modular form $H$ \eq{second_def_H} can be viewed as arising from a meromorphic Jacobi form $\psi$ given in (\ref{polar:M24}). 
Using the relation between $\psi$ and the $K3$ elliptic genus (\ref{eq:psi_EGK3}), as well as the identity $(\theta_{2,1}-\th_{2,-1})(\t,z) =-i \theta_1(\t,2z)$, we obtain 
 the following relation between the elliptic genus of $K3$ (cf. \S\ref{sec:EG}) and the mock modular form $H$: 
\be\label{splitEGM24}
{\bf EG}(\t,z;K3) =  \frac{\th_1^2(\t,z)}{\eta^3(\t)} \,\left(24 \,\m(\t,z) + H(\t)\right)~, 
\ee
where 
\be\label{def:mu}
\m(\t,z) = {i \over \theta_{1}(\t,2z)}  {\rm Av}^{(2)}\left[{y+1\over y-1}\right] = \frac{-i y^{1/2}}{\th_{1}(\t,z)}\,\sum_{\ell=-\inf}^\inf \frac{(-1)^{\ell} y^\ell q^{\ell(\ell+1)/2}}{1-y q^\ell}~.
\ee

Note that while none of the two summands at the right-hand side of  (\ref{splitEGM24}) transforms modularly, their modular anomalies cancel and the left-hand side is a perfectly well-behaved Jacobi form, as discussed in \S\ref{sec:modular} and \S\ref{sec:EG}. In particular, a simple way to derive the shadow of $H$ is by studying the modular properties of  the Appell--Lerch sum $\mu(\t,z)$ \cite{Eguchi2009a,Zwegers2008}. 
We will see in \S\ref{sec:M24moonshine}-\ref{sec:umK3} the two interesting physical interpretations of the above splitting (\ref{splitEGM24}) of ${\bf EG}(K3)$, one in terms of the characters of ${\cal N}=4$ superconfomal algebra and one in terms of the elliptic genus of du Val singularities. 

The aforementioned observation on the conspicuous relation between the first few coefficients of the mock modular form $H$ and certain representations of $M_{24}$ led to the suspicion that there exists 
a $\Z$-graded, infinite-dimensional  $M_{24}$-module  $K= \bigoplus_{n=1}^\inf  K_n$ underlying $H$, namely $H(\t) =q^{-\frac{1}{8}}\big( -2  + \sum_{n=1}^\inf q^n\,({\rm dim}({K_n} )\big)$. A natural question is thus whether  the other summand in the splitting of the Jacobi form ${\bf EG}(K3)$ (\ref{splitEGM24}) harbors an action of $M_{24}$ as well. A simple guess arises from the fact that $M_{24}$ is a subgroup of the permutation group $S_{24}$ and as a result has a defining permutation representation ${\bf R}$, of dimension 24. 
A natural proposal for the ``twined'' version of (\ref{splitEGM24}) 
is therefore 
\be\label{K3_EG_Fou}
	 \phi_g(\t,z) 
 		=  \frac{\th_1^2(\t,z)}{\eta^3(\t)} \,\big((\tr_{\bf R} {g } ) \,\m(\t,z) + H_g(\t)\big)~,
	\ee
	where $H_g$ denotes the graded characters of the $M_{24}$ module $K$. 
Following the spirit of monstrous moonshine, we say that there is a non-trivial moonshine connection if all such $\phi_g$ transform nicely as Jacobi forms under some $\Gamma_g \subseteq \SL_2(\ZZ).$ Physical considerations reviewed in \S\ref{sec:orbifolds} moreover suggest that $ \Gamma_0(|g|) \subseteq \Gamma_g$.  

Fortunately, the possibility for this type of Jacobi forms is very limited and we are constrained to consider 
\be
\phi = c\, \phi_{0,1} + F \,\phi_{-2,1}~,
\ee
where $\phi_{0,1}$ and $\phi_{-2,1}$ are given in (\ref{phi01}), $c\in \CC$, and $F$ is a weight two modular form for $\Gamma_g$, possibly with a non-trivial multiplier system when $c=0$.  
The dimension of the space of possible $F$ is often small for the $\SL_2(\Z)$-subgroup $\Gamma_g$ we are interested in. For instance, when $\Gamma_g = \SL_2(\Z)$ the only possible weight two form is $F=0$. Hence, knowing the first few of the Fourier coefficients of $\phi_g$,  dictated by our guesses for the first few $M_{24}$-representations, is often sufficient to fix the whole function. 
As a result, not long after the original observation \cite{Eguchi2010}, candidates for the McKay--Thompson series were proposed for all conjugacy classes $[g]\subset M_{24}$ in \cite{cheng2010k3 ,Gaberdiel2010,Gaberdiel2010a,Eguchi2010a}, and they take the form
\be\label{m24_weakJacform}
{\phi}'_g(\t,z)= \frac{\tr_{\bf R} {g } }{12}\, \phi_{0,1}(\t,z) + \til T_g(\t) \, \phi_{-2,1}(\t,z)~,\;\ee 
where the functions $\tilde{T}_g(\t)$ are weight 2 modular forms explicitly specified in the references given above and  collected in Table 2 of \cite{cheng2011largest}. 
More precisely, these $\phi_g$ for any $g\in M_{24}$ are weak Jacobi form of weight zero and index one satisfying the elliptic invariance $\f_g\lvert_{1} (\l,\m)  = \phi_g$ for all $(\l,\m)\in \ZZ^2$ (cf. \eq{elliptic}), and transform as
$$  
{\phi}'_g(\t,z)=\r_{n_g|h_g}(\gamma) \,e\left(-\frac{c z^2}{c\t+d}\right) {\phi}'_g\big({a\t+b\over c\t+d},{z\over c\t+d}\big)~ ,
$$
for $\gamma \in\Gamma_0(|g|)$, where the multiplier $\r_{n_g|h_g}$ is summarised in  \cite{cheng2011largest}. 

In terms of the weak Jacobi forms (\ref{m24_weakJacform}), the main statement of Mathieu moonshine is the following. 
\begin{conj}\label{conj_mock}
There exists a naturally defined $\Z$-graded, infinite-dimensional $M_{24}$ module $K= \bigoplus_{n=1}^\inf  K_n$ such that for any $g\in M_{24}$, the graded character
\be\label{rep_mock}
H_g(\t) :=q^{-\frac{1}{8}}\big( -2  + \sum_{n=1}^\inf q^n\,(\tr_{K_n} g)\big)
\ee
satisfies $\phi_g = \phi_g'$, where $\phi_g$ is as defined in \eq{K3_EG_Fou} and $\phi_g'$ is the explicitly given weak Jacobi form \eq{m24_weakJacform}.
Moreover, the representations $K_n$ are even in the sense that they can all be written in the form $K_n=k_n\oplus k_n ^{\,\ast}$ for some $M_{24}$-representations $k_n$ and their dual representation $k_n ^{\,\ast}$.
\end{conj}

A proof of the key fact in the above conjecture, namely the existence of an $M_{24}$-module $K=\bigoplus_{n=1}^\inf K_n $ such that \eq{rep_mock} holds, has been attained in \cite{gannon2016much}. However, a construction of the module $K$, analogous to the construction of $V^\natural$ by Frenkel--Lepowsky--Meurman in the case of monstrous moonshine, is still absent. Therefore in no way do we know why $K$ should be ``natural''. 
As explicit data, the first few Fourier coefficients of the $q$-series $H_g(\t)$ and the corresponding $M_{24}$-representations are given in \cite{cheng2011largest}. 

Note that the above implies that there is a $M_{24}$-supermodule  underlying all terms in the $q$-, $y$-expansion of the $K3$ elliptic genus.
It is hence tempting to endow the McKay--Thompsen series $\phi_g$ with the physical interpretation as twined elliptic genera of $K3$ CFT (cf. \eq{def:twined_elliptic_genus}). This (im)possibility will be extensively discussed in \S\ref{chap:moonshineK3}.
Finally, note that the modular form property of $\phi_g$, as well as the mock modular property of $\mu(\t,z)$, immediately lead to the fact that $H_g$ are also mock modular forms. Explicitly, they are given by 
\be\label{h_g_explicit}
H_g(\t) = \frac{\tr_{\bf R} {g } }{24} H(\t) - \frac{\til T_g(\t)}{\eta(\t)^3}~,
\ee
and they are weight 1/2 mock modular forms with shadows given by $(\tr_{\bf R} {g } )\eta^3(\t)$, generalising the mock modular property of $H(\t)$  discussed around (\ref{completion_H}).

\subsection{Umbral moonshine} 
\label{sec:umbral}

\begin{figure}
\begin{center}
\caption{The construction of umbral moonshine. 
\label{UM_construction}
}
\vspace{10pt}
\includegraphics[scale=0.9]{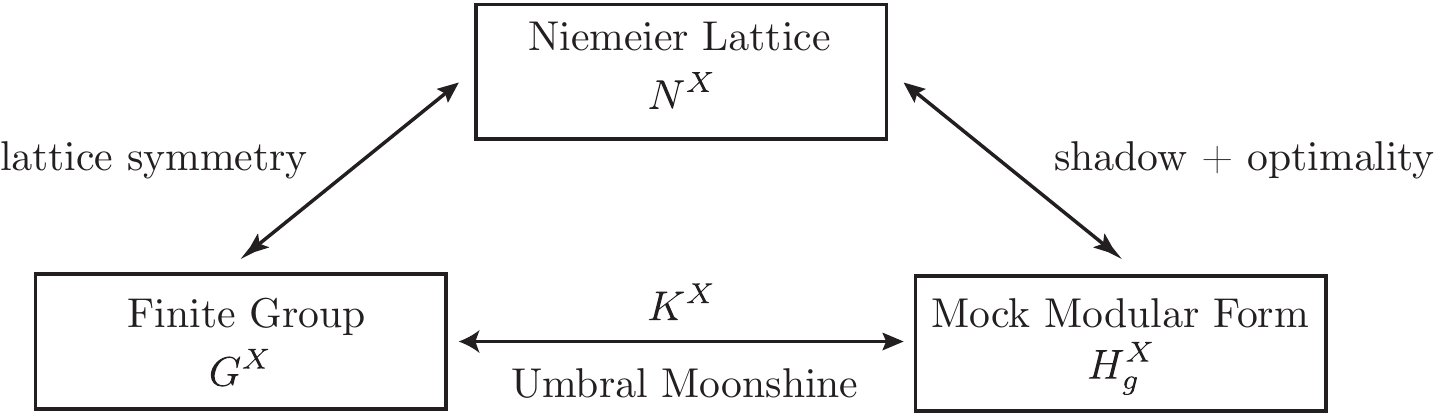}
\end{center}
\end{figure}

A few years after the discovery of Mathieu moonshine,  it was realised that it is in fact just one instance of a larger system of moonshine, called  ``umbral moonshine" \cite{UM,cheng2014umbral}. 
There are in total 23 instances of umbral moonshine, which admit a uniform description (see Figure \ref{UM_construction}).  
The main statement of umbral moonshine is as follows. 

\begin{conj}\label{conj_mock}
Let $G^X$ be one of the 23 finite groups specified in (\ref{eqn:def_um_grp}), $m$ be the corresponding positive integer specified in \eq{eqn:Cox_index}, and $I^X$ be the specific subset of $\{1,2,\dots,m-1\}$ described in \eq{def:setIX}. 
Then there exists a naturally defined bi-graded, infinite-dimensional $G^X$-module $$K^X=\bigoplus_{r\in I^X} \bigoplus_{\substack{D\leq 0\\ D=r^2\xmod{4m}}}  K_{r,D}^X$$ such that for any $g\in G^X$ and for any $r\in I^X$, the graded character (``corrected'' by a polar term $ -2 q^{-\frac{1}{4m}}$ as in below) coincides with the component $H^X_{g,r}$ of a vector-valued mock modular forms $H^X_g = (H^X_{g,r})_{r\in I^X}$: 
\be\label{rep_mock}
H^X_{g,r} = -2 q^{-\frac{1}{4m}} \d_{r,1} \  + \sum_{\substack{D\leq 0\\ D=r^2\!\!\xmod{4m}}}^\inf q^{-D/4m}\,(\tr_{K_{r,D}^X} g). 
\ee
\end{conj}
In what follows we will briefly describe the specification of the main players, the finite groups $G^X$ and the mock modular forms $H^X_g$, in the above conjecture. See Figure \ref{UM_construction}.

The starting point of this uniform construction are the 23 Niemeier lattices $N^X$ introduced in \S\ref{sec:sporadic_lattices}. Recall that they are uniquely labelled by their root systems. We will denote by $X$ the root systems, by $N^X$ the corresponding Niemeier lattices, and by $G^X$ the finite groups arising from the automorphisms Aut$(N^X)$ via  \eq{def:niemeier_grp}: 
\be\label{eqn:def_um_grp}
G^X := {\rm Aut}(N^X)/ {\rm Weyl}(X)
\ee
These are the finite groups relevant for umbral moonshine and we will refer to them as the {\em umbral groups}.

On the modular side, we use the root system $X$ to specify certain mock modular forms  related to the finite group $G^X$. 
To explain how this is done, first recall that the McKay--Thompson series $T_g$  in monstrous moonshine and the mock modular forms $H_g$ in Mathieu moonshine have very special properties.
First, once their (mock) modular data (consisting of the group $\Gamma_g$, the weight, and the multiplier)  are specified, the functions are completely determined by the analyticity property of how they grow near the cusps $i\infty\cup \QQ$.  
Second, they have ``optimal growth'' in the following sense.  These functions 1)  are bounded at all cusps that are not $\Gamma_g$-equivalent to $i\infty$ and 2) have the slowest possible growth near $i\infty$ that is compatible with the modular data. 
For instance, in the case of monstrous moonshine it is elementary to see that a modular form satisfying condition 1) for $\Gamma_g\supset \langle T \rangle$  must behave like $q^{-n}(1+ O(q))$ for some  integer $n$ near the cusp $i\infty$. As a result, the condition 2) states that $n=1$, which is indeed the case for the moonshine functions $T_g$. 
Another way to state the above is to say that the functions in monstrous and Mathieu moonshine can be written (up to a constant) in terms of Rademacher sums over the minimal polar term in the expansion near $i\infty$   \cite{DunFre_RSMG,cheng2012rademacher}. See also \S\ref{sec:mm} for a discussion on Rademacher sums.

The functions of umbral moonshine turn out to have analogous uniqueness properties, and the relevant concept here is the notion of optimal mock Jacobi forms. We will first focus on the case $g=e$ and $\Gamma_g = \SL_2(\ZZ)$. 
Let $\psi=\sum_r h_r\theta_{m,r}$ be a mock Jacobi form of weight one and index $m$. We say it is an optimal mock Jacobi form if 
\begin{gather}\label{eqn:intro:opt}
h_r(\tau)=O(q^{-\frac1{4m}})
\end{gather} 
as $\Im(\t)\to \infty$, for each $r\in\ZZ/2m$. 
For instance, the function $\psi^{3E_8}$ defined in \S\ref{subsec:MMF} is an optimal mock Jacobi since it has index 30 and $H_1^{3E_8}(\t)= -2q^{-{1\over120}}( 1+O(q))$, while $H_7^{3E_8}$ vanishes at $\Im(\t)\to \infty$ (cf. \eq{eqn:3E8}). Similarly, $\psi^{24A_1}$ is an index 2 optimal mock Jacobi form.

At weight one, the space of such optimal mock Jacobi forms turns out to be very restricted: 
the mock modular transformation property together with the pole structure of the functions near the cusps are sufficient to determine the whole $q$-series. 
In particular, they can be obtained as simple Rademacher sums involving only the polar parts as input. 
Such forms are even more scarce if we want them to have non-transcendental Fourier coefficients. 
Note that this must be the case for the function to play a role in moonshine, since the graded dimensions are necessarily integers and of course non-transcendental. 
In \cite{omjt} it is shown that if $\psi$ is such a form, it must lie in a 34-dimensional space, irrespective of its index. 
Moreover, inside this 34-dimensional space there are 39 special elements (which span the 34-dimensional space) distinguished by their special symmetries.
Recall that  Atkin--Lehner symmetries normalising $\Gamma_0(m)$ are specified by a subgroup $K$ of the group of exact divisors Ex$_m$ (cf. \eq{eqn:groupAL}). 
Given such a pair $m$ and $K$, we say that an index $m$ mock Jacobi form $\psi$ is {\em $K$-symmetric} if 
\be\label{def:ALsymJac}
\psi = \sum_{r ~ \rm{mod} ~2m} h_r\theta_{m,r}= \sum_{r ~ \rm{mod} ~2m} h_{r}\theta_{m,a(n)r}~~{\text{ for all }} n\in K~,
\ee
where, for a given $n$, we define $a(n)$ to be the unique element in $\ZZ/2m$ satisfying
\[
a(n) = \begin{cases} 1\xmod{2m/n} \\  -1\xmod{2n}\end{cases}. 
\]
Note that the symmetry is an involution, since $a^2=1\xmod{2m}$.
For instance, the mock Jacobi form $\psi^{3E_8}$ introduced in (\ref{eqn:3E8}) is invariant under the action of $K=\{1,6,10,15\}<{\rm{Ex}}_{30}$, corresponding to $a(n)=1,11,19,29$. 
The surprising result in \cite{omjt} then states that a non-vanishing $K$-symmetric index $m$ optimal mock Jacobi form at weight one has non-transcendental coefficients if and only if the corresponding $SL_2(\RR)$ subgroup $\Gamma^{m+K}$ defines a genus zero quotient in the upper-half plane. Recall that these genus zero groups also play an important role in monstrous moonshine. 
Note that we necessarily  need to have $m\not\in K$ (referred to as the ``non-Fricke'' property) for the mock Jacobi form to be non-vanishing, since at weight one has $\sum_{r } h_r\theta_{m,r}=-\sum_{r } h_r\theta_{m,-r}$ and $a(m)=-1$.
There are just 39 such non-Fricke genus zero groups $\Gamma^{m+K}<\SL_2(\RR)$ and we will denote the corresponding unique optimal mock Jacobi form, with the normalisation
\be\label{def:normalisation_um}
h_1 = -2 q^{-\frac1{4m}} (1+O(q) )~,\ee
by $\psi^{m+K}$. 
In fact, these 39 distinguished optimal mock Jacobi forms $\psi^{m+K}$ turn out to have Fourier coefficients that are not only non-transcendental, but also integral. Moreover, 23 among the 39 have positive coefficients in the following sense. By writing (cf. \eq{def:thetahat})
\[
\psi = \sum_{1\leq r\leq m-1}h_r \tilde \th_{m,r}~,\] $h_r$ has the expansion
\be \label{def:positivity}
h_r = \begin{cases} -2q^{-1/4m}+ \sum_{n\geq 0}c_{r,n} q^{n/4m}  ,~& {\rm{if} }~r^2=1\xmod{4m}  \\ 
\sum_{{n\geq 0}}c_{r,n} q^{n/4m}  ,~& \rm{otherwise }\end{cases}~,
\ee
with 
\(
c_{r,n} \in \ZZ_{\geq0}
\).
This positivity property makes it possible for it to be the graded dimensions of finite group representations\footnote{It is believed \cite{omjt} that the remaining 16  optimal mock Jacobi forms $\psi^{m+K}$ with positive and negative integral coefficients have also umbral type moonshine attached to them, but with additional supermodule structure that accounts for the minus sign.}. To sum up,  for any index, there are 23 special mock Jacobi forms of weight one for $\SL_2(\ZZ)$ distinguished by 
\vspace{-5pt}
\begin{enumerate}
\item the optimality growth condition (\ref{eqn:intro:opt}), 
\vspace{-8pt}
\item the Atkin--Lehner symmetries \eq{def:ALsymJac},  
\vspace{-8pt}
\item the normalisation (\ref{def:normalisation_um}), 
\vspace{-8pt}
\item the positivity and integrality of the coefficients \eq{def:positivity}.
\end{enumerate}

The interesting observation is that these $\psi^{m+K}$ with positivity properties are in one-to-one correspondence with the 23 Niemeier root systems $X$!
To explain this correspondence, first recall the ADE classification of the modular invariant combination of $\widehat{A_1}$ affine Lie algebras   \cite{Cappelli:1987xt}, which has led to a classification of ${\cal N}=2$ superconformal minimal models with spectral flow symmetries. Their classification gives rise to a square matrix $\Omega^Y$ of size $2m$ for each simply-laced root system $Y$, where $m$ coincides with the Coxeter number of $Y$. Moreover, the term $(\Omega^Y_{r,r}-\Omega^Y_{r,-r})$ coincides with the multiplicity of $r$ as a Coxeter exponents (the degrees of the invariant polynomials shifted by one) of $Y$ and takes values in $\{0,1,2\}$. 
The above can be generalised to a union of simply-laced root systems with the same Coxeter number (recall that this is indeed the case for Niemeier root systems)  $X=\cup_i Y_i$ by defining $\Omega^X= \sum_i \Omega^{Y_i}$. 
Then the mock Jacobi form $\psi^X = \psi^{m+K}$, with theta-decomposition $\psi^X=\sum_r H^X_r\theta_{m,r}$, corresponding to the Niemeier root system $X$, display the following relations to $X$.
\vspace{-6,5pt}
\begin{enumerate}
\item The Coxeter number of $X$ coincides with the index of the $\psi^X$,
\vspace{-0,5pt}
\be\label{eqn:Cox_index}
m = {\rm Cox}(X)~.
\ee
\vspace{-25pt}
\item  The matrix $\Omega^X$ and the form $\psi^X=\psi^{m+K}$ have the same Atkin--Lehner symmetries: $(\Omega^X)_{r,r'}=(\Omega^X)_{r,a(n)r'}$ for all $n\in K$. Using these symmetries, it is  convenient to define a set $I^X$ of the orbits of the Atkin--Lehner symmetry group acting on $\{1,\dots,m-1\}$ (in $\ZZ/2m$), labelling the independent components $H_r^X$ of the vector-valued mock modular form $(H^X_r)$ and leading to 
\be\label{def:setIX}
\psi^X = \sum_{r\in I^X} H^X_r \sum_{n\in K} \tilde \theta_{m,a(n)r} ~. 
\ee
\vspace{-15pt}
\item The shadow of $\psi^X$ is determined by $\Omega^X$. More precisely,  the completion of $\psi^X$ is specified by the skew-holomorphic Jacobi form $\sigma = \sum_r \overline{\theta^1_{m,r}} \O^{X}_{r,r'} \theta_{m,r'}$ (cf. \eq{eqn:jac:mck-hhat} and the preceeding text).
\end{enumerate}
\vspace{-6,5pt}
The description of $\psi^{24A_1}$ and $\psi^{3E_8}$ in \S\ref{subsec:MMF} provides examples of the above. 
The mock Jacobi form $\psi^X$ then gives us the vector-valued mock modular forms $H^{X} = (H^X_r)$ which will play the role of the graded dimensions of the module for the umbral group $G^X$. In other words, we have $H^X_r=H^X_{e,r}$ in Conjecture \ref{conj_mock}.
For the case $X=24A_1$, this is the Mathieu moonshine function $\psi^{24A_1}$ we discussed in \S\ref{sec:modular} and \S\ref{subsec:Mathieu}. 
Another simple example is $X=12A_2$, where $\psi^{X} =\sum_{r=1,2} H^X_r\tilde \theta_{3,r}$ (so $m=3$ and $I^X=\{1,2\}$), with
\begin{gather}\begin{split}
H^X_1(\t)& =   2 q^{-1/12} (-1 + 16 \,q+ 55 \,q^2+ 144  \,q^3+\dots)\\ 
H^X_2(\t)& =  2 q^{8/12} (10 + 44 \,q+ 110 \,q^2+\dots) . 
\end{split}
\end{gather}
At the same time, the symmetries of the corresponding Niemeier lattice gives $G^X \cong 2.M_{12}$. The relation between the finite group $G^X$ and the  vector-valued mock modular form $H^X$ can be observed from the fact that the group $2.M_{12}$ has irreducible representations of dimensions $16,55, 144$ as well as $10,44,110$.

After specifying the mock Jacobi forms for $\SL_2(\ZZ)$, in order to describe the moonshine relation we also need a set of mock Jacobi forms $\psi^X_g= \sum_r H_{g,r}^X\theta_{m,r}$, one for each conjugacy class $[g]\subset G^X$,  for subgroups of $\SL_2(\ZZ)$. 
The mock modular forms $H^X_g=(H^X_{g,r})$ will then play the role of graded characters of the umbral moonshine module, as described in Conjecture \ref{conj_mock}. 
This can be achieved in a way largely analogous to the $\SL_2(\ZZ)$  case, though additional subtleties do emerge and extra care needs to be taken. We refer to \cite{cheng2018weight} for more details. 

Once the mock modular forms $H^X_g$ are specified, it is trivial to verify the existence of the $G^X$-module $K_{r,D}^X$ in Conjecture \ref{conj_mock} term by term, namely one $D$ at a time. Furthermore, the existence of the whole umbral module $K^X=\bigoplus_r \bigoplus_D K_{r,D}^X$ has been proven mathematically using properties of (mock) modular forms \cite{gannon2016much,DuncanGriffinOno}. However, the construction, or even an understanding of the exact nature of $K^X$, is not yet obtained in general. 
Construction of $K^X$ has so far only been achieved for  certain particularly simple cases of umbral moonshine, corresponding to Niemeier root systems 
$3 E_8$ \cite{duncan2017umbral}, $4A_6$ and  $2A_{12}$ \cite{Duncan:2017bhh},  $4D_6$, $3D_8$, $2D_{12}$ and $D_{24}$ \cite{Cheng:2017grj}, as well as 
$6D_4$ \cite{Anagiannis:2017src}. 
The construction in \cite{duncan2017umbral} relies on special identities satisfied by the mock modular forms $H^{3E_8}_{g,r}$ relating it to a lattice-type sum, while in \cite{Duncan:2017bhh,Cheng:2017grj} the modules are constructed using the interpretation of the meromorphic Jacobi forms associated to $\Psi^X_{g}$
as the twined partition function of certain vertex operator algebras (or chiral CFTs). 
In \cite{Anagiannis:2017src} the module for the $6D_4$ case of umbral moonshine is constructed by exploiting the relation between  the (twined) $K3$ elliptic genus, umbral and Conway moonshine, which we will explain in \S\ref{sec:umK3}. Note that, so far, this is the only constructed module for which the corresponding umbral group (when embedded in $\Co_0$) does not fix a $4$-plane in the $24$-dimensional representation of $\Co_0$. The significance of this will be discussed in \S \ref{sec:symK3}.

Moreover, generalised umbral moonshine, analogous to the generalised monstrous moonshine discussed in \S\ref{sec:mm}, has been established in \cite{GUM}, hinting that some elements  of CFT/modular tensor category structure should be present at the umbral moonshine module $K^X$. 
Despite these results, it is fair to say that a uniform construction of the umbral module, reflecting the uniform description of umbral moonshine, is currently one of the biggest challenges in the study of moonshine. 

We will end our review on umbral moonshine by noting a special property, called {\em discriminant property}, of umbral moonshine. 
It relates the discriminants $D$ (the power of individual terms $q^{-D/4m}$ in the $q$-series $H^X_r$) and the number field generated by the characters of representations showing up in $K_{r,D}^X$. For instance, in the case $X=24A_1$, the $M_{24}$-representation underlying the $q^{7/8}$ term in $H_g= H^{24A_1}_{g,1}$ is $K^{24A_1}_{-7} = \rho \oplus \rho^\ast$, where $\rho$ is a 45-dimensional irreducible representation and $\rho^\ast$ is the dual representation. At the same time, $\tr_\rho g$ (and hence also  $\tr_{\rho^\ast} g$) generates the field $\QQ(\sqrt{-7})$. Analogous relations continue for larger $q$-power as long as $\QQ(D)=\QQ(-7)$, and similar properties hold uniformly for all 23 cases of umbral moonshine.  
At present there is no physical understanding of this surprising and profound-looking property.

\subsection{Thompson moonshine}
The Thompson sporadic group $\Th$ is a subgroup of the Monster that has order $|{\rm \it \Th}|\sim9\cdot10^{16}$ and contains $48$ conjugacy classes. 
Thompson moonshine describes a special relation between the representation theory of $\Th$ and certain weakly holomorphic modular forms of weight $1/2$. 
It was first discussed in \cite{Harvey:2015mca} and is the first instance of the so-called {\em skew-holomorphic moonshine} \cite{skew_hol_moonshine} relating skew-holomorphic Jacobi forms and finite group representations. See \S\ref{subsec:shjacobiforms} and  \S\ref{sec:other_physics} for more mathematical and physical information on skew-holomorphic Jacobi forms.

Just like Mathieu and umbral moonshine, Thompson moonshine relates finite groups connected to the Monster (cf. \S\ref{sec:sporadic_lattices}) and weight 1/2 modular objects. However, there are a few differences. 
First, the modular objects in Thompson moonshine are modular while in umbral moonshine they are mock modular with non-vanishing shadows. 
Second, there is a super-structure in the Thompson moonshine module, i.e. the representations come with signs. See Conjecture \ref{conj:thompson}. 
That said, this extra $\ZZ_2$-grading structure is very simple and completely controlled  by $m \xmod 4$, where $m$ is the natural $\ZZ$-grading corresponding to the exponents of $q$ in the moonshine function. 
Finally, we remark that Thompson moonshine also enjoys a discriminant property analogous to that of umbral moonshine, as discussed at the end of \S \ref{sec:umbral}.

To describe Thompson moonshine, we start with the following conjecture \cite{Harvey:2015mca}: 
{\conj \label{conj:thompson} There exists a $\ZZ$-graded $\Th$-supermodule (cf. \eq{def:supermodule})
\begin{equation}
W=\bigoplus_{m\geq-3\atop m\equiv0,1\bmod4}^\infty W_m~
\end{equation}
such that the graded dimensions of $W_m$ are related to a certain weight $1/2$ weakly holomorphic modular form $\mathcal{F}_3$ (\ref{eq:F3}) by
\begin{equation}
\mathcal{F}_3(\tau)=\sum_{m=-3\atop m\equiv0,1\bmod4}{\str}_{W_m}(1)q^m~.
\end{equation}
Furthermore, the associated McKay-Thompson series 
\begin{equation}
\mathcal{F}_{3,g}(\tau)=\sum_{m=-3\atop m\equiv0,1\bmod4}{\str}_{W_m}(g)q^m~,
\end{equation}
are  given by certain weakly holomorphic weight $1/2$ modular forms, 
with $\mathcal{F}_{3,e}=\mathcal{F}_3$.
Finally, $W_m$ is even under the superspace $\ZZ_2$-grading if $m\geq0$ and $m\equiv0\bmod4$ or when $m=-3$, and is odd otherwise. 
}\\

\noindent Let's discuss the elements of this conjecture. The function $\mathcal{F}_3$ is defined by
\begin{equation}
\label{eq:F3}
\mathcal{F}_3(\tau):=2f_3(\tau)+248\theta(\tau)~,
\end{equation}
where $\theta(\tau)$ is the Jacobi theta function discussed in \S\ref{subsec:modular_forms}. We next describe the specification of  $f_3$. 

Define the Kohnen-plus space $M_{1/2}^+$ as the set of holomorphic functions that transform like $\theta$, and additionally satisfy the condition $c(n)=0~\forall~n\neq0,1\bmod4$ in their Fourier expansions $\sum c(n)q^n$. It can be shown that in fact $M_{1/2}^+$ is one-dimensional, spanned by $\theta$ itself. However, if we allow for poles at the cusps, we can get a nontrivial space $M_{1/2}^{!,+}$ of weakly holomorphic modular forms of weight $1/2$ in the Kohnen-plus space. In \cite{Zag_TrcSngMdl}, Zagier constructed an infinite-dimensional basis for $M_{1/2}^{!,+}$, spanned by the (unique) functions
\begin{equation}
f_d(\tau)=q^{-d}+\sum_{n>0}A(n,d)q^n~,~~d\equiv0,3\bmod4~,
\end{equation}
where $f_0=\theta$. Here we are mainly interested in the case $d=3$. 
Using uniqueness, it is easy to show that $f_3$ can be explicitly specified as
\begin{equation}
\label{eq:by_hand_f3}
f_3(\tau)=-\frac{1}{20}\left(\frac{[\theta(\tau),E_{10}(4\tau)]}{\Delta(4\tau)}+608\theta(\tau)\right)=q^{-3}-248q+26752q^4-\dots~,
\end{equation}
where $E_{10}=E_4E_6$ is the weight 10 Eisenstein series  and $\Delta$ is the weight 12 cusp form introduced in \S\ref{subsec:modular_forms}. 
The \emph{Rankin-Cohen bracket} is defined by $[f,g]:=kfDg-lgDf$, with $D=\frac{1}{2\pi i}\frac{d}{d\tau}$ and $k,l$ being the weights of the modular forms $f,g$, respectively. 

Thompson moonshine was initiated by the observation that the first few Fourier coefficients of $f_3$ can be expressed in terms of the dimensions of irreducible representations of $\Th$. For example $1=1$, $248=248$ and $26752=27000-248$. It is worth noting that this connection between $f_3$ and $\Th$ had already appeared in the context of generalised monstrous moonshine \cite{DixonL.1988Batb,BorcherdsRichard1995AfoO,MR2904095}, cf. \S\ref{sec:mm}.  Specifically, the centraliser of the class $3C$ of $\MM$ is isomorphic to $\mathbb{Z}_3\times\Th$ \cite{conway_norton}, and consequently the twisted module $V^\natural_{3\text{C}}$ carries a natural action of $\Th$. The corresponding twisted generalised character is given by the so-called \emph{Borcherd's lift} of $f_3$,
\begin{equation}
Z(3\text{C},e;\tau)=q^{-\frac{1}{3}}\prod_{n>0}(1-q^n)^{A(n^2,3)}~.
\end{equation}
Notice however that the above connection to generalised monstrous moonshine is limited only to the coefficients $A(n^2,3)$ of $f_3$, which does not explain why the rest of its coefficients also exhibit a relation to $\Th$. 
To do better, the insight of  \cite{Harvey:2015mca} is to ``correct'' the function by  including additional contributions from the theta function $\theta(\t)$, and consider instead   
\begin{equation}
\mathcal{F}_3(\tau):=2f_3(\tau)+248\theta(\tau)=2q^{-3}+248+2\cdot27000q^4-2\cdot85995q^5+\dots~, 
\end{equation}
which belongs to the Kohnen plus space. 
Note that the alternating signs correspond to the superspace structure mentioned in Conjecture \ref{conj:thompson}.

The McKay--Thompson series ${\cal F}_{3,g}$ of Thompson moonshine have a structure very similar to the above. 
Schematically, they can be most naturally viewed as a sum of two contributions
\be\label{thompson_mckaythompson}
{\cal F}_{3,g} = {\text{Projected Rademacher Sum}} + \th{\text{-corrections}}~,
\ee
where the first term is the Rademacher sum of the unique polar term $2q^{-3}$ for the suitable group $\Gamma_g$ and multiplier system, projected onto the Kohnen plus space, and the second term again involves linear combinations of $\theta(d^2\t)$, where $d$ are certain integers compatible with the group $\Gamma_g$. We refer to \cite{Harvey:2015mca}, and in particular \cite{GriffinMichael2016Apot} for more details. 

It is interesting to note that the  modular objects in Thompson moonshine enjoy a close relation to the so-called traces of singular moduli, which are important quantities  in number theory. 
As mentioned in the final example of \S\ref{subsec:MMF}, the generating functions of the traces of singular moduli turn out to have interesting modular properties \cite{Zag_TrcSngMdl}.
In particular, we have 
\begin{equation}
A(n,3)=\frac{1}{\sqrt{n}}\sum_{Q\in\mathcal{Q}_{-3n}/\PSL_2(\ZZ)}\chi_{n,-3}(Q)J(\alpha_Q)~,~~n\in\ZZ_+~,
\end{equation}
coinciding with the Fourier coefficients of the modular form $f_3$ which participates in Thompson moonshine. 
In the above, $\chi_{n,-3}$ denotes a certain genus character whose precise definition can be found in \cite{Zag_TrcSngMdl}. 
An analogous relation to  traces of singular moduli can be established also at higher level, when one only keeps track of $\Gamma_0(N)$-equivalences of quadratic forms.

\section{Moonshine at weight three-halves}
\label{ch:moonshineweight32}

This is an arena for moonshine that is currently in rapid development. As a result we will be very brief and just highlight a few key features and prospects of the recent developments. 

After the fruitful explorations in moonshine at weight one-half, it is natural to wonder about the landscape of moonshine in the dual weight, $w=2-\frac{1}{2}=\frac{3}{2}$. 
Weight $3\over 2$ mock modular forms are much less constrained than their weight $1\over 2$ counterparts, and they 
 turn out to provide a fertile ground to study the connections to finite groups, already giving us the {\em O'Nan moonshine} \cite{onan} and the {\em class number moonshine} \cite{class_nr_moonshine}, the latter involving the Hurwitz-Kronecker class number mock modular form described in \eq{hurwitz_mock}. 
 
It is more likely than not that there are more treasures in weight 3/2. Here we limit ourselves to commenting on the two established moonshine examples. 
We will point out three conspicuous properties of the two examples of moonshine at weight three-halves\footnote{In fact, these properties make the weight 3/2 connections so different from the moonshine examples at weight zero and weight one-half discussed thus far, such that one might  wonder whether we should still use the term ``moonshine'' for them.}.

\myparagraph{Supermodules.}
In monstrous and Conway moonshine discussed in \S\ref{cha:weight0}, the underlying modules are ``real'' modules (as opposed to supermodules  \eq{def:supermodule}) and the group characters involved are ordinary traces (as opposed to supertraces).
The same is true in Mathieu and umbral moonshine, up to the polar terms in \eq{rep_mock} which have negative coefficients but do not carry any non-trivial information on group representations. 
In Thompson moonshine we encounter supermodules, but only in a rather trivial way: the supermodule is purely even or purely odd depending on whether the corresponding $q$-power is $+1$ or $-1$ mod 4. In particular, in weight one-half we do not encounter 
 moonshine examples in which Fourier coefficients of mock modular forms are given by differences of traces of group representations, i.e. in every homogeneous component the supermodule is always either even or odd.

In weight three-halves this is no longer true and the super-ness of the supermodules is an intrinsic and unavoidable property of the two moonshine examples we discuss.

\myparagraph{Cusp forms.}
In the moonshine examples we discussed so far, the (mock) modular objects involved are very distinguished in that they are essentially uniquely determined by their modularity and analyticity properties.  
In particular,  the graded characters of monstrous and Conway moonshine are Hauptmoduls and  can be written as, up to an additive constant, Rademacher sums with a single polar term $q^{-1}$. The graded characters of umbral (including Mathieu) and Thompson moonshine are also (projected) Rademacher sums up to addition of theta functions (cf. \eq{thompson_mckaythompson} and \cite{cheng2018weight}). The latter can arguably be viewed as a minor correction since the coefficients of theta functions are vanishing at almost all exponents (vanishing unless the exponenets are square numbers up to a specific rescaling) and they do not carry any interesting number theoretic information. 

This crucial uniqueness property,  which serves to distinguish mock modular forms for weight zero and weight one-half moonshine,  
is no longer the relevant criterion  in weight three-halves moonshine. The mock modular forms in O'Nan and class number moonshine involve weight 3/2 cusp forms in a very pronounced way. (See the end of \S\ref{subsec:modular_forms} for the definition of cusp forms.)

This property opens the door to connecting with arithmetic geometry, a totally different area in mathematics. 
The main ingredients of this connection are: 1) The  Waldspurger-type relation among coefficients of weight 3/2 newforms and the special $L$-values of the corresponding weight two newforms under the so-called Shimura correspondence \cite{MR783554}. 2) The modularity theorem relating the $L$-function of elliptic curves and that of certain weight two newforms. 3) The (strong form of) Birch and Swinnerton-Dyer conjecture (and its proven elements) relating special $L$-values of the elliptic curves and their important arithmetic data including the order of the Tate-Shafarevich group. See \cite{MR833192,MR954295,MR3148103}. 4) The congruences among the moonshine modular objects stemming from the congruences among the characters which is a basic property of finite group representations. 
Combining the above ingredients, the connections between weight 3/2 cusp forms and finite groups  lead to an infinite sequence of statements on the $p$-part of the Tate-Shafarevich group, organised by the finite group structure. Moreover, in class number moonshine the torsion subgroups of the elliptic curves (and other Abelian varieties in general) attached to certain modular data play an important role in the construction of moonshine mock modular forms. Exploring the implication of the finite group structure in mock modular forms in arithmetic geometry is an interesting new mathematical direction in the study of moonshine.

\myparagraph{New finite groups.}
The above two properties ,  regarding the finite group and the modular form sides of the moonshine connection respectively, makes it possible for weight three-halves moonshine to encompass new types of finite groups into the realm of moonshine. 

O'Nan moonshine, as the name suggests, connects a set of weight 3/2 mock modular forms to the O'Nan group. The O'Nan group was discovered in 1976 in the midst of the intense developments related to the classification of finite groups, and is one of the 6 pariah groups not related to the Monster. The class number moonshine, on the other hand, delivers a family of infinitely many weight 3/2 mock modular forms arising from Hurwitz--Kronecker class numbers \eq{def:hurwitz}, which accommodates representations of {\em infinitely many groups}. 
Recall that, apart from the sporadic groups there are three infinite families in the classification of finite simple groups, and one of them is the cycle groups $\ZZ_p$ of prime order $p$. 
It turns out that for each prime number $p$ one has a mock modular form ${\cal H}_{p,1}$  of level $4$,  a scalar multiple of the class number mock modular form ${\cal H}$ \eq{hurwitz_mock},
and a mock modular form ${\cal H}_{p,p}$ of level $4p$, such that they coincide with the two graded characters of a $\ZZ$-graded supermodule for  $\ZZ_p$. 
Apart from this infinite family of groups, it has also been shown in \cite{class_nr_moonshine} that mock modular forms at different levels (given by 4 times the order of the group elements) conspire to encode supermodules for interesting Mathieu groups.

These instances of weight three-halves moonshine greatly broaden the horizon of  connections between modular objects and finite groups. All instances of moonshine at weight zero and one-half are connected to groups related to the Monster, but the O'Nan and the class number examples show that this does not have to be the case. In these examples it was demonstrated that a pariah sporadic group as well as one of the 3 infinite families of non-sporadic finite simple groups are also related to (mock) modular forms in a similar way! This suggests the tantalising possibility that moonshine-type connections are in fact ubiquitous and an intrinsic property in the landscape of modular objects and finite groups, and that the latter is an important organising structure of the former. 

Last but not least, these new connections, involving supermodules of infinitely many groups, pose interesting challenges to physics. Can physics still provide the finite group modules and an explanation for these connections? If so, what are the relevant physical systems that can give rise to these supermodules?

\newpage

\part{Moonshine and string theory}

String theory has been playing a key role in moonshine from its earliest years, starting with the CFT construction of the monstrous moonshine module and  the string-inspired proof of monstrous moonshine by Borcherds (cf. \S\ref{sec:mm}).  String theory is also expected to play an interesting role in the understanding of Mathieu and umbral moonshine, and we hope that moonshine can teach us something novel and important about certain aspects of string compactification. 
More generally speaking, the connection to physics is one of the key reasons why many string theorists and mathematicians find moonshine exciting. 
In this part we will discuss some of these connections. 
In \S\ref{sec:physics_monstrous} we outline the relation between monstrous moonshine and certain (1+0)-dimensional string compactification. In \S\ref{chap:moonshineK3}, the main section of this part, we summarise the known relations between various moonshines and $K3$ compactifications. 
Finally we briefly mention some other connections discussed recently in \S\ref{sec:other_physics}.

\section{Monstrous moonshine}
\label{sec:physics_monstrous}

In this section we briefly describe a recent string theory construction, developed in   \cite{paquette2016monstrous,Paquette:2017xui}, involving the monstrous moonshine module $V^\natural$ and the associated twisted module $V^\natural_g$ discussed in \S\ref{sec:mm}.
Although the striking genus zero property of monstrous moonshine has been mathematically proven, its physical interpretation has remained elusive for a long time. The recent work  \cite{paquette2016monstrous,Paquette:2017xui} sheds light on this important property by considering certain heterotic string theory compactifications down to $0+1$ dimensions. 

In \cite{paquette2016monstrous,Paquette:2017xui}, one takes as the starting point a compactification down to $1+1$ dimensions where the  internal CFT, which has central charge $(24,12)$, is given by $V^\natural\times V_{}^{s\natural}$. 
The second factor is the Conway module introduced in \S\ref{sec:conway_moonshine}, which plays mostly the role of the spectator whose main function is to provide the right-moving $\mathcal{N}=1$ supersymmetry of the world-sheet theory. This model turns out to have ${\cal N}=(0,24)$ spacetime supersymmetry in two-dimensions.  We further compactify the remaining spatial direction on a circle. This additional compactification is crucial for computing the ``BPS index'' described below and for the construction of more general models. 
More generally, for each $g\in\MM$ one constructs a CHL-like model by orbifolding the theory by a symmetry which acts on $V^\natural$ as $g$ and on the circle as a shift of order dictated by certain properties of the action of $g$ on $V^\natural$ (and coincides with the order of $g$ in the simplest cases). To avoid subtleties involving gravitational anomalies, this theory should be considered at the zero-string coupling limit.

In \cite{paquette2016monstrous} it is shown that the (first-quantized) BPS states of the above models form a module of the generalised Kac-Moody algebras $\mathfrak{m_g}$, which is the monster algebra constructed by Borcherds for $g= e$ and the ``twisted" monster algebras  constructed by Carnahan  for $g\neq e$  \cite{Carnahan2008,MR2904095,carnahan2012generalized}. 
See also the discussion on generalised monstrous moonshine in \S\ref{sec:mm}. 
We are interested in the (twined) BPS index counting spacetime BPS states in these theories. Denote the relevant Hilbert space by $\mathcal{H}^g_{\text{BPS}}$, we define 
\begin{equation}
Z_{(g,h)}(T,U):=\tr_{\mathcal{H}^g_{\text{BPS}}}\left((-1)^F \, h \, e^{2\pi iTW}e^{2\pi iUM}\right)~,
\end{equation}
where $g,h$ are commuting elements of the Monster. In the above,  $T,U\in \HH$ are complexified chemical potentials for the winding and momentum quantum numbers $W,M$ respectively, with the imaginary parts scaling according to the inverse temperature and radius as 
\be
\im T \sim {\beta R}~,~~ \im U \sim {\beta \over R}~.
\ee
Using the Fock space construction of  $\mathcal{H}^g_{\text{BPS}}$ by allowing for an arbitrary number of free strings, as well as the twisted denominator formula, relating an infinite product to an infinite sum and generalising the Koike--Norton--Zagier formula \eq{eq:replicability_J}, one obtains
\begin{equation}\label{BPSindexmonster}
\begin{split}
&Z_{(e,g)}(T,U)=\left(T_{(e,g)}(T)-T_{(e,g)}(U)\right)^{24} \\
&Z_{(g,e)}(T,U)=\left(T_{(e,g)}(T)-T_{(g,e)}(U)\right)^{24}~.
\end{split}
\end{equation}
Note that $T_{(e,g)}=T_g$ is the McKay--Thompson series of the monstrous moonshine.

The above expressions lead to a physical interpretation of the modularity and  the genus zero property of monstrous moonshine. 
First, it was shown in \cite{paquette2016monstrous}, using the techniques developed in \cite{Persson:2015jka}, that the self-duality group (i.e., a subgroup of the T-duality group mapping between the same CHL model but generically at different values of the moduli)  involves two subgroups of $\SL_2(\RR)$, acting on $T$ and $U$ independently in \eq{BPSindexmonster}.
In particular, the $\SL_2(\RR)$ subgroup acting on $T$ is nothing but $\Gamma_g$, the associated genus zero group in monstrous moonshine (cf. Conjecture \ref{conj:conwa_norton}).
As a result, the physical invariance of the BPS index under self-duality group  immediately leads to the $\Gamma_g$-invariance modularity property of the moonshine function $T_{g}(T)$. 
Second, recall that the Hauptmodul property of $T_{(e,g)}$ can be recast in the claim that they have only a single pole at $\tau\rightarrow i\infty$ and its images under $\Gamma_g$. In terms of the supersymmetric index, this means that $Z_{(g,e)}(T,U)^{1/24}$ and its T-duality images will diverge as $T\rightarrow i\infty$, for fixed $U$. Using the T-duality
\[
Z_{(e,g)}(T,U)=-Z_{(e,g)}\left(-\frac{1}{U},-\frac{1}{T}\right)
\]
and the generalised moonshine relation between $T_{(g^ah^c,g^bh^d)}(\tau)$ and $T_{(g,h)}\left(\frac{a\tau+b}{c\tau+d}\right)$ (cf. \S\ref{sec:mm}), 
this property gets translated to the statement that for all co-prime integers $a$ and $b$,    $Z_{(g^a,g^b)}(T,U)$ is  divergent in the limit $U\to i\infty$ with $T$ fixed, if and only if    this limit is related to the limit of $T\to i\infty$ with $U$ fixed by a self-duality of the CHL model  associated to $g\in \MM$. Now we will argue why this is true from physical considerations. Let us interpolate between the two limits by varying the radius $R$, while fixing the inverse temperature $\beta$ to be large. 
At the low temperature limit,  the divergence at the limit $T\to i\infty$ with $U$ fixed is caused by the ground state contribution $e^{\beta R}$, which blows up as $R\to \infty$. The small radius limit is convergent unless there is a phase transition and a new ground state with contribution $e^{\beta / R}$ emerges at the other side of the critical line. If this happens, one can show that there is an $SU(2)$ symmetry at the self-dual radius limit on the critical line, which in particular contains a T-duality transformation. In other words, the two sides of the critical line are related by a self-duality contained in $\Gamma_g$, and this proves the Hauptmodul property.

\section{Moonshine and $K3$ string theory}
\label{chap:moonshineK3}

As mentioned in \S\ref{sec:Moonshine at weight one-half}, Mathieu moonshine was first noticed in the context of $K3$ sigma models and was later recognised as being one of the twenty-three cases of umbral moonshine. As we will discuss shortly, all instances of umbral moonshine, not just Mathieu moonshine, enjoy a  close relation to $K3$ elliptic genus, the quantity that captures the BPS states of  $K3$ sigma models, and more generally BPS states of superstring theory compactifications involving $K3$ surfaces. This tentative connection to $K3$ string theories is what makes Mathieu and umbral moonshine so interesting for many string theorists. 

This is because $K3$ manifolds have the special property that they are the only non-trivial (i.e. non-toroidal) Calabi--Yau manifolds in two (complex) dimensions, and the resulting string theory compactifications are much more manageable than compactifications on generic Calabi--Yau three-folds due to higher amount of supersymmetries, and contain much richer structure and information than toroidal compactifications which are for many purposes too simple to be useful.  
As a result, $K3$ compactifications have been playing a prominent role throughout the development of string theory, being the playground for developing revolutionary new ideas including black hole microstates counting \cite{StromingerPhys.Lett.B379:99-1041996} and AdS/CFT correspondence \cite{MaldacenaAdv.Theor.Math.Phys.2:231-2521998}.
Furthermore, they also feature in important dualities relating different string theories. 

In this section we will review the observed relation between moonshine and $K3$ sigma models, and outline the challenges to be overcome in order to understand the origin of this relation and to find a physical system which provides the much wanted uniform construction of the umbral moonshine module. 
Finally, we will summarise the different ideas that aim to address these challenges, along with their current status.

 \subsection{Mathieu moonshine and $K3$ elliptic genus}
\label{sec:M24moonshine}

As mentioned in \S\ref{sec:EG}, the Calabi-Yau property of $K3$ together with the Jacobi form property dictates that its elliptic genus is given by $2\f_{0,1}$ in terms of the familiar weak Jacobi form \eq{phi01}. 
Recall that $K3$ is the only Calabi-Yau two-fold besides the four-torus. It possesses the special property that it is not only K\"ahler but also admits a hyper-K\"ahler structure. As a result, the $U(1)$ symmetry of the ${\cal N}=2$ superconformal algebra discussed in \S\ref{sec:EG} 
can be extended to $SU(2)$, and the sigma model has enhanced ${\cal N}\!=4$ superconformal symmetry for both the left and the right movers. 
This leads to a specific decomposition of the elliptic genus of an ${\cal N}=4$ SCFT that we will now explain. 

Since the underlying CFT admits an action by the ${\cal N}=4$ superconformal algebra, the Hilbert space decomposes into a direct sum of (unitary) irreducible representations of this algebra. Hence the elliptic genus (\ref{elliptic_genus}) can be written as a sum of characters of representations of this algebra with some multiplicities. A natural embedding of the $U(1)$ current algebra of the ${\cal N}=2$ superconformal algebra into the $SU(2)$ current algebra of the ${\cal N}=4$ superconformal algebra is obtained by choosing $J_0^3 \sim J_0$. As a result, the ${\cal N}=4$ highest weight representations are again labelled by two quantum numbers $h,\ell$, corresponding to the operators $L_0$, $J_0^3$ respectively, and the character of such an irreducible representation $V_{h,\ell}$ is defined as
$$
{\it ch}_{h,\ell}(\t,z) := \tr_{V_{h,\ell}} \left( (-1)^{J_0}y^{J_0} q^{L_0-c/24}\right) .
$$
For central charge $c=6$, there are two supersymmetric (also called `BPS' or `massless') representations in the Ramond sector, and they have the quantum numbers 
$$
h=\tfrac{1}{4}\;,\; \ell = 0 ,\tfrac{1}{2} \;.
$$
Their characters  have been computed to be \cite{Eguchi1987,Eguchi1988,Eguchi1988a}
\bea\label{massless_characters}
{\it ch}_{\frac{1}{4},0}(\t,z) &=& \frac{\th_1^2(\t,z)}{\eta^3(\t)} \m(\t,z)\, ,\\\notag
{\it ch}_{\frac{1}{4},\frac{1}{2}}(\t,z) &=& q^{-\frac{1}{8}} \frac{\th_1(\t,z)}{\eta^3(\t)} - 2 \frac{\th_1(\t,z)^2}{\eta^3(\t)} \m(\t,z)\, ,
\eea 
where $\m(\t,z)$ 
is the so-called Appell-Lerch sum defined in \eq{def:mu}. 

Notice that the supersymmetric representations have non-vanishing Witten index 
$$ch_{\frac{1}{4},0}(\t,z=0) =1\,, \quad ch_{\frac{1}{4},0}(\t,z=0) =-2\, .$$ 
The non-supersymmetric (or `non-BPS' or `') representations have 
$$
h =\tfrac{1}{4} + n\, ,\quad \ell =\tfrac{1}{2}\, ,\quad\text{} \quad n = 1,2,\dots,
$$
and their characters are given by
\be\label{massive_characters}
ch_{\frac{1}{4}+n,\frac{1}{2}}(\t,z) = q^{-\frac{1}{8}+n} \frac{\th_1^2(\t,z)}{\eta^3(\t)} \, .
\ee
By definition, their contribution to the Witten index vanishes, $ch_{\frac{1}{4}+n,\frac{1}{2}}(\t,z=0) = 0$. This is why the Witten index only receives contribution from quantum states that are of the form (massless$_L\times{\rm massless}_R$) in terms of ${\cal N}=4$ representations, while 
 the elliptic genus  receives contribution from (massless$_L\times{\rm massless}_R$) as well as (massive$_L\times{\rm massless}_R$).

We are now ready to describe the first physical interpretation of the specific way (\ref{splitEGM24}) of  decomposing the $K3$ elliptic genus, which makes the connection to $M_{24}$ manifest. 
From  (\ref{splitEGM24}) and \eq{massless_characters}-\eq{massive_characters}, we see that ${\bf EG}(K3)$ can be expressed as 
\be
{\bf EG}(K3) =  20\, ch_{\frac{1}{4},0} -2 \, ch_{\frac{1}{4},{1\over 2}}  + \sum_{n=1}^\infty c_H(n)\,ch_{\frac{1}{4}+n,\frac{1}{2}}
\ee
where 
\be 
H(\t) =q^{-1/8}\left(-2 +\sum_{n=1}^\infty c_H(n)q^n\right) =2 q^{-\frac{1}{8}} \left( -1 + 45\, q+ 231\, q^2 + 770\,q^3\dots \right)
\ee
is the mock modular form given in \eq{second_def_H}. In particular, the multiplicities $c_H(n)$ of the massive multiplets are equal to the dimensions of the homogeneous component of the $M_{24}$-module $K$ according to Mathieu moonshine Conjecture \ref{conj_mock}.

As mentioned already in \S\ref{subsec:Mathieu}, 
it is then tempting to identify $\f_g$ \eq{K3_EG_Fou}, encoding the McKay--Thompson series of Mathieu moonshine $H_g$, as the elliptic genus twined by certain symmetries of the $K3$ non-linear sigma model whose action preserves the ${\cal N}=4$ SCA. While this is possible for some $g\in M_{24}$, this interpretation is proven to fail for some other group elements. We will discuss this in \S \ref{sec:symK3}.

\subsection{Umbral moonshine and $K3$ elliptic genus}
\label{sec:umK3}

As alluded to before, the ${\cal N}=4$ decomposition is not the only possible interpretation for the splitting (\ref{splitEGM24}) of the elliptic genus, into the part coming directly from the moonshine mock modular form and the contribution from the Appell-Lerch sum $\mu$. 
A second interpretation, in terms of the elliptic genus of singular spaces labelled by ADE root systems, yields the same result in this case but the crucial difference is that it can be readily generalised to the other 22 cases of umbral moonshine.

To explain this, we first associate a function $\eg(X;\tau,z)$ to each of the 23 root systems $X$ in the following way. 
Note that a common definition of the $K3$ surface is that it has at worst 
du Val type surface singularities, i.e. singularities of the complex plane of the form $\mathbb C^2/G$, where $G$ is a finite subgroup of $SU(2)_{\mathbb C}$. These singularities have an ADE classification, in accordance with the famous McKay correspondence. A conformal field theory description of string theory with these ADE singularities as the target space was given in \cite{Ooguri:1995wj}. The form of their elliptic genus was investigated in a number of papers, including \cite{Troost:2010ud,Eguchi:2010cb,Ashok:2011cy,Ashok:2013pya,Murthy:2013mya,Cheng:2014zpa,Harvey:2014nha}.
Let 
$\eg(Y;\tau,z)$ denote the holomorphic part of the elliptic genus of the singularities corresponding to the root system $Y$. 
(See also the remark at the end of \S\ref{sec:EG}.) Extending the definition to unions of the root systems, $\eg(\cup_i Y_i;\tau,z)= \sum_i \eg( Y_i;\tau,z)$, we obtain a definition of $\eg(X;\tau,z)$ for each of the 23 Niemeier root systems $X$. 
For instance, 
\be
\eg(A_1;\tau,z) = ch_{\frac{1}{4},0}(\t,z) = \frac{\th_1^2(\t,z)}{\eta^3(\t)} \m(\t,z), 
\ee
and $\eg(24A_1;\tau,z) = 24\frac{\th_1^2(\t,z)}{\eta^3(\t)} \m(\t,z)$ is the first term in the splitting of $\eg(K3)$ in \eq{splitEGM24}. 

Recall that, for each of the 23 Niemeier lattices $N^X$, uniquely determined by its root system $X$, umbral moonshine associates a finite group $G^X$ and a weight one mock Jacobi form $\psi^X_g = \sum_r H^X_{g,r} \tilde\theta_{m,r}$ for each $g\in G^X$. 
 In \cite{Cheng:2014zpa}, it was shown that there are 23 ways of splitting $\eg(K3)$ into two part: 
\be\label{um_splitEG}
{\bf EG}(K3;\tau,z) = \eg(X;\tau,z) + {\theta_1^2(\tau,z) \over 2\eta^6(\tau)} \left({1\over 2\pi i}{\partial\over \partial w} \psi^X(\tau,w)\right) \Big\lvert_{w=0}
\ee 
where $\psi^X =\psi^X_e$ is the optimal mock Jacobi form given in 
\S\ref{sec:umbral}, encoding the graded dimension of the umbral moonshine module, and the first term is the singularity elliptic genus discussed above. 
As a result, the splitting \eq{splitEGM24} of the elliptic genus can also be interpreted as coming from two contributions, one from the singularity elliptic genus corresponding to the root system $X=24A_1$, and one from the umbral moonshine function $\psi^{24A_1}$.
Different from the ${\cal N}=4$ character interpretation, this interpretation of the splitting is applicable for all 23 cases of umbral moonshine. In other words, \eq{um_splitEG} is really 23 equalities, corresponding to all 23 Niemeier lattices $N^X$.

This interpretation suggests a twined version of the function ${\bf EG}(K3;\tau,z)$, obtained by twining the right-hand side of the  equality  \eq{um_splitEG}:
\be\label{um_twining}
\phi_g^{X}(\tau,z) := \eg_g(X;\tau,z) + {\theta_1^2(\tau,z) \over \eta^6(\tau)} \left({1\over 2\pi i}{\partial\over \partial w} \psi^X_g(\tau,w)\right) \Big\lvert_{w=0}~,
\ee 
where $g$ can be any element of $G^X$. 
The first term on the right-hand side is the (holomorphic part of the) twined elliptic genus of the corresponding singularity sigma model. This can be computed explicitly, since we know the  $G^X$--action on the Niemeier root system $X$, which translates into an  $G^X$--action on the singularity CFT that preserves the  superconformal structure,  and leads to a definition of its twined elliptic genus. The second term, on the other hand, is directly provided by the McKay--Thompson series $\psi^X_g$ of umbral moonshine. 
It is easy to check that this definition of $\phi_g^X$ for $N=N^{24A_1}$ leads to the same weak Jacobi forms as in \eq{K3_EG_Fou} for the case of Mathieu moonshine. 
Just like in the case of Mathieu moonshine, 
it is then tempting to interpret all the  functions $\phi_g^{X}$ arising from umbral moonshine as in \eq{um_twining}, as the elliptic genus twined by certain symmetries of the $K3$ non-linear sigma model whose action preserves the ${\cal N}=4$ SCA. 

For some of the $g\in G^X$, such an interpretation is not availble, as we will discuss in \S\ref{sec:symK3}. 
In particular, only subgroups of $G^X$ which preserve a four-plane,  i.e. a  four-dimensional oriented subspace within $N^X\otimes_{\ZZ} \RR$, will acquire a role as a symmetry group of $K3$ sigma models. 
For later use, for a given Niemeier lattice $N=N^X$ with non-trivial root system, we will denote the set of Jacobi forms arising from \eq{um_twining} by 
\be\label{def:umset}
\Phi(N^X):= \{\phi_g^X~\lvert~g~{\text{is a four-plane preserving element of } G^X}\}~.\ee

\subsection{Conway moonshine and $K3$ elliptic genus}
\label{subsec:conwayK3}

Recall the construction of the twisted Conway module $V_{\text{tw}}^{s\natural}$ described in \S \ref{sec:conway_moonshine}. Given a fixed $n$-dimensional subspace in $\Lambda \otimes_{\ZZ} \CC$, there are different ways to build $U(1)$ currents by considering bilinear combinations of the $24$ fermions \cite{Cheng:2014owa,MR3465528}. 
In this theory, one can construct a $U(1)$ current $J$ of level 2 from fermions associated to a four-plane, i.e. a subspace of (real) dimension four.
Given such a four-plane $\Pi$, fixing the $U(1)$ current and the SVOA structure, breaks the symmetry of $V^{s\natural}_{\text{tw}}$ from $\Co_0$  to its subgroup $G_\Pi$ preserving $\Pi$.
Conversely, given a four-plane preserving $G_\Pi< \Co_0$, one can construct a $U(1)$ current $J$ such that $V^{s\natural}_{\text{tw}}$,  when equipped with a module structure for  $J$ and for the ${\cal N}=1$ superconformal algebra, has symmetry $G_\Pi$.

Interestingly, the $U(1)$-charged graded character of $V^{s\natural}_{\text{tw}}$ coincides with ${\bf EG}(K3)$ (up to a sign)\cite{MR3465528}. More generally, one can consider the  $U(1)$-graded character of the twisted Conway module
twined by a four-plane preserving element of $\Co_0$ (see \S \ref{sec:conway_moonshine} for notation): 
\begin{equation}
\f_g^\Lambda:=-\tr_{V^{s\natural}_{\text{tw}}}\left[\mathfrak{z}~\hat{g}~y^{J_0}q^{L_0-\frac{c}{24}}\right],
\end{equation}
where $J_0$ is the zero mode of the $U(1)$ current. 
Explicitly, it is given by
\begin{equation}
\label{eq:phi_g}
\begin{split}
\f_g^\Lambda(\tau,\zeta)&=\frac{1}{2}\left[\frac{\theta_3(\tau,\zeta)^2}{\theta_3(\tau,0)^2}\frac{\eta_{-g}(\tau/2)}{\eta_{-g}(\tau)}-\frac{\theta_4(\tau,\zeta)^2}{\theta_4(\tau,0)^2}\frac{\eta_{g}(\tau/2)}{\eta_{g}(\tau)}\right. \\
&~~~~~~~~~~\left.-\frac{\theta_2(\tau,\zeta)^2}{\theta_2(\tau,0)^2}C_{-g}\eta_{-g}(\tau)-\frac{\theta_1(\tau,\zeta)^2}{\eta(\tau)^6}D_g\eta_g(\tau)\right]~,
\end{split}
\end{equation}
where most special functions were defined in \eqref{eta_g_and_C_g}. The function $D_g$ is defined by
\begin{equation}
D_g:=\n\sideset{}{'}\prod_{i=1}^{12}\left(1-\l_i^{-1}\right)=\sideset{}{'}\prod_{i=1}^{12}\left(\n_i-\n_i^{-1}\right).
\end{equation}
where in $\sideset{}{'}\prod$ one skips the two pairs of eigenvalues associated with the fixed four-plane, for which $\l_i^{\pm1}=1$. 
Notice that $D_g$ is non-vanishing if and only if it fixes exactly a four-plane and not more. 
In the latter case, $D_g$ is determined up to a sign by the eigenvalues of $g$, since we are free to exchange what we call $\l_i$ and $\l_i^{-1}$. As a result, for exactly four-plane preserving elements there are in fact two choices of $\f_g^\Lambda$, depending on the choice of the sign of $D_{g}$, and we define
\begin{equation}
\begin{split}\label{eq:phi_epsilon}
\phi^\Lambda_{g,\e}(\tau,\zeta)&:=\frac{1}{2}\left[\frac{\theta_3(\tau,\zeta)^2}{\theta_3(\tau,0)^2}\frac{\eta_{-g}(\tau/2)}{\eta_{-g}(\tau)}-\frac{\theta_4(\tau,\zeta)^2}{\theta_4(\tau,0)^2}\frac{\eta_{g}(\tau/2)}{\eta_{g}(\tau)}\right. \\
&~~~~~~~~~~\left.-\frac{\theta_2(\tau,\zeta)^2}{\theta_2(\tau,0)^2}C_{-g}\eta_{-g}(\tau)-\e~\frac{\theta_1(\tau,\zeta)^2}{\eta(\tau)^6}\left|D_g\right|\eta_g(\tau)\right]~,
\end{split}
\end{equation}
where $\e=\pm1$ encodes the sign ambiguity of $D_g$.

In all cases, it was shown in \cite{MR3465528} that the $\f_{g,\e}^\Lambda$ are Jacobi forms of weight $0$ and index $1$, for some $\Gamma_g\subseteq \SL(2,\ZZ)$, for every $g\in\Co_0$ that fixes at least a four-plane. For future use, denote the set of all these functions by
\begin{equation}\label{def:conwayEG}
\Phi(\Lambda):=\left\lbrace \phi_{g,+}^\Lambda, \phi_{g,-}^\Lambda~|~g~\text{is a four-plane preserving element of} \Co_0\right\rbrace~.
\end{equation}

We will mention a few observations about this set of functions  and those  arising from umbral moonshine, the $\Phi(N^X)$ defined in \eq{def:umset}. 
First,  these sets of functions notably have a large overlap: quite often Conway and umbral moonshine give rise to the same weak Jacobi form, albeit  via very different ways. An important difference is that a weak Jacobi form $\phi^X_g$ can be defined for \underline{all} $g\in G^X$ via umbral moonshine \eq{um_twining}, while in the Conway case one has to restrict to four-plane preserving elements (in order to obtain a Jacobi form of index one). 
Second, the functions that only appear in $\Phi(\Lambda)$ tend to be realised as twined elliptic genera of $K3$ sigma models on the sub-loci in the moduli space where the model admits a description as $T^4$-orbifold. Note also that the geometry of $T^4$-orbifold connects nicely to the free field orbifold construction of the Conway module. In fact,  $V_{\text{tw}}^{s\natural}$ is isomorphic to the $\Z_2$ orbifold of  a certain $T^4$ model, when one (roughly speaking) ``forgets" about the distinction between left- and right-movers \cite{MR3465528,Taormina:2017zlm,Creutzig:2017fuk}. 
Third, the functions that only arise from umbral and not from Conway moonshine have the property that they must correspond to a symmetry which acts differently on the left- and right-movers of a $K3$ sigma model \cite{ChengMirandaC.N.2016KSTL}. 
We will discuss the connections between the moonshine symmetry groups and the stringy $K3$ symmetry groups in the next subsection.

\subsection{Symmetries of $K3$ string theory and moonshine}
\label{sec:symK3}

In the previous subsections we have seen a close mathematical relation between the $K3$ elliptic genus and the functions of umbral and Conway moonshine. It is therefore natural to  compare the moonshine groups and the symmetries of $K3$ sigma models. 
In fact, it was known that symmetries of $K3$ surfaces have a close relation to sporadic groups. 
In particular, a celebrated theorem by Mukai \cite{Mukai} established the role of $M_{23}$, one of the 26 sporadic groups and a subgroup of $M_{24}$, as the finite group that ``organises'' the symmetries of $K3$ surfaces. 
More precisely, in \cite{Mukai} it was shown that there is of a bijection between (isomorphism classes of) $M_{23}$ subgroups with at least five orbits and (isomorphism classes of) finite groups of $K3$ symplectomorphisms. A generalisation of this classical result to ``stringy $K3$ geometry'' was initiated by Gaberdiel, Hohenegger, and Volpato,   who in  \cite{Gaberdiel2011} classified the symmetry groups of all (non-singular) $K3$ sigma model using lattice techniques in a method closely following the simplified proof of the Mukai theorem by Kondo \cite{Kondo}. 
From the spacetime  (D-branes) point of view, the results of \cite{Gaberdiel2011} can be viewed as classifying symplectic autoequivalences (symmetries) of derived categories on $K3$ surfaces \cite{MR3586514}. See also \cite{Cheng:2015kha} for a related discussion on symmetries of appropriately defined moduli spaces relevant for curve counting on $K3$.
In \cite{ChengMirandaC.N.2016KSTL} the classification was extended to theories corresponding to the singular loci in the moduli space of $K3$ sigma models. This is necessary if one is interested in the full string theory  and not just the sigma models. 
Despite the fact that the type IIA worldsheet theory behaves badly along these loci, the full type IIA string theory is not only completely well-defined but  also possesses special physical relevance in connection to non-Abelian gauge symmetries \cite{Aspinwall:1995zi}.
The spacetime gauge group is enhanced from  $U(1)^{24}$ to some non-Abelian group at these loci, and the ADE type gauge group  is given by the ADE type singularity of the $K3$ surface \cite{Witten:1995ex,Aspinwall:1995zi}. The existence of such loci with enhanced gauge symmetries  in the moduli space, though not immediately manifest from the world-sheet analysis in type IIA, is clear from the  the dual description in terms of heterotic $T^4$ compactification.

To state the classification result, 
let us first review some general properties of sigma models on K3 (see \cite{Aspinwall:1996mn,Nahm:1999ps}). The moduli space of non-singular non-linear sigma models on $K3$ with $\mathcal{N}=(4,4)$ supersymmetry is given by an open subset in
\be \label{moduli_space}{\cal M}=(SO(4)\times O(20))\backslash O^+(4,20)/O^+(\Gamma^{4,20})\ ,
\ee where $(SO(4)\times O(20))\backslash O^+(4,20)$ is the Grassmannian of positive four-planes $\Pi$,  a four-dimensional oriented positive-definite subspace within $\RR^{4,20}\cong \Gamma^{4,20}\otimes_\ZZ \RR$, and $\Gamma^{4,20}$ is the even unimodular lattice with signature $(4,20)$.
In other words, choosing a point in the moduli space ${\cal M}$ is equivalent to choosing a four-plane inside $\Gamma^{4,20}\otimes_\ZZ \RR$. 
The complement in $\M$ of this open subset is the set of singular four-planes $\Pi$ (i.e. orthogonal to a root $v\in \Gamma^{4,20}$, $v^2=-2$) and correspond to certain singular limits of sigma models on K3. The whole space
${\cal M}$ is also the moduli space of type IIA string theory at a fixed finite value of $g_s$.  
In the singular cases, the full supersymmetry-preserving symmetries of the corresponding type IIA string theory contains a continuous non-Abelian gauge symmetry which we will mod out to obtain the discrete symmetry group $G_{\rm IIA}$. As a result, we have
\be
G_{\rm IIA}(\Pi) = {\rm Stab}(\Pi)/{\rm W} 
\ee
where ${\rm Stab}(\Pi)$ is the subgroup of $O^+(\Gamma^{4,20})$ which stabilises the four-plane $\Pi$ pointwise, and  ${\rm W}$ is the Weyl group generated by reflections with respect to all root vectors orthogonal to $\Pi$, if there are any.

In  \cite{ChengMirandaC.N.2016KSTL} (Corollary 4) it was then shown that the groups $G_{\rm IIA}$, realised somewhere in ${\cal M}$, are in bijection with the four-plane preserving subgroups of the twenty-four finite groups $G_N$ associated to the twenty-four even self-dual lattices of rank twenty-four (cf. \eq{def:niemeier_grp}), including the twenty-three umbral groups $G^X$ and  the Conway group $\Co_0$.
In particular, it means that generically the umbral group $G^X$, and in particular $G^{24A_1}\cong M_{24}$, {\underline{cannot}} be the symmetry group of any individual $K3$ sigma model as they are not  four-plane preserving subgroups of $\Co_0$. 

After classifying the symmetry groups as abstract groups, we would like to know how they act on the spectrum of supersymmetric states. 
This information is captured by the  elliptic genera twined by these symmetries, as defined in \eq{def:twined_elliptic_genus}.
In  \cite{ChengMirandaC.N.2016KSTL}, it was conjectured (among other finer conditions) that any twined $K3$ elliptic genus realised anywhere in the moduli space ${\M}$ coincides with at least one of the functions in one of the twenty-four sets of functions $\Phi(N)$ defined in \eq{def:umset} and in \eq{def:conwayEG}, where $N$ is either one of the twenty-three Niemeier lattices $N^X$ or the Leech lattice $\Lambda$.
Conversely, it was conjectured that {\it all} elements of $\Phi(N)$  play a role in capturing the symmetries of BPS states of $K3$. 
Strong evidence for the above conjecture was recently found in \cite{Paquette:2017gmb} by considering string theories closely related (by orbifolding) to type IIA string theory compactified on $K3 \times T^2$, and by requiring  the analyticity structure of the BPS-counting functions is compatible with the physics of wall-crossing. 
It is however not yet understood why umbral moonshine functions should have such a close relationship to $K3$ sigma model twined elliptic genera. See \S\ref{subsec:phys_Other directions} for more discussions.

\subsection{Beyond perturbative string theory}
\label{subsec:phys_Other directions}

The most important lesson from the previous subsection is that supersymmetry-preserving symmetries  of $K3$ CFTs are closely related to the symmetries of umbral moonshine, but fall short of explaining umbral moonshine:  many umbral groups simply cannot be realised as symmetries of $K3$ sigma model anywhere in the moduli space. 
In particular, while the conjectures discussed in \S\ref{sec:symK3} 
state that all umbral and Conway moonshine functions associated to four-plane preserving group elements play a role as the twining genera of $K3$ sigma models, and in fact capture them completely, the physical relevance of the umbral (including Mathieu) moonshine functions corresponding to group elements preserving only a two-plane remains unclear. 
There is at the moment no consensus regarding what the precise physical setup relevant for  umbral moonshine is. At the same time, it is precisely this necessity to go beyond perturbative string theory, and the possibility of having moonshine involved in non-perturbative setups such as stringy black holes or in a novel type of supersymmetric structure, that makes the possible connections between umbral moonshine and string theory so exciting. 
Here we briefly discuss a few possibilities that have been discussed to different degrees in the literature.

\myparagraph{Combining symmetries.}
The idea is to find a way to combine symmetries realised at different points in the moduli space and in this way generate a larger group, a ``meta'' group which is not restricted to a specific point in the modili space, which also contains two-plane preserving elements. 
This possibility was first raised as a question 
``Is it possible that these automorphism groups at isolated points in the moduli
space of $K3$ surface are enhanced to $M_{24}$ over the whole of moduli space when we consider the elliptic genus?''
in the original article \cite{Eguchi2010}. 
It is motivated by the fact that 
the elliptic genus receives only contributions from BPS states and is invariant across the moduli space. 
As a result, the protection of supersymmetry can lead to a ``meta'' object admitting the action of the ``meta symmetry group'' that we are after. 
Concrete steps towards realising this idea in the context of Kummer surfaces, i.e. $K3$ surfaces obtained as torus orbifolds $T^4/\ZZ_2$,  were taken in \cite{Taormina:2011rr,Taormina:2013mda,cheng2015landau} and further explored in \cite{GaberdielMatthiasR2017Mmas}. In particular, it was shown that  it is possible to combine the geometric symmetries of three specific  Kummer surfaces, given by the groups
\begin{equation}
G_0:=\ZZ_2^4\rtimes(\ZZ_2\times\ZZ_2)~,~~G_1:=\ZZ_2^4\rtimes A_4~,~~G_2:=\ZZ_2^4\rtimes S_3~, 
\end{equation}
into the so-called octad group $G=\ZZ_2^4\rtimes A_8$. This is a maximal subgroup of $M_{24}$ but too large to be the automorphism group of a single $K3$ sigma model. The work of  \cite{GaberdielMatthiasR2017Mmas} supplied evidence that  $G$ indeed acts on the twisted sector BPS states that survive after moving away from the Kummer loci in the moduli space. This suggests the possibility  that $G$, a maximal subgroup of $M_{24}$, indeed acts as an ``overarching" symmetry within the Kummer moduli space. 
Finally, this idea is also explored in 
 \cite{UMLG}  in the context of Landau--Ginzburg models flowing to $K3$ signa models in the IR. 
A challenging aspect of this approach is that it is technically very hard to follow a path in the moduli space to a generic point which does not correspond to a torus orbifold or admits a Landau--Ginzburg UV description. As a result it is not straightforward to combine symmetries which are realised in a generic $K3$ sigma model.

\myparagraph{Lower-dimensional compactifications.}
A second approach is to consider string compactifications  
where larger groups are realised at  given points in the moduli space as symmetry groups of the full string theory, and not just for the perturbative string states and not just the BPS sector. 
For theories with 16 supercharges, this is only possible for compactifications with less than six non-compact dimensions. This idea was first discussed in \cite{cheng2010k3}, where the implications of Mathieu moonshine for four-dimensional BPS black holes, arising in type II superstring theory compactified on $K3\times T^2$, were explored. It was futher developed in \cite{Persson:2013xpa}, and has led to nice new insights into string dualities \cite{Persson:2015jka}.

More recently, it was shown that in three there are points in 
the moduli space of string theory compactifications to three dimensions which admit the Niemeier groups as discrete symmetry groups \cite{kachru20173d}. 
 In the type IIA frame, these are given by compactifications on $K3\times T^3$. 
The action of these symmetry groups on the 1/2-BPS states of the theory has been analyzed \cite{kachru20173d}, and  
 it would be interesting to understand the action on the 1/4-BPS states. However, it is currently unclear how  functions related to the umbral moonshine McKay--Thompson series $\psi^X_g$ can be obtained in this physical setup. Recently, a further compactification on $K3\times T^4$ was also explored in \cite{Zimet:2018dev}.

\myparagraph{Heterotic $K3$ compactifications.}
The plausibility of the relevance of this setup lies in the fact that its moduli space contains the moduli space of ${\cal N}=(4,4)$ $K3$ CFT as a sub-locus. 
It is hence imaginable that moving away from the sub-loci will allow us to see more symmetries. 
At a technical level, we have less means to compute the spectrum due to smaller number of supersymmetries. 
This route has been somewhat explored in \cite{cheng2013mathieu,Harrison:2013bya}, but a lot remains to be done.

\myparagraph{Five-brane dynamics.}
The connection between umbral moonshine and the double scaled little string theory, describing the perturbative dynamics in the presence of NS five-branes in type IIB superstring theory, has been investigated in  \cite{Harvey:2013mda,Harvey:2014cva}. 
This idea is natural in that it incorporates new non-perturbative elements (NS fivebranes) into the physical setup, and that the double scaled little string theories admit an $ADE$ classification which also plays an important role in the construction of umbral moonshine.
It would be interesting to understand the finite group symmetries present in this setup.

\myparagraph{Relations to certain VOA.}
As we have seen in the previous subsections, umbral and Conway moonshine are related, as they are both closely related to $K3$ sigma models. 
In \cite{Cheng:2014owa}, a  variant of the Conway moonshine module is shown to exhibit an action of a variety of two-plane preserving subgroups of $\Co_0$, including $M_{23}$, and yields as twining genera a set of weak Jacobi forms of weight zero and index two. 
In addition, the mock modular forms which display $M_{23}$ representations appear to be very closely related to the mock modular forms which play a role in $M_{24}$ moonshine. It might be fruitful to gain a deeper physical understanding of this apparent connection and see if one can ``tweak'' the better understood Conway module in some ways to accommodate the umbral modules.

\myparagraph{Total unknown.}
As none of the above approaches has led to a definite answer so far, a possible scenario is that the connections we observed between $K3$ elliptic genera and umbral moonshine are just a coincidence and the physical context (if any) of umbral moonshine lies completely somewhere else.

Finding the correct physical setup in which the umbral moonshine modules can be constructed in a natural and uniform way is  an active research area and is currently the holy grail in the study of umbral moonshine.

\section{Other connections}
\label{sec:other_physics}

\myparagraph{MSW strings.}
Another physical setup which delivers examples of modular objects connected to finite groups is M5 branes wrapping divisors of Calabi--Yau three-folds \cite{Cheng:2017dlj}.  These give rise to effective strings, the so-called MSW strings, with (0,4) worldsheet supersymmetry \cite{Maldacena:1997de}. 
In \cite{Cheng:2017dlj}, it was  pointed out that the generalised elliptic genera of the MSW string theories are examples of skew-holomorphic Jacobi forms (cf. \S\ref{subsec:shjacobiforms}), a type of modular object playing a starring role in the family of moonshine generalising the Thompson moonshine, but whose role in physics has thus far not been highlighted. 
In particular, the generalised elliptic genera for one or two M5 branes wrapping the surfaces $\PP^2$, del Pezzo 8, and half-$K3$ were examined in \cite{Cheng:2017dlj}. 
In the first case, 
it was known that two M5 branes wrapping the surface $\PP^2$ leads to mock skew-holomorphic forms that are closely related to the generating function of the class number mock modular form \eq{hurwitz_mock}. This connection suggests the possibility that M5-branes on $\PP^2$ give a starting point from which we may pursue a geometric understanding of Mathieu moonshine for (rescaled) Hurwitz class numbers described in \S\ref{ch:moonshineweight32}.

For the other cases, one finds that a weight $3/2$ modular form governing the generalised elliptic genus of a single M5 brane wrapping the del Pezzo 8 and half-$K3$ surfaces is $f^{(1)}= E_4\eta^{-5}$. (See \S\ref{subsec:modular_forms} for the definition of $E_4$ and $\eta$. )  In \cite{Cheng:2017dlj} it was shown that $f^{(1)}$ is the graded superdimension of a supermodule for the Mathieu group $M_{12}$, and the corresponding graded characters $f^{(1)}_g$ for all $g\in M_{12}$ are explicitly given. It would be very interesting to have a geometric and string theoretic understanding of this $M_{12}$-supermodule.

\newpage
\addcontentsline{toc}{section}{References}

\bibliographystyle{utphys}
\bibliography{main_bib}

\end{document}